\input harvmac.tex
\input epsf.tex
\input amssym
\input ulem.sty

\input miniltx

\def\Gin@driver{dvips.def}
\input graphicx.sty

\resetatcatcode



\let\includefigures=\iftrue
\let\useblackboard=\iftrue
\newfam\black

\def\figin{\epsfcheck\figin}\def\figins{\epsfcheck\figins}
\def\epsfcheck{\ifx\epsfbox\UnDeFiNeD
\message{(NO epsf.tex, FIGURES WILL BE IGNORED)}
\gdef\figin##1{\vskip2in}\gdef\figins##1{\hskip.5in}
\else\message{(FIGURES WILL BE INCLUDED)}%
\gdef\figin##1{##1}\gdef\figins##1{##1}\fi}
\def\DefWarn#1{}
\def\figinsert{\goodbreak\midinsert}
\def\ifig#1#2#3{\DefWarn#1\xdef#1{fig.~\the\figno}
\writedef{#1\leftbracket fig.\noexpand~\the\figno} %
\figinsert\figin{\centerline{#3}}\medskip\centerline{\vbox{\baselineskip12pt
\advance\hsize by -1truein\noindent\footnotefont{\bf
Fig.~\the\figno:} #2}}
\bigskip\endinsert\global\advance\figno by1}


\includefigures
\message{If you do not have epsf.tex (to include figures),}
\message{change the option at the top of the tex file.}
\input epsf
\def\figin{\epsfcheck\figin}\def\figins{\epsfcheck\figins}
\def\epsfcheck{\ifx\epsfbox\UnDeFiNeD
\message{(NO epsf.tex, FIGURES WILL BE IGNORED)}
\gdef\figin##1{\vskip2in}\gdef\figins##1{\hskip.5in}
\else\message{(FIGURES WILL BE INCLUDED)}%
\gdef\figin##1{##1}\gdef\figins##1{##1}\fi}
\def\DefWarn#1{}
\def\figinsert{\goodbreak\midinsert}
\def\ifig#1#2#3{\DefWarn#1\xdef#1{fig.~\the\figno}
\writedef{#1\leftbracket fig.\noexpand~\the\figno}%
\figinsert\figin{\centerline{#3}}\medskip\centerline{\vbox{
\baselineskip12pt\advance\hsize by -1truein
\noindent\footnotefont{\bf Fig.~\the\figno:} #2}}
\endinsert\global\advance\figno by1}
\else
\def\ifig#1#2#3{\xdef#1{fig.~\the\figno}
\writedef{#1\leftbracket fig.\noexpand~\the\figno}%
\global\advance\figno by1} \fi

\def\figin{\epsfcheck\figin}\def\figins{\epsfcheck\figins}
\def\epsfcheck{\ifx\epsfbox\UnDeFiNeD
\message{(NO epsf.tex, FIGURES WILL BE IGNORED)}
\gdef\figin##1{\vskip2in}\gdef\figins##1{\hskip.5in}
\else\message{(FIGURES WILL BE INCLUDED)}%
\gdef\figin##1{##1}\gdef\figins##1{##1}\fi}
\def\DefWarn#1{}
\def\figinsert{\goodbreak\midinsert}
\def\ifig#1#2#3{\DefWarn#1\xdef#1{fig.~\the\figno}
\writedef{#1\leftbracket fig.\noexpand~\the\figno} %
\figinsert\figin{\centerline{#3}}\medskip\centerline{\vbox{\baselineskip12pt
\advance\hsize by -1truein\noindent\footnotefont{\bf
Fig.~\the\figno:} #2}}
\bigskip\endinsert\global\advance\figno by1}

\def \la {\langle}
\def \ra {\rangle}

\def \pa {\partial}

\def \eps {\epsilon}
\def\vev#1{\left\langle #1 \right\rangle}
\def\OO{{\cal OO}}

\catcode`\@=11
\def\slash#1{\mathord{\mathpalette\c@ncel{#1}}}
\overfullrule=0pt
\def\AA{{\cal A}}

\def\CC{{\cal C}}
\def\DD{{\cal D}}

\def\LL{{\cal L}}

\def\OO{{\cal O}}

\def\WW{{\cal W}}

\def\eps{\epsilon}

\def\underrel#1\over#2{\mathrel{\mathop{\kern\z@#1}\limits_{#2}}}

\catcode`\@=12


\def\vev#1{\left\langle #1 \right\rangle}

\def \sinh{{\rm sinh}}
\def \cosh{{\rm cosh}}

\def\csch{{\rm csch}}


\def\p{{\partial}}

\def\tt{{\tilde t}}

\def\LL{{\cal L}}

\def\bx{{\bar x}}
\def\DD{{\cal D}}
\def\bp{{\bar p}}
\def\bz{{\bar z}}

\def\bw{{\bar w}}

\def\be{{\bar e}}
\def\eps{{\epsilon}}

\lref\CamanhoAPA{
  X.~O.~Camanho, J.~D.~Edelstein, J.~Maldacena and A.~Zhiboedov,
  ``Causality Constraints on Corrections to the Graviton Three-Point Coupling,''
JHEP {\bf 1602}, 020 (2016).
[arXiv:1407.5597 [hep-th]].
}

\lref\GiddingsGJ{
  S.~B.~Giddings and R.~A.~Porto,
  ``The Gravitational S-matrix,''
Phys.\ Rev.\ D {\bf 81}, 025002 (2010).
[arXiv:0908.0004 [hep-th]].
}

\lref\PenedonesNS{
  J.~Penedones,
  ``High Energy Scattering in the AdS/CFT Correspondence,''
[arXiv:0712.0802 [hep-th]].
}

\lref\PenedonesVOO{
  J.~Penedones,
  ``TASI lectures on AdS/CFT,''
[arXiv:1608.04948 [hep-th]].
}

\lref\DiFrancescoNK{
  P.~Di Francesco, P.~Mathieu and D.~Senechal,
  ``Conformal Field Theory,''
}

\lref\PenedonesUE{
  J.~Penedones,
  ``Writing CFT correlation functions as AdS scattering amplitudes,''
JHEP {\bf 1103}, 025 (2011).
[arXiv:1011.1485 [hep-th]].
}

\lref\KabatPZ{
  D.~N.~Kabat,
Comments Nucl.\ Part.\ Phys.\  {\bf 20}, 325 (1992).
[hep-th/9204103].
}

\lref\VenezianoER{
  G.~Veneziano,
JHEP {\bf 0411}, 001 (2004).
[hep-th/0410166].
}

\lref\CostaMG{
  M.~S.~Costa, J.~Penedones, D.~Poland and S.~Rychkov,
  ``Spinning Conformal Correlators,''
JHEP {\bf 1111}, 071 (2011).
[arXiv:1107.3554 [hep-th]].
}

\lref\ZhiboedovBM{
  A.~Zhiboedov,
  ``A note on three-point functions of conserved currents,''
[arXiv:1206.6370 [hep-th]].
}

\lref\OsbornCR{
  H.~Osborn and A.~C.~Petkou,
  ``Implications of conformal invariance in field theories for general dimensions,''
Annals Phys.\  {\bf 231}, 311 (1994).
[hep-th/9307010].
}

\lref\LiITL{
  D.~Li, D.~Meltzer and D.~Poland,
  ``Conformal Collider Physics from the Lightcone Bootstrap,''
JHEP {\bf 1602}, 143 (2016).
[arXiv:1511.08025 [hep-th]].
}

\lref\CostaDW{
  M.~S.~Costa, J.~Penedones, D.~Poland and S.~Rychkov,
  ``Spinning Conformal Blocks,''
JHEP {\bf 1111}, 154 (2011).
[arXiv:1109.6321 [hep-th]].
}

\lref\MaldacenaWAA{
  J.~Maldacena, S.~H.~Shenker and D.~Stanford,
  ``A bound on chaos,''
JHEP {\bf 1608}, 106 (2016).
[arXiv:1503.01409 [hep-th]].
}

\lref\MaldacenaIUA{
  J.~Maldacena, D.~Simmons-Duffin and A.~Zhiboedov,
  ``Looking for a bulk point,''
[arXiv:1509.03612 [hep-th]].
}

\lref\LiuTY{
  H.~Liu and A.~A.~Tseytlin,
  ``On four point functions in the CFT / AdS correspondence,''
Phys.\ Rev.\ D {\bf 59}, 086002 (1999).
[hep-th/9807097].
}

\lref\DolanDV{
  F.~A.~Dolan and H.~Osborn,
  ``Conformal Partial Waves: Further Mathematical Results,''
[arXiv:1108.6194 [hep-th]].
}

\lref\HijanoZSA{
  E.~Hijano, P.~Kraus, E.~Perlmutter and R.~Snively,
  ``Witten Diagrams Revisited: The AdS Geometry of Conformal Blocks,''
JHEP {\bf 1601}, 146 (2016).
[arXiv:1508.00501 [hep-th]].
}

\lref\SimmonsDuffinUY{
  D.~Simmons-Duffin,
  ``Projectors, Shadows, and Conformal Blocks,''
JHEP {\bf 1404}, 146 (2014).
[arXiv:1204.3894 [hep-th]].
}

\lref\PenedonesNS{
  J.~Penedones,
  ``High Energy Scattering in the AdS/CFT Correspondence,''
[arXiv:0712.0802 [hep-th]].
}

\lref\KomargodskiGCI{
  Z.~Komargodski, M.~Kulaxizi, A.~Parnachev and A.~Zhiboedov,
  ``Conformal Field Theories and Deep Inelastic Scattering,''
[arXiv:1601.05453 [hep-th]].
}

\lref\GribovNW{
  V.~N.~Gribov,
  ``The theory of complex angular momenta: Gribov lectures on theoretical physics,''
}

\lref\PaulosFAP{
  M.~F.~Paulos, J.~Penedones, J.~Toledo, B.~C.~van Rees and P.~Vieira,
  ``The S-matrix Bootstrap I: QFT in AdS,''
[arXiv:1607.06109 [hep-th]].
}


\lref\AmatiUF{
  D.~Amati, M.~Ciafaloni and G.~Veneziano,
  ``Classical and Quantum Gravity Effects from Planckian Energy Superstring Collisions,''
Int.\ J.\ Mod.\ Phys.\ A {\bf 3}, 1615 (1988)..
}

\lref\AmatiTN{
  D.~Amati, M.~Ciafaloni and G.~Veneziano,
  ``Can Space-Time Be Probed Below the String Size?,''
Phys.\ Lett.\ B {\bf 216}, 41 (1989)..
}

\lref\GiddingsBW{
  S.~B.~Giddings, D.~J.~Gross and A.~Maharana,
  ``Gravitational effects in ultrahigh-energy string scattering,''
Phys.\ Rev.\ D {\bf 77}, 046001 (2008).
[arXiv:0705.1816 [hep-th]].
}

\lref\GaoGA{
  S.~Gao and R.~M.~Wald,
  ``Theorems on gravitational time delay and related issues,''
Class.\ Quant.\ Grav.\  {\bf 17}, 4999 (2000).
[gr-qc/0007021].
}

\lref\ShapiroUW{
  I.~I.~Shapiro,
  ``Fourth Test of General Relativity,''
Phys.\ Rev.\ Lett.\  {\bf 13}, 789 (1964)..
}

\lref\CiafaloniESA{
  M.~Ciafaloni and D.~Colferai,
  ``Rescattering corrections and self-consistent metric in Planckian scattering,''
JHEP {\bf 1410}, 85 (2014).
[arXiv:1406.6540 [hep-th]].
}

\lref\DodelsonUOA{
  M.~Dodelson and E.~Silverstein,
  ``Longitudinal nonlocality in the string S-matrix,''
[arXiv:1504.05537 [hep-th]].
}

\lref\CornalbaXK{
  L.~Cornalba, M.~S.~Costa, J.~Penedones and R.~Schiappa,
  ``Eikonal Approximation in AdS/CFT: From Shock Waves to Four-Point Functions,''
JHEP {\bf 0708}, 019 (2007).
[hep-th/0611122].
}

\lref\CornalbaXM{
  L.~Cornalba, M.~S.~Costa, J.~Penedones and R.~Schiappa,
  ``Eikonal Approximation in AdS/CFT: Conformal Partial Waves and Finite N Four-Point Functions,''
Nucl.\ Phys.\ B {\bf 767}, 327 (2007).
[hep-th/0611123].
}

\lref\CornalbaZB{
  L.~Cornalba, M.~S.~Costa and J.~Penedones,
  ``Eikonal approximation in AdS/CFT: Resumming the gravitational loop expansion,''
JHEP {\bf 0709}, 037 (2007).
[arXiv:0707.0120 [hep-th]].
}

\lref\GaryAE{
  M.~Gary, S.~B.~Giddings and J.~Penedones,
  ``Local bulk S-matrix elements and CFT singularities,''
Phys.\ Rev.\ D {\bf 80}, 085005 (2009).
[arXiv:0903.4437 [hep-th]].
}

\lref\CornalbaAX{
  L.~Cornalba, M.~S.~Costa and J.~Penedones,
  ``Deep Inelastic Scattering in Conformal QCD,''
JHEP {\bf 1003}, 133 (2010).
[arXiv:0911.0043 [hep-th]].

\lref\HartmanLFA{
  T.~Hartman, S.~Jain and S.~Kundu,
  ``Causality Constraints in Conformal Field Theory,''
JHEP {\bf 1605}, 099 (2016).
[arXiv:1509.00014 [hep-th]].
}}

\lref\HartmanDXC{
  T.~Hartman, S.~Jain and S.~Kundu,
  ``A New Spin on Causality Constraints,''
JHEP {\bf 1610}, 141 (2016).
[arXiv:1601.07904 [hep-th]].
}

\lref\HofmanAWC{
  D.~M.~Hofman, D.~Li, D.~Meltzer, D.~Poland and F.~Rejon-Barrera,
  ``A Proof of the Conformal Collider Bounds,''
JHEP {\bf 1606}, 111 (2016).
[arXiv:1603.03771 [hep-th]].
}

\lref\FaulknerMZT{
  T.~Faulkner, R.~G.~Leigh, O.~Parrikar and H.~Wang,
  ``Modular Hamiltonians for Deformed Half-Spaces and the Averaged Null Energy Condition,''
JHEP {\bf 1609}, 038 (2016).
[arXiv:1605.08072 [hep-th]].
}

\lref\HartmanLGU{
  T.~Hartman, S.~Kundu and A.~Tajdini,
  ``Averaged Null Energy Condition from Causality,''
[arXiv:1610.05308 [hep-th]].
}

\lref\FitzpatrickYX{
  A.~L.~Fitzpatrick, J.~Kaplan, D.~Poland and D.~Simmons-Duffin,
  ``The Analytic Bootstrap and AdS Superhorizon Locality,''
JHEP {\bf 1312}, 004 (2013).
[arXiv:1212.3616 [hep-th]].
}

\lref\KomargodskiEK{
  Z.~Komargodski and A.~Zhiboedov,
  ``Convexity and Liberation at Large Spin,''
JHEP {\bf 1311}, 140 (2013).
[arXiv:1212.4103 [hep-th]].
}

\lref\HofmanAR{
  D.~M.~Hofman and J.~Maldacena,
  ``Conformal collider physics: Energy and charge correlations,''
JHEP {\bf 0805}, 012 (2008).
[arXiv:0803.1467 [hep-th]].
}

\lref\AfkhamiJeddiNTF{
  N.~Afkhami-Jeddi, T.~Hartman, S.~Kundu and A.~Tajdini,
  ``Einstein gravity 3-point functions from conformal field theory,''
[arXiv:1610.09378 [hep-th]].
}

\lref\CaronHuotVEP{
  S.~Caron-Huot,
  ``Analyticity in Spin in Conformal Theories,''
[arXiv:1703.00278 [hep-th]].
}

\lref\HeemskerkPN{
  I.~Heemskerk, J.~Penedones, J.~Polchinski and J.~Sully,
  ``Holography from Conformal Field Theory,''
JHEP {\bf 0910}, 079 (2009).
[arXiv:0907.0151 [hep-th]].
}

\lref\CornalbaFS{
  L.~Cornalba,
  ``Eikonal methods in AdS/CFT: Regge theory and multi-reggeon exchange,''
[arXiv:0710.5480 [hep-th]].
}

\lref\CostaCB{
  M.~S.~Costa, V.~Goncalves and J.~Penedones,
  ``Conformal Regge theory,''
JHEP {\bf 1212}, 091 (2012).
[arXiv:1209.4355 [hep-th]].
}

\lref\BrowerEA{
  R.~C.~Brower, J.~Polchinski, M.~J.~Strassler and C.~I.~Tan,
  ``The Pomeron and gauge/string duality,''
JHEP {\bf 0712}, 005 (2007).
[hep-th/0603115].
}

\lref\HartmanLFA{
  T.~Hartman, S.~Jain and S.~Kundu,
  ``Causality Constraints in Conformal Field Theory,''
JHEP {\bf 1605}, 099 (2016).
[arXiv:1509.00014 [hep-th]].
}

\lref\AharonyBM{
  O.~Aharony, S.~Minwalla and T.~Wiseman,
  ``Plasma-balls in large N gauge theories and localized black holes,''
Class.\ Quant.\ Grav.\  {\bf 23}, 2171 (2006).
[hep-th/0507219].
}

\lref\KavirajCXA{
  A.~Kaviraj, K.~Sen and A.~Sinha,
  ``Analytic bootstrap at large spin,''
JHEP {\bf 1511}, 083 (2015).
[arXiv:1502.01437 [hep-th]].
}

\lref\GiddingsXY{
  S.~B.~Giddings and V.~S.~Rychkov,
  ``Black holes from colliding wavepackets,''
Phys.\ Rev.\ D {\bf 70}, 104026 (2004).
[hep-th/0409131].
}

\lref\RattazziPE{
  R.~Rattazzi, V.~S.~Rychkov, E.~Tonni and A.~Vichi,
  ``Bounding scalar operator dimensions in 4D CFT,''
JHEP {\bf 0812}, 031 (2008).
[arXiv:0807.0004 [hep-th]].
}

\lref\SimmonsDuffinWLQ{
  D.~Simmons-Duffin,
  ``The Lightcone Bootstrap and the Spectrum of the 3d Ising CFT,''
JHEP {\bf 1703}, 086 (2017).
[arXiv:1612.08471 [hep-th]].
}

\lref\CornalbaXK{
  L.~Cornalba, M.~S.~Costa, J.~Penedones and R.~Schiappa,
  ``Eikonal Approximation in AdS/CFT: From Shock Waves to Four-Point Functions,''
JHEP {\bf 0708}, 019 (2007).
[hep-th/0611122].
}

\lref\CornalbaXM{
  L.~Cornalba, M.~S.~Costa, J.~Penedones and R.~Schiappa,
  ``Eikonal Approximation in AdS/CFT: Conformal Partial Waves and Finite N Four-Point Functions,''
Nucl.\ Phys.\ B {\bf 767}, 327 (2007).
[hep-th/0611123].
}

\lref\CornalbaZB{
  L.~Cornalba, M.~S.~Costa and J.~Penedones,
  ``Eikonal approximation in AdS/CFT: Resumming the gravitational loop expansion,''
JHEP {\bf 0709}, 037 (2007).
[arXiv:0707.0120 [hep-th]].
}

\lref\CornalbaQF{
  L.~Cornalba, M.~S.~Costa and J.~Penedones,
  ``Eikonal Methods in AdS/CFT: BFKL Pomeron at Weak Coupling,''
JHEP {\bf 0806}, 048 (2008).
[arXiv:0801.3002 [hep-th]].
}

\lref\FitzpatrickDM{
  A.~L.~Fitzpatrick and J.~Kaplan,
  ``Unitarity and the Holographic S-Matrix,''
JHEP {\bf 1210}, 032 (2012).
[arXiv:1112.4845 [hep-th]].
}

\lref\CasiniBF{
  H.~Casini,
  ``Wedge reflection positivity,''
J.\ Phys.\ A {\bf 44}, 435202 (2011).
[arXiv:1009.3832 [hep-th]].
}

\lref\StreaterVI{
  R.~F.~Streater and A.~S.~Wightman,
  ``PCT, spin and statistics, and all that,''
Princeton, USA: Princeton Univ. Pr. (2000) 207 p..
}

\lref\GrinsteinQK{
  B.~Grinstein, K.~A.~Intriligator and I.~Z.~Rothstein,
  ``Comments on Unparticles,''
Phys.\ Lett.\ B {\bf 662}, 367 (2008).
[arXiv:0801.1140 [hep-ph]].
}

\lref\RychkovIQZ{
  S.~Rychkov,
  ``EPFL Lectures on Conformal Field Theory in D $\geq$ 3 Dimensions,''
[arXiv:1601.05000 [hep-th]].
}

\lref\SimmonsDuffinGJK{
  D.~Simmons-Duffin,
  ``TASI Lectures on the Conformal Bootstrap,''
[arXiv:1602.07982 [hep-th]].
}

\lref\GiddingsXY{
  S.~B.~Giddings and V.~S.~Rychkov,
  ``Black holes from colliding wavepackets,''
Phys.\ Rev.\ D {\bf 70}, 104026 (2004).
[hep-th/0409131].
}

\lref\LuscherEZ{
  M.~Luscher and G.~Mack,
  ``Global Conformal Invariance in Quantum Field Theory,''
Commun.\ Math.\ Phys.\  {\bf 41}, 203 (1975)..
}

\lref\AldayGDE{
  L.~F.~Alday, A.~Bissi and E.~Perlmutter,
  ``Holographic Reconstruction of AdS Exchanges from Crossing Symmetry,''
[arXiv:1705.02318 [hep-th]].
}

\lref\AldayHTQ{
  L.~F.~Alday and A.~Bissi,
  ``Unitarity and positivity constraints for CFT at large central charge,''
[arXiv:1606.09593 [hep-th]].
}

\Title{
\vbox{\baselineskip8pt
}}
{\vbox{
\centerline{Bulk Phase Shift, CFT Regge Limit}
\vskip.1in
\centerline{and Einstein Gravity}
}}

\vskip.1in
 \centerline{
Manuela Kulaxizi,$^1$ Andrei Parnachev,$^{1}$ Alexander Zhiboedov$^2$} \vskip.1in 
\centerline{\it $^1$
School of Mathematics, Trinity College Dublin, Dublin 2, Ireland}
\centerline{\it $^2$
Department of Physics, Harvard University, Cambridge, MA 02138, USA}

\vskip.7in \centerline{\bf Abstract}{ 
\vskip.2in
The bulk phase shift, related to a CFT four-point function,  describes two-to-two scattering at fixed impact parameter in the dual AdS spacetime. We describe its properties for a generic CFT and then focus on large $N$ CFTs with classical bulk duals.
We compute the bulk phase shift for vector operators using Regge theory.
We use causality and unitarity to put bounds on the bulk phase shift. 
The resulting constraints bound three-point functions of two vector operators and the stress tensor in terms of the gap of the theory. Similar bounds should hold for any spinning operator in a CFT. 
Holographically this implies that in a classical gravitational theory any non-minimal coupling to the graviton, as well as any other particle with spin greater than or equal to two, is suppressed by the mass of higher spin particles.
}
 
\Date{May 2017}

\listtoc\writetoc
\vskip 1.5in \noindent

\newsec{Introduction}

\noindent A useful way to chart distinct physical regimes in a gravitational theory is to study scattering as a function of energy $S$ and impact parameter $L$ \refs{\AmatiUF\AmatiTN-\GiddingsBW}, see also figure 1 in \GiddingsGJ  . In flat space the relevant quantity, namely the phase shift $\delta(S,L)$, is related to the Fourier transform of the four-point  scattering amplitude $A(s,t=-\vec q^2)$ with respect to the transferred momentum $\vec q$, see also \refs{\CiafaloniESA,\DodelsonUOA}. In AdS the phase shift is given by the convolution of the four-point function of local operators with proper external wave functions \refs{\CornalbaXK\CornalbaXM\CornalbaZB\GaryAE-\CornalbaAX}. 
In this paper we adopt the approach of \CornalbaAX\ where the phase shift $\delta(S,L)$ was shown to be given by a certain double Fourier transform of the Lorentzian four-point function to be discussed in detail below.

One motivation to study the phase shift comes from its relation to causality. Indeed, in AdS/CFT boundary observers can exchange signals either through the bulk or along the boundary. Microscopic CFT causality guarantees that the signals sent along the boundary are the fastest. In Einstein theory an analogous statement in the bulk is guaranteed due to the theorem of Gao and Wald \GaoGA. More generally, signals in gravitational theories are delayed with respect to the asymptotic causal structure \ShapiroUW. This fact could be used to constrain effective theories of gravity \CamanhoAPA. 

Another motivation is to have a more direct connection between the physics of bulk scattering and CFT correlators. This connection could be potentially useful in both directions. One can hope to use bulk intuition to get some insight into the CFT dynamics and conversely use non-perturbative CFT methods to say something nontrivial about physics in the bulk.

In this paper we will be interested in regimes where the interaction is weak $\delta(S,L) \ll 1$. The behavior of $\delta(S,L)$ at large energies $S \gg 1$ describes propagation of energetic particles through AdS and as such should respect the asymptotic causal structure. For interacting CFTs in $d>2$ at large distances the interaction is generically controlled by the graviton exchange \refs{ \FitzpatrickYX,\KomargodskiEK}.\foot{In free theories the dependence on impact parameter is trivial, namely at high energies it is just a function of $S$. This is an extreme example where locality is absent even on super-AdS scales.}

 One can show that this implies the following behavior for the phase shift:
\eqn\largedistance{
\delta(S,L) =c S e^{-(d-1) L} + ... \ , ~~~ L \gg 1
}
which is simply the large impact parameter $L$ asymptotic of the graviton propagator in the impact parameter space $H^{d-1}$.\foot{For CFTs with scalar operators of dimensions ${d - 2 \over 2} < \Delta < d-2$ this might not be true. To avoid this we can imagine considering a target that does not couple to such light scalars. 
Also note that in the language of double twist operators ${\cal O}(\pa^2)^n \pa_{\mu_1} ... \pa_{\mu_s} O$ \largedistance\ corresponds to $s \gg n \gg 1$ \refs{\PenedonesNS,\KavirajCXA,\AldayHTQ}.} The coefficient $c$ is fixed in terms of the three-point coupling between the propagating probe and the graviton. Asymptotic causality implies that $c \geq 0$, namely that gravity is attractive. When applied to scalar particles this is automatically satisfied due to unitarity. When applied to particles with spin this condition implies the
positivity of the energy flux  \HofmanAR . Recently this statement was proved using very different-looking methods \refs{\HartmanLFA\HartmanDXC\HofmanAWC\KomargodskiGCI\FaulknerMZT-\HartmanLGU}.

For special theories the simple behavior \largedistance\ of the phase shift persists for much smaller impact parameters $L> L_*$. These are theories with a large central charge $\la T_{\mu \nu} T_{\rho \sigma}\ra \sim c_T \gg 1$ and a large gap $\Delta_{gap} \gg 1$ in the spectrum of higher spin operators \HeemskerkPN .
We are interested precisely in  this regime, namely large $S$ and $L$ such that the graviton exchange dominates
\eqn\exchangedom{
\delta(S,L) = S \delta(L) + ... , ~~~ L \geq L_* \ .
}
This regime was considered in \CamanhoAPA\ where it was shown that positivity of $\delta(L) \geq 0$ for $L \geq L_*$ implies stronger bounds on the coupling of matter with spin to the graviton. In the limit $L_* \to 0$ the only allowed coupling is the one of Einstein gravity. Moreover, \CamanhoAPA\ argued that classically only higher spin particles could fix the problem. The same problem was addressed using CFT methods in \AfkhamiJeddiNTF\ under the assumption that double trace operators could be neglected. Here we find that they are non-negligible in the Regge limit but drop out when we consider scattering in the impact parameter space.

There are two alternative ways in which the phase shift $\delta(S, L)$ could be computed in a CFT with a holographic dual: either using the shock wave geometry, or using the four-point correlation function. 
In the AdS analysis of \CamanhoAPA\ only the shock wave method was used; this was a computation  done entirely on the gravity side.
The purpose of the present paper is to provide a CFT computation of the phase shift, which is similar to the scattering amplitude computation  in the case of flat space. 
A purely CFT description of the phase shift will facilitate computing corrections and will advance our understanding of the  implications of \CamanhoAPA\ for  generic CFTs. In particular, using CFT methods we can compute the correction to the phase shift due to the tower of higher spin particles. This is achieved using conformal Regge theory \CostaCB\ and does not have a simple bulk counterpart. 

The rest of this paper is organised as follows.

In section 2 we review the construction of \CornalbaAX\ for the phase shift $\delta(S,L)$. To define it, it is convenient to describe the wave function of each operator in its own Poincare patch. The corresponding conformal transformation is standard in the context of Regge theory and appeared in several works, for example \refs{\CornalbaFS,\HofmanAR,\CornalbaAX}.

In section 3 we consider a simple example of the Witten exchange diagram and discuss its different limits. In particular we describe in detail the structure of the correlator before and after the Fourier transform and the role played by the double trace operators. The result for the phase shift in this case is what one gets from the shock wave computation in the bulk \CamanhoAPA .

In section 4 we use Conformal Regge Theory to compute the high energy limit of the phase shift $\delta(S,L)$, which is related to the Regge limit of the four point function in CFT \CostaCB. 
We consider a four-point function which involves spin one operators $J_{\mu}$. We give an expression for the phase shift, assuming that the Regge limit is dominated by a Regge pole.\foot{This is believed to be the case in large $N$ CFTs.}
Next, we compute the contribution of the stress tensor to the phase shift;  the result 
is effectively the same as the spinning Shapiro time delay of \CamanhoAPA\ that was previously computed using shock waves in AdS. 
This time, however, we can compute the correction to it due to the tower of higher spin particles and explicitly see how the gravity computation breaks down.

In section 5 we study constraints on the bulk phase shift and leading Regge trajectory arising from unitarity and causality. In the infinite gap limit we obtain the constraints of \CamanhoAPA. Similarly to previous work, we derive a parametric bound on $\la J_{\mu} T_{\rho \sigma} J_{\nu} \ra$ when the gap is finite.  This is related to the fact that the minimal impact parameter for which one can run the argument to constrain the three-point coupling depends on the energy $S_*$ for which our approximation is reliable, {\it i.e.,} ${L_*^2 \over \Delta_{gap}^2} \sim {\log S_* \over \Delta_{gap}^2}$.
Considering impact parameters of order ${1 \over \Delta_{gap}}$ requires $S \gg \Delta_{gap}^2$. In this way we get\foot{If we were to use the eikonal methods we would get $\log S_* \sim \log c_T$, where $c_T$ is the central charge of the CFT.}
\eqn\boundA{
 |\beta_4| \lesssim {\log \Delta_{gap} \over \Delta_{gap}^2} \ ,
}
where $\beta_4$ is the combination of the coefficients in the three-point function $\la J_{\mu} T_{\rho \sigma} J_{\nu} \ra$ which corresponds to the non-minimal coupling to the graviton. We also obtain the corresponding expressions for non-conserved, spin one operators. The bound for any spinning operator should be similar.

\newsec{Bulk Phase Shift}

In this section we introduce  the bulk phase shift. It describes $2 \to 2$ scattering at fixed impact parameter in AdS. In Minkowski space the phase shift $\delta(s, \vec b)$ is given by the Fourier transform of the scattering amplitude ${1 \over s}A(s,t=- \vec q^2)$ with respect to the transferred momentum, $\vec q$. Here both the transfer momentum, $\vec q$, and the impact parameter $\vec b$, belong to the transverse space $R^{d-2}$. We describe the analogous AdS kinematics below.

Let us first review $AdS_{d+1}$ space in the embedding coordinates:
\eqn\ads{
- (Z^{-1})^2 - (Z^0)^2 + \sum_{i = 1}^{d} (Z^i)^2 = - 1. 
}
The conformal group generators are simply the $SO(2,d)$ rotations $\hat M_{AB} =- i  \left( Z_A \pa_{Z^B} - Z_B \pa_{Z^A} \right)$, while, $\delta Z_A = i \hat M_{A B} Z^B$. The embedding coordinates of \ads\ are related to the usual, AdS, global coordinates by
\eqn\globalco{
Z^{0} + i Z^{-1} = e^{i \tau} \cosh \rho,~~~ Z^i = n^i \sinh \rho , ~~~ \vec n^2 = 1, ~~~ i = 1, ... , d .
}
To approach the boundary we take $\rho \to \infty$. Then $(\tau, \vec n)$ label points in the boundary CFT. In the embedding space points on the boundary correspond to $Z^2 = 0$ together with the identification $Z \sim \lambda Z$.

\ifig\kinematics{Scattering kinematics. The red dots mark origins of the four Poincare patches. Each operator is inserted in the vicinity of the corresponding origin with the wavefunction which is a plane wave. The scattering occurrs in $H^{d-1}$ which is marked by a blue line. It stands for the intersection of two Poincare horizons which are depicted in shaded gray. Each Poincare horizon is generated by future or past null geodesics emanated from points $x_i$.} {\epsfxsize1.5in\epsfbox{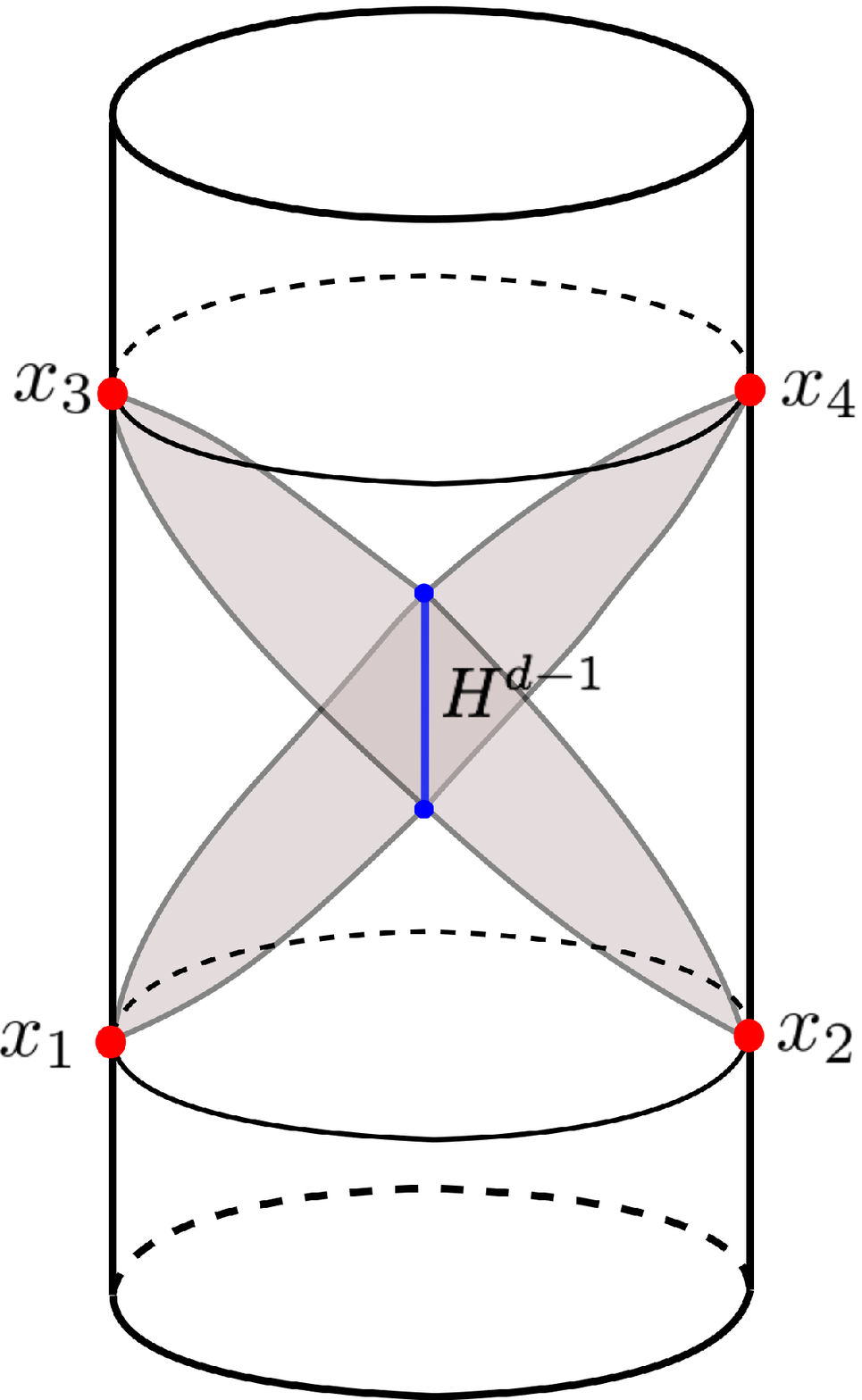}}

We imagine a situation where two excitations start at $\tau = - {\pi \over 2}$ on the antipodal points on the sphere, collide at $\tau = 0$,\foot{Notice that this statement does not assume bulk locality. The fact that $\tau =0$ is the first instance when excitations can interact follows from boundary causality \MaldacenaIUA.} and reach the boundary at $\tau = {\pi \over 2}$ at the antipodal points on the sphere see \kinematics.

More precisely, the four points on the boundary are 
\eqn\fourpoints{\eqalign{
{\rm Point} ~ 1: \ \ \ Z^{-1} &= - 1 ,~ Z^1 = - 1 , \cr
{\rm Point} ~ 2: \ \ \ Z^{-1} &= - 1 ,~ Z^1 = 1 , \cr
{\rm Point} ~ 3: \ \ \ Z^{-1} &= 1 ,~ Z^1 = -1 , \cr
{\rm Point} ~ 4: \ \ \ Z^{-1} &= 1 ,~ Z^1 = 1 , \cr
}}
where we only wrote the non-zero $Z^A$ components. For the points above, $Z^2 = 0$, since all of them belong to the AdS boundary.

It is useful to introduce different Poincare coordinates around the boundary points above. It helps to understand symmetries of the problem. Choosing a Poincare patch for each point, corresponds to splitting $R^{2,d}$ into $M^2 \times M^d$ \refs{\CornalbaAX,\PenedonesNS} , where $M^2$ is specified by a pair of orthonormal null vectors $K^A$ and $\bar K^A$, such that $K^2 = \bar K^2 = 0$ and $- 2 K.\bar K = 1$, then we have
\eqn\poincarepatch{
Z^A = \bar K^A + y^2 K^A + y^A
}
where $y^A$ are the usual coordinates in Minkowski space. Again for \poincarepatch\ we have  $Z^2 =0$ as it should be. By contracting both sides with $K_A$ we get the gauge condition
\eqn\gaugecondition{
- 2 Z . K = 1.
}
If we are to introduce the AdS Poincare coordinates, then $Z.K =0$ corresponds to the Poincare horizon.

Symmetry transformations in a given Poincare patch are related to the transformations in global AdS as follows \refs{\PenedonesNS , \DiFrancescoNK}
\eqn\symmetries{\eqalign{
M_{\mu \nu} &= \hat M_{\mu \nu} , \cr
P_{\mu} =- 2  K^A \hat M_{A \mu} &, ~~~ K_{\nu}  =- 2 \bar K^A \hat M_{A \nu}, \cr
D &= 2 K^{A} \bar K^{B} \hat M_{A B} \ .
}}
Here the indices $\mu , \nu$ span $M^d$. 

We are now ready to describe the Poincare patches associated to points \fourpoints . It is convenient to describe them using \gaugecondition\ which take the form
\eqn\incomingpoincare{\eqalign{
{\cal P}_1 : Z^{-1} + Z^{1} &= -1,  ~~~{\cal P}_{2} : Z^{-1} - Z^{1} = -1 , \cr
{\cal P}_{4} : Z^{-1} + Z^{1} &= 1,  ~~~{\cal P}_{3} : Z^{-1} - Z^{1} = 1. 
}}

Past Poincare horizons of the patches ${\cal P}_3$ and ${\cal P}_4$ and future Poincare horizons of ${\cal P}_1$ and ${\cal P}_2$ intersect along $Z^{-1} = Z^{1} = 0$. This is $H^{d-1}$
\eqn\hyperpolicimp{
- (Z^{0})^2 + \sum_{i = 2}^{d} (Z^i)^2 = - 1
}
and describes the transverse impact parameter space in which we separate the colliding particles.

We are interested in the wave functions for the boundary operators such that the state in the bulk is localized in $H^{d-1}$. As explained in \HofmanAR\ these are just plane waves in the corresponding Poincare patches. Indeed, let us consider for example the wave function of an eigenstate of the bulk isometry associated to the four-momentum $P_{\mu}$
\eqn\generator{
P_{\mu} =2 i \left( K.Z \pa_{Z^{\mu}} - Z_{\mu} K . \pa_{Z} \right) \ .
}
The Poincare horizon corresponds to $K.Z = 0$ where the generator simplifies and the momentum eigenstate has the wave function
\eqn\eigenstate{
\phi(Z) \sim e^{- i P \bar K . Z} \delta^{(d-1)}(\vec Z - \vec W)
}
where $\vec Z$ refers to $Z^i$ with $i = 2, .. , d$. The eigenvalue is 
\eqn\eigenvalue{
P_{\mu} \phi(Z) = P W_\mu \phi(Z) .
}
In terms of boundary CFT this corresponds to the four-momentum $p^{\mu} = P W^{\mu}$, where $p^2 = - P^2$ and $W^{\mu}$ is the unit vector \hyperpolicimp\ $W^2 = -1 $ that specifies the direction of the four-momentum. We will consider time-like momenta $p^2<0$.

Notice that the isometry that we call $P_{\mu}$ depends on the choice of $K^{A}$ and is different for Poincare patches $1,3$ and $2,4$. For example, translations in patches $1,3$, that are identified as $P_{\mu , 1} = - P_{\mu , 3}$, become special conformal transformations in patches $2,4$ and vice versa. On the other hand, Lorentz transformations $x \to \Lambda x$ are the same in all four Poincare patches.

For two particles localized at different points on $H^{d-1}$ the geodesic distance between them is equal to
\eqn\geodesic{
\cosh L = W_1 . W_2 = - {p_1 . p_2 \over \sqrt{-p_1^2} \sqrt{-p_2^2}} , ~~~ 0 \leq L < \infty \ .
}
We will refer to $L$ as the impact parameter and it is the same $L$ that appeared in the introduction. 

Similarly, we can introduce energy of the collision
\eqn\energyofthecol{
S = P_1 P_2 = \sqrt{ p_1^2 p_2^2} \ .
}
On the scales small compared to the AdS radius this coincides with the usual Mandelstam $s$. 

We now imagine elastic scattering of two particles at fixed impact parameter space in AdS, see \CornalbaAX . In light of kinematics described above a natural object to consider is the following. We start with the four-point function
\eqn\sccorryA{  
\la  \phi\left( y_4 \right) \psi \left(y_3 \right) \psi \left( y_2 \right) \phi\left(y_1 \right)  \ra  = { \AA(u,v) \over y_{14}^{2 \Delta_\phi} y_{23}^{2 \Delta_\psi}}    
}
where as usual
\eqn\defuvA{   
u= {y_{12}^2 y_{34}^2\over y_{13}^2 y_{24}^2} =z \bar z, \qquad v= {y_{14}^2 y_{23}^2\over y_{13}^2 y_{24}^2} = (1-z) (1-\bar z) ,
}
and operators are ordered as written.

To introduce the phase shift we introduce corresponding Poincare patches
\eqn\finiteform{\eqalign{
x_i &= (x_i^+ , x_i^-, \vec x_i) = - {1 \over y_i^+} (1 , y_i^2, \vec y_i ), ~~~ i = 1,4 \  , \cr
x_j &= (x_j^+ , x_j^-, \vec x_j) = - {1 \over y_j^-} (1 , y_j^2, \vec y_j ), ~~~ j = 2,3 ~ ,
}}
where the subscript stands for the operator insertion $y_i$ with $i= 1, .., 4$. This coordinate transformation is common in considerations of the Regge limit. After doing the coordinate transform we set $x_{1,4} =\pm {x \over 2}$, $x_{2,3} = \pm {\bar x \over 2}$. Notice that in this way the correlator \sccorryA\ is time-ordered.

For reader's convenience let us present the explicit transformation formulas
\eqn\substitution{\eqalign{
y_1^+ &= - y_4^+ =- {2 \over x^+}  , ~~~y_2^- = - y_3^- =- {2 \over \bar x^+} \ ,\cr
y_1^- &= - y_4^- = - {x^2 \over 2 x^+} ,~~~y_2^+ = - y_3^+ = - {\bar x^2 \over 2 \bar x^+}  \ , \cr
\vec y_1 &= \vec y_4 = {\vec x \over x^+} ,~~~ \vec y_2 = \vec y_3 = {\vec{\bar{x}} \over \bar x^+} .
}}
In these formulas we take $x^{\mu}$ and $\bar x^{\mu}$ to be future-directed time-like vectors.  This implies that $x^+, \bar x^+ > 0$ and $x^2 < 0$ and $\bar x^2 < 0$. Moreover, the spacelike $y_{ij}^2$ in the original coordinates corresponds to timelike $x^2$ after the transformation, e.g. $y_{14}^2 = -  {4 x^2 \over (x^+)^2}$.
Below we will consider more general $x$ and $\bar x$ which could be both spacelike and timelike.

The conformal mapping  \finiteform\ yields
\eqn\sccorrA{\eqalign{
A(x, \bar x)&= \la \phi\left(- {x \over 2} \right) \psi \left( - {\bar x \over 2} \right) \psi \left({\bar x \over 2} \right) \phi\left({x \over 2} \right)  \ra \cr
&= \left( {x^+ \over 2} \right)^{-2 \Delta_{\phi} } \left( {\bar x^+ \over 2} \right)^{-2 \Delta_\psi} \la \phi (y_4) \psi(y_3) \psi(y_2) \phi(y_1)\ra = {\AA(u,v) \over (-x^2)^{ \Delta_\phi} (-{\bar x}^2)^{\Delta_\psi} }  ,
}}
where in the LHS each operator is inserted in its own Poincare patch as described above.
The cross ratios in terms of new variables take the form
 \eqn\crossratiosB{\eqalign{
 u &= \left( {1 + {x^2 \bar x^2 \over 16} - {x.\bar x \over2} \over 1 + {x^2 \bar x^2 \over 16} + {x.\bar x \over2} } \right)^2  , \cr
 v &= {x^2 \bar x^2 \over \left(1 + {x^2 \bar x^2 \over 16} + {x.\bar x \over2} \right)^2} .
 }}
 
 Next, as reviewed above, we do the transformation to the impact parameter space
 \eqn\transformimpact{
 B(p , \bar p) =\int d^d x d^d \bar x e^{i p.x} e^{i \bar p. \bar x} A(x, \bar x).
 }
 In doing the Fourier transform we have to specify the $i \eps$ prescription. The prescription is dictated by the ordering of operators and in this case takes the form $x^2 \to x^2 - i \eps  x^0 $ and similarly for $\bar x^2$ .

 Let us understand the symmetries of $B(p , \bar p)$. The transform is manifestly invariant under Lorentz transformations, thus, it can only depend on $p^2$, $\bar p^2$ and $p . \bar p$. Moreover, we have the following transformation of $A(x, \bar x)$
 \eqn\transformA{
 A(\lambda x, {1 \over \lambda} \bar x) = \lambda^{2 (\Delta_{\psi} - \Delta_{\phi} )} A(x, \bar x)
 }
 which translates using \transformimpact\ into
 \eqn\symmetry{
 B(\lambda p , {1 \over \lambda} \bar p) =\lambda^{-2 (\Delta_{\psi} - \Delta_{\phi} )} B(p, \bar p) .
 }
 
This simmetry together with Lorentz invariance implies that
\eqn\definitionphase{
B(p, \bar p) = B_0(p,\bar p) \left(  1 + i \delta(S, L)  \right)
}
where $S$ and $L$ were defined in  \energyofthecol\ and \geodesic. $B_0(p,\bar p)$ is the Fourier transform of the disconnected part $\la \phi \phi \ra \la \psi \psi \ra$
\eqn\fourierdisconnected{\eqalign{
B_0(p,\bar p) &= \int d^d x d^d \bar x e^{i p.x} e^{i \bar p. \bar x} {1 \over (-x^2 + i \eps x^0)^{ \Delta_\phi} (-{\bar x}^2 + i \eps \bar x^0)^{\Delta_\psi} } , \cr
&=\theta(p^0) \theta(-p^2) \theta(\bar p^0) \theta(-\bar p^2) e^{- i \pi (\Delta_{\phi} + \Delta_{\psi})}C(\Delta_{\phi}) C(\Delta_{\psi})  (-p^2)^{\Delta_{\phi} - {d \over 2}} (-\bar p^2)^{\Delta_{\psi} - {d \over 2}} , \cr
C(\Delta) &= {2^{d+1-2\Delta} \pi^{1+{d \over 2}} \over \Gamma(\Delta) \Gamma(\Delta  - {d - 2 \over 2})}
}}

The unity in \definitionphase, therefore, describes a free propagation in AdS. The phase shift $\delta(S,L)$, on the other hand, describes interaction. We will be interested in the scattering of energetic probes, namely $S \gg 1$. The physical picture in this case depends on the impact parameter $L$.

In this paper we focus on impact parameters for which the dominant contribution to the phase shift is due to the graviton exchange
\eqn\regimesofinterest{
\delta(S,L) = c S  \Pi_{d-1}(L) + ...
}
where $\Pi_{d-1}(L)$ is the propagator in $H^{d-1}$ and $...$ stands for subleading corrections. 
For a generic CFT we expect $L_* > R_{AdS}$, but for theories with large gap or local bulk dual we could have $L_* \ll R_{AdS}$.  

In the present paper we analyze \regimesofinterest\ using CFT methods. The high energy limit $S \gg 1$ of the AdS phase shift $\delta(S,L)$ is related to the Regge limit and we assume that the relevant singularity in the $J$ plane is a pole. In this way the corrections to the simple gravity formula \regimesofinterest\ are dictated by the properties of the leading Regge trajectory.

\subsec{Phase Shift for Spinning Operators}

Below we will be interested in correlation functions of operators with spin as well. For simplicity we focus on the case of two spin one operators and two scalars. A natural object to consider in this case is the following
\eqn\naturalobjectSpin{
A^{m n}(x, \bar x) =\la J^{m}\left(- {x \over 2} \right) \psi \left( - {\bar x \over 2} \right) \psi \left({\bar x \over 2} \right) J^{n}\left({x \over 2} \right)  \ra , 
}
where the polarization tensors are defined in the corresponding Poincare patch. To relate to the original coordinates we get the following expression for the polarization tensors
\eqn\polarizationtensors{
\eps_{\mu} (y_1) = {\pa x^m \over\pa  y_1^\mu} , ~~~ \eps_{\nu} (y_4) = {\pa x^n \over \pa y_4^\mu} .
}
So that in the original coordinates we have
\eqn\correlationfunction{
A^{m n}(x, \bar x) =\left({x^+ \over 2} \right)^{-2 (\Delta_{J} + 1)} \left({\bar x^+ \over 2} \right)^{-2 \Delta_\phi} \eps_{\mu}(y_1) \eps_{\nu}(y_4) \la J^{\mu}(y_4) \phi(y_3) \phi(y_2) J^{\nu}(y_1)\ra .
}

Next we consider the Fourier transform as above
\eqn\fouriertransform{
B^{m n}(p , \bar p) = \int d x d \bar x \, e^{i p x} e^{i \bar p . \bar x} A^{m n}(x, \bar x) \ .
}
As before it will be convenient to separate the contribution of the disconnected piece $B_0^{mn}$. We set
\eqn\Bovec{
B_0^{mn}(p,\bp)=\left(\eta^{mn}-g(\Delta){p^m p^n\over p^2}\right)\,\,B_0(p,\bp) \,,
}
where $g(\Delta)=2 {\Delta_J-d/2\over \Delta_J-1}$ and study the bulk phase shift matrix $\delta^{mn}(p,\bp)$ defined by
\eqn\deltadef{
{B^{m n}(p , \bar p)-B_0^{mn}(p,\bp)\over B_0(p,\bp)}= i \delta^{mn}(p,\bp)\,,
}
where $\eta^{mn}$ denotes the Minkowski metric.

\newsec{Conformal Blocks and Particle Exchanges in AdS}

In this section we consider a simple example of a particle exchange in AdS that allows us to illustrate some ideas which we develop further in the paper in more generality. We review notions of the Regge and the bulk point limits. We compare the behavior of the corresponding Witten diagram to the behavior of a single conformal block in these limits.\foot{The result for the Witten diagram in the Regge limit does not depend on the details of the three-point coupling in AdS \CostaCB.}
The block and the AdS diagram differ by the contribution of an infinite series of double trace operators, see e.g. \HijanoZSA , which is non-negligible in the regime of our interest.
We then transform the result to the impact parameter space and reproduce the high energy limit of the phase shift that is obtained from the bulk considerations \CamanhoAPA. 

The result is that both the block and the Witten diagram develop the same singularity in the Regge limit. However, the bulk point limit behavior is very different. The conformal block is singular in the bulk point limit, whereas the Witten diagram is regular.
The difference is important to us because as we review below the result for the Regge limit of the correlator that is produced by a Regge pole is similar to the exchange diagram and is regular in the bulk point limit. This fact will be important for the discussion of bounds on the Regge limit coming from causality and unitarity.

Next, we consider the transform to the impact parameter space in AdS. This was done in \CornalbaAX\ and we review it below. The Regge limit is related to the high energy behavior of the phase shift. The crucial observation is that upon the transformation to the impact parameter space the contribution of double trace operators effectively cancels and the phase shift is controlled solely by the quantum numbers of the particle that is being exchanged in AdS. In this way the simple result of the bulk computation is reproduced.

\subsec{Basic Kinematics}

Consider the four point function of scalar operators
\eqn\sccorryAa{\eqalign{  
\vev{ \phi\left( y_4 \right) \psi \left(y_3 \right) \psi \left( y_2 \right) \phi\left(y_1 \right) } & = { \AA(u,v) \over y_{14}^{2 \Delta_\phi} y_{23}^{2 \Delta_\psi}  }\cr
u= {y_{12}^2 y_{34}^2\over y_{13}^2 y_{24}^2} =z \bar z, &\qquad v= {y_{14}^2 y_{23}^2\over y_{13}^2 y_{24}^2} = (1-z) (1-\bar z) \,.
}}
It is convenient to discuss the kinematics in terms of the partial amplitude $\AA(z,\bz)$ or rather
\eqn\fourpoint{
G(z,\bar z)\equiv \la \phi(\infty)  \psi (1) \psi (z, \bar z) \phi(0) \ra =\left([1-z][1-\bar z])\right)^{-\Delta_\psi}\,\,\AA(z,\bar z)\,.
}
where $\psi(\infty)=\lim_{y\rightarrow\infty} y^{2\Delta_\psi}\psi(y)$.

The correlator $G(z, \bar z)$ admits an expansion in terms of conformal blocks in three different channels \refs{\HartmanLFA, \HofmanAWC}
\eqn\blockss{\eqalign{
{\rm s-channel:} ~~~G(z,\bar z) = (z \bar z)^{- {1 \over 2} (\Delta_\phi + \Delta_{\psi}) } \sum_{{\cal O}_{\Delta, J}} \lambda_{\phi \psi {\cal O}_{\Delta, J}} \lambda_{\psi \phi {\cal O}_{\Delta, J}} g_{\Delta, J}^{\Delta_{\phi \psi}, - \Delta_{\phi \psi}}(z, \bar z) \ , \cr
~~~ \Delta_{\phi \psi} = \Delta_{\phi} - \Delta_{\psi},~~~ |z|<1 .\cr
}}
\eqn\blockst{\eqalign{
{\rm t-channel:} ~~~G(z,\bar z) = \left([1-z][1- \bar z]\right)^{- \Delta_\psi } \sum_{ {\cal O}_{\Delta, J} } \lambda_{\phi\phi {\cal O}_{\Delta, J}} \lambda_{\psi \psi {\cal O}_{\Delta, J}} g_{\Delta, J}^{0,0}(1-z,1-\bar z) , \cr 
~~~ |1-z|<1 .
}}
\eqn\blocksu{\eqalign{
{\rm u-channel:} ~~~G(z,\bar z) = (z \bar z)^{{1 \over 2} (\Delta_\phi + \Delta_{\psi}) } \sum_{ {\cal O}_{\Delta, J} } \lambda_{\phi \psi {\cal O}_{\Delta, J}} \lambda_{\psi \phi {\cal O}_{\Delta, J}} g_{\Delta, J}^{\Delta_{\phi \psi}, - \Delta_{\phi \psi}}({1 \over z}, {1 \over \bar z}) \ , \cr
~~~ |z|>1 ,\cr
}}
where the sum runs over an infinite set of primary operators that appear in the corresponding OPE channel. The expressions for conformal blocks can be found for example in \DolanDV.

There are several limits of this correlator that are usually discussed. These include light-cone, Regge and bulk point limits.
We first review the behavior of a single conformal block in each of these limits.  Let us describe the limits of interest in detail. 

 The simplest limit to consider is the light-cone limit,
\eqn\lightcone{
{\rm Lightcone :} ~~~~~ z \to 1, ~~ \bar z - {\rm fixed} .
}
In the $t$-channel OPE this limit is controlled by the lowest twist operators that contribute as $(1-z)^{{\tau \over 2}}$, where $\tau = \Delta - J$. 
 
To approach the Regge limit we continue $z$ around $0$ and then send both $z, \bar z \to 1$
\eqn\Regge{
{\rm Regge:} ~~~~~ z e^{- 2 \pi i} , \bar z \to 1, ~~~ {z \over \bar z} - {\rm fixed} \ ,
}
where $z e^{- 2 \pi i}$ indicates that we have to analytically continue $G(z,\bar z)$ before taking the limit. To analyze the Regge limit, it is convenient to introduce the following coordinates following \CostaCB\ 
\eqn\continuation{
1-z = \sigma e^{\rho}, ~~~ 1-\bar z = \sigma e^{- \rho} .
}
The Regge limit corresponds to $\sigma \to 0$ limit with $\rho$ held fixed. Notice as well that our choice of points corresponds to $\sigma<0$. In this limit both the single trace and double trace operators are important. 

By the bulk point limit we loosely mean 
\eqn\bulkpoint{
{\rm Bulk ~ point:} ~~~~~ z \to \bar z ,
}
or $\rho \to  0$ in the notations of \continuation .
When talking about \bulkpoint\ we have to specify on which sheet and along which path we approach $z=\bar z$ point. We will be interested in \bulkpoint\ after analytic continuation \Regge. The limit considered in \MaldacenaIUA\ corresponds to the same contrinuation for $z$ as above in the consideration of Regge limit supplemented by the continuation $\rho \to \rho \pm i \pi$ which makes $z$ and $\bar z$ negative see.

\subsec{Limits of the $t$-channel Conformal Block}

Imagine that in  \fourpoint\ there is an operator with dimension $\Delta$ and spin $J$ being exchanged in the t-channel $\phi \phi \sim {\cal O}_{\Delta, J} \sim \psi \psi$. We would like to evaluate the contribution of one t-channel block in the Regge limit. Notice that the t-channel OPE does not converge in the Regge limit and at the end we will have to use Regge theory to analyze the $t$-channel physics.

Let us first consider this exercise in $d=2,4$. The conformal blocks in the $t$-channel expansion \fourpoint\ are given by\foot{For convenience we use $z_{here} = 1-z$ in \fourpoint . We hope it will not cause any confusion. }
\eqn\conformalblocks{\eqalign{
k_\beta (z) &= \ _2  F_1 \left( {\beta \over 2}, {\beta \over 2}, \beta , z \right) \ , \cr
g_{\Delta, J}^{d=2} (z, \bar z) &= z^{{\Delta - J \over 2}} \bar z^{{\Delta - J \over 2}}  \left( z^{J} k_{{\Delta+J \over 2}}(z) k_{{\Delta-J \over 2}}(\bar z) + \bar z^{J} k_{{\Delta+J \over 2}}(\bar z) k_{{\Delta-J \over 2}}(z)  \right) \ , \cr
g_{\Delta, J}^{d=4} (z, \bar z)&= {z^{{\Delta - J \over 2}} \bar z^{{\Delta - J \over 2}} \over z - \bar z} \left( z^{J+1} k_{{\Delta+J \over 2}}(z) k_{{\Delta- J - 2 \over 2}}(\bar z) - \bar z^{J+1} k_{{\Delta+J \over 2}}(\bar z) k_{{\Delta-J-2 \over 2}}(z)  \right) .
}}

Notice the behavior of the blocks before analytic continuation. In the light-cone limit $z \to 0$, $\bar z$ fixed, the block behaves as $g_{\Delta, J}(z, \bar z) \sim z^{\Delta - J \over 2}$ and is controlled by the twist $\tau = \Delta - J$ of the operator. In the limit $z, \bar z \to 0$ the block behaves as $g_{\Delta, J}(z, \bar z) \sim z^{{\Delta + J \over 2}} \bar z^{{\Delta - J \over 2}} + c.c  $ and is controlled by the dimension of the operator $\Delta$. In the bulk point limit $z \to \bar z$ on the principal sheet the blocks are regular.

To approach the Regge limit it is useful to recall the analytic continuation of the hypergeometric function\foot{See (4.27) in \HartmanLFA .}
\eqn\analyticcont{
~~~k_\beta (1-z e^{- 2 \pi i}) = k_\beta (1 - z) + 2 \pi i {\Gamma(\beta) \over \Gamma({\beta \over 2})^2} \ _2 F_1 ({\beta \over 2}, {\beta \over 2}, 1, z) .
}

Using this identity we can find the leading order behavior of the conformal block in $d=2$ and $d=4$. The result is
\eqn\asymptotic{\eqalign{
g_{\Delta, J}^{d=2} &=i  {4^{\Delta + J - 1} \Gamma({\Delta + J-1 \over 2}) \Gamma({\Delta + J+1 \over 2}) \over \Gamma({\Delta+J \over 2})^2}{1 \over  \sigma^{J-1} } e^{- (\Delta - 1) \rho} + ... , \cr
g_{\Delta, J}^{d=4} &=i {4^{\Delta + J - 1} \Gamma({\Delta + J-1 \over 2}) \Gamma({\Delta + J+1 \over 2}) \over \Gamma({\Delta+J \over 2})^2} {1 \over  \sigma^{J-1} } { e^{- (\Delta - 3) \rho} \over e^{2 \rho} - 1} \ ,
}}
where we suppressed the constant of proportionality which is irrelevant for us. Let us pause for a moment and contemplate the result \asymptotic . 

First, these examples show the generic feature of the behavior of the conformal block in the Regge limit
\eqn\Reggelimitblock{
{\rm Regge :} ~~~ g_{\Delta, J} (\sigma, \rho) =i {f^{\Delta,J}(\rho) \over \sigma^{J-1}} \ .
}
This is what makes it hard to approach the Regge limit using the standard OPE technique. The contribution of a given primary is controlled by its spin $J$ and operators with higher spin dominate.

Second, the behavior in the bulk point limit $\rho \to 0$ depends on the number of dimensions $d$. With some hindsight the behavior of the block is
\eqn\bulkpointblock{
{\rm Bulk~ point :} ~~~ \lim_{\rho \to 0}f^{\Delta,J}(\rho) \sim {1 \over \rho^{d-3}} \ ,
}
where in $d=3$ the singularity becomes logarithmic.

The expression in general $d$ could be found using the Casimir equation \DolanDV.\foot{This has two solutions which correspond to the usual conformal block and its shadow.  The two can be distinguished by the transformation properties under $\rho \to \rho + 2 \pi i$ at large $\rho$ \SimmonsDuffinUY. The usual conformal block transforms with the $e^{- 2 \pi i \Delta}$ phase, whereas the shadow block picks $e^{2 \pi i (\Delta - d)}$. This allows one to pick the correct solution.} The result is that $f^{\Delta,J}(\rho) = c_{\Delta, J} \Pi_{\Delta - 1}(\rho)$, where $\Pi_{\Delta - 1}(\rho)$ is the Euclidean propagator in hyperbolic space $H^{d-1}$ \ \foot{The propagator satisfies the following equations 
\eqn\propagatorB{\eqalign{
{1 \over \sqrt{g}} \pa_i \sqrt{g} g^{i j} \pa_j \Pi_{\Delta - 1} (\rho) &- (\Delta -1) (\Delta +1 -d ) \Pi_{\Delta - 1} (\rho) = \delta_{H^{d-1}}(\rho) \cr
{1 \over \sqrt{g}} \pa_i \sqrt{g} g^{i j} \pa_j \Pi_{\Delta - 1} (\rho)  &= \ddot \Pi_{\Delta - 1}(\rho)+ (d-2) \coth \rho \dot \Pi_{\Delta - 1}(\rho)  \ , \cr
d s^2_{H^{d-1}} &= d \rho^2 + \sinh^2 \rho \ d \Omega_{S^{d-2}}^2 .
}}
where dots stand for the $\pa_{\rho}$ derivatives.
}
\eqn\propagator{\eqalign{
\Pi_{\Delta - 1}(\rho) &= { \pi^{1 - {d \over 2}} \Gamma(\Delta -1) \over 2 \Gamma(\Delta - {d -2\over 2})} \ e^{-(\Delta -1) \rho} \ _{2} F_1({d \over 2} - 1, \Delta - 1, \Delta - {d \over 2} + 1, e^{- 2 \rho}) , \cr
c_{\Delta, J} &={4^{\Delta + J - 1} \Gamma({\Delta + J-1 \over 2}) \Gamma({\Delta + J+1 \over 2}) \over \Gamma({\Delta+J \over 2})^2}  {2 \Gamma(\Delta - {d -2\over 2}) \over \pi^{1 - {d \over 2}} \Gamma(\Delta -1) } \ .
}}

We see that each conformal block is singular at $\rho = 0$. This singularity is purely kinematical. It could be thought as coming from an infinite volume of $H_{d-1}$ in AdS which describes the center of the collision and not from the small impact parameter. The full correlator is regular in this limit \MaldacenaIUA. 

The Regge limit of the stress tensor conformal block with external spinning operators was analyzed in detail in \AfkhamiJeddiNTF. This conformal block alone, however, does not control the Regge limit of the correlator as we will see shortly.

\ifig\witten{Exchange Witten diagram. We consider the contribution of the exchange Witten diagram to the four-point function.} {\epsfxsize1.4in\epsfbox{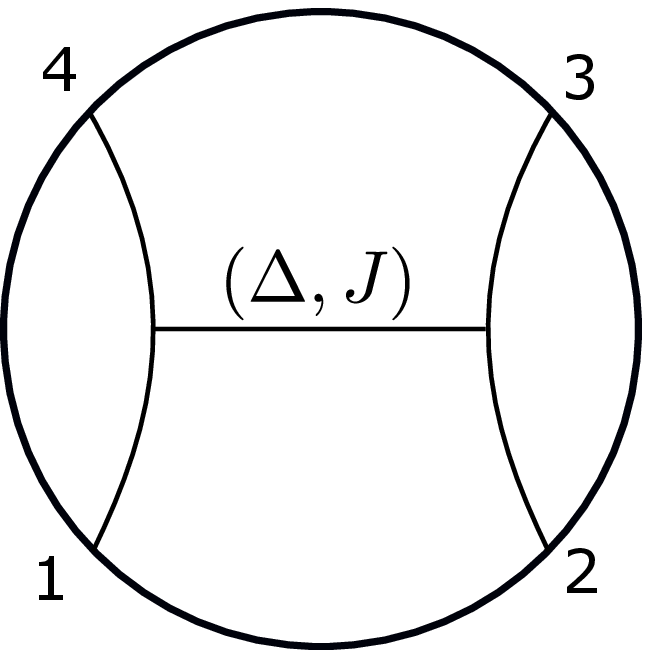}}

\subsec{Exchange Witten diagram}

We would like to contrast the behavior we found above for one conformal block with the behavior of the corresponding Witten diagram \witten. We imagine an exchange of a particle with quantum numbers the same as above, namely dimension $\Delta$ and even spin $J$. As discussed in detail 
in \HijanoZSA\ an exchange in the bulk when decomposed in terms of conformal blocks \fourpoint\ contains 
two types of contributions: single-trace operator conformal block $g_{\Delta, J}$; corrections to the anomalous dimensions of double trace operators that, schematically, could be written as $\phi \pa^{2 n} \pa_{\mu_1} ... \pa_{\mu_{s} } \phi $ and $\psi \pa^{2 n} \pa_{\mu_1} ... \pa_{\mu_{s} }  \psi $ with $s \leq J$.

The precise result for the Witten diagram depends on the details of the three-point coupling in the bulk. However, in the Regge limit the answer is universal (see for example, the appendix A.3 of \CostaCB). The relevant expressions in \CostaCB\ are given in terms of the Mellin amplitude associated to a given Witten diagram, but they can easily be rewritten as follows
\eqn\exchangeRegge{\eqalign{
{\rm Regge :} ~~~ \AA^{AdS}(\sigma, \rho)  =i {f^{AdS}(\rho) \over \sigma^{J-1}}
}}
where $\AA^{AdS}(\sigma,\rho)$ denotes the contribution of the Witten diagram depicted in Fig.2 to the reduced amplitude $\AA(\sigma,\rho)$ and\foot{For the odd spins we would also have the factor $e^{-i \pi J}$ depending on the ordering of the external operators.}
\eqn\exchangeReggeB{\eqalign{
f^{AdS}(\rho)&=c_0 \int_{- \infty}^{\infty} d \nu \  \gamma(\nu) \gamma(- \nu) {1 \over  \nu^2 + (\Delta - {d \over 2})^2 }  \Omega_{i \nu} (\rho) \ , \cr
\gamma(\nu) &=\Gamma \left( 2 \Delta_\phi + J + i \nu - {d \over 2} \over 2 \right) \Gamma \left( 2 \Delta_\psi + J + i \nu - {d \over 2} \over 2 \right) \, , \cr
c_0 &=  \lambda_{\psi \psi {\cal O}_{\Delta, J}} \lambda_{\phi \phi {\cal O}_{\Delta, J}} 4^{1+J} \pi^{{d \over 2}} K_{\Delta,J} ,
}} 
where the harmonic function $\Omega_{i \nu}(\rho)$ in the hyperbolic space $H^{d-1}$ is defined in appendix A and $K_{\Delta, J}$ could be found, for example, in eq (41) of \CostaCB. 

It is important to note that $\Omega_{i\nu}(\rho)$ is regular at $\rho = 0$. To see this one can simply expand $\Omega_{i \nu} (\rho)$ at small $\rho$
\eqn\expansionOmega{
\Omega_{i \nu}(\rho) \sim \Gamma \left({d -2\over 2} + i \nu \right) \Gamma \left({d -2\over 2} - i \nu \right) \nu \ \sinh[ \pi \nu] \left(1 - \rho^2 {\nu^2 + {(d-2)^2 \over 4} \over 2 (d-1)} + ...\right)
}
where further corrections are again polynomial in $\nu^2$ and $\rho^2$. We can now estimate the behavior of the integrand \exchangeReggeB\ at large $\nu$, we have
\eqn\largenu{\eqalign{
\gamma(\nu) \gamma(- \nu) &\sim e^{- \pi | \nu|} , \cr
\Gamma \left({d -2\over 2} + i \nu \right) \Gamma \left({d -2\over 2} - i \nu \right) \nu \ \sinh[ \pi \nu] &\sim \nu^{d-2} .
}}
Due to the exponential suppression produced by the $\gamma(\nu) \gamma(- \nu)$ the integral converges at large $\nu$ and the result is an expansion of the type $\sum_{n=0}^{\infty} c_n \rho^{2n}$. In contrast to \bulkpointblock\ we have in this case
\eqn\bulkpointWitten{
{\rm Bulk~ point :} ~~~ \lim_{\rho \to 0} f^{AdS}(\rho) \sim 1 \ .
}

To connect with the previous section notice that 
 \eqn\identity{
 \Omega_{i \nu}(\rho) = {i \nu \over 2 \pi} \left( \Pi_{i \nu + {d \over 2}-1}(\rho) - \Pi_{- i \nu + {d \over 2}-1} (\rho) \right) .
 }
One can, in principle, substitute this into \exchangeReggeB\ and  compute the integral.
The contribution due to the pole at $\nu^2 = (\Delta - {d \over 2} )^2$ in \exchangeReggeB\ gives precisely the conformal
block of the exchanged operator \asymptotic. 
Note that it is singular in the $\rho\rightarrow0$ limit.
At the same time, contributions due to the double trace poles contribute corresponding conformal blocks as well
(with some non-vanishing coefficients).
The total result is regular in the $\rho \to 0$ limit as shown above.
Hence, the double trace operators precisely cancel the singular $\rho\rightarrow0$ behavior of the stress-tensor conformal block.  More generally this cancelation follows from the s-channel OPE, see section 6.3 in \MaldacenaIUA .

In contrast to a single conformal block the exchange Witten diagram generates a consistent solution to the large $N$ crossing \HeemskerkPN . We can think of the effective Lagrangian which contains two scalar fields dual to $\phi$ and $\psi$ and the cubic coupling of this field to the dual of the exchanged operator ${\cal O}_{\Delta, J}$.  It also serves as a correct model for the Regge pole in a CFT as we will review in the next section. 

Another little comment is to notice that if we consider the bulk point singularity region by analytically continue \exchangeRegge\ $\rho \to \rho \pm i \pi$ it does become singular for small $\rho$ in accord with general arguments presented in \MaldacenaIUA . To see it explicitly let us consider $d=4$. In this case $\Omega_{i \nu}(\rho) \sim {\nu \sin \nu \rho \over \sinh \rho}$. This develops a singularity at $\rho = 0$ upon continuation $\rho \to \rho \pm i \pi$.

\subsec{Impact Parameter Space}

As explained in the previous section, to make a more direct connection with the bulk it is useful to transform the correlator to the impact parameter space \CornalbaAX. The starting point is the correlation function \sccorrA\ obtained from \sccorryA\ via \substitution. To determine the reduced amplitude holographically, we simply need to add the contribution of the disconnected piece to the Witten diagram result \exchangeRegge\ of the previous section. Before performing the Fourier transform we must relate $(\sigma,\rho)$ or $(z,\bar z)$ with $(x,\bx)$ in the Regge limit which corresponds to small $(x,\bx)$. With the help of \crossratiosB\  one finds that 
\eqn\crossratios{
(1-z) (1- \bar z) = \sigma^2 =  x^2 \bar x^2 + ... , ~~~ (1-z) + (1-\bar z) \simeq 2 \sigma \cosh \rho =  2 x . \bar x + ... ,
}
where $(x, \bar x)$ label points in Minkowski space and $...$ in the formula above stands for terms that are higher order in $(x,\bar x)$.\foot{We use mostly plus convention.} As a result, the reduced amplitude in the Regge limit, including the contribution of the disconnected piece, is
\eqn\reggea{\eqalign{
\AA(x,\bar x) &= 1+ i c_0  \int_{- \infty}^{\infty} d \nu \  \gamma(\nu) \gamma(- \nu) {1 \over  \nu^2 + (\Delta - {d \over 2})^2 }  {\Omega_{i \nu} (\rho) \over (- \sqrt{x^2 \bar x^2})^{J-1}} , \cr
\cosh \rho &= - {x. \bar x \over \sqrt{x^2 \bar x^2}} \ .
}}

In the formula above both $x$ and $\bar x$ are future-oriented time-like vectors. Next we perform the Fourier transform \transformimpact\ with $A(x,\bx)$ obtained from \sccorrA\ after substitutution of the partial amplitude $\AA(x,\bx)$, from the formula above. To perform the integral, it is useful to recall the following identity, for the derivation see appendix A,
\eqn\identityJ{
{\gamma(\nu) \gamma(- \nu)  \Omega_{i \nu} (e. \bar e) \over (-x^2 )^{\Delta_O + {J-1 \over 2}} (-\bar x^2  )^{\Delta_\phi + {J-1 \over 2}}} ={2^{2-a-b} e^{- i {\pi \over 2}(a + b)} \over \pi^{d-2}} \int_{M^+} d^d p d^d \bar p {e^{- i x.p } e^{- i \bar x. \bar p } \over (-p^2)^{{d -a \over 2} } (-\bar p^2)^{{d-b \over 2} }} \Omega_{i \nu} (L) \
}
where $a = 2 \Delta_\phi + J - 1$, $b = 2 \Delta_{\psi} + J - 1 $ and $M^+$ stands for the future Milne wedge, defined as follows
\eqn\milnewedge{
M^+ = \left\{ p^{\mu}: ~ p^0 > 0, ~ p^2 < 0 \right\} \ .
}

Using the representation \identityJ\ it is trivial to do the Fourier transform. Notice that the factors $(-1)^{J-1}$ cancel out. We get
\eqn\resultphaseshift{\eqalign{
\delta(S,L)  ={c_0 \pi^{d+2} 4^{2 +d - J -  \Delta_\phi - \Delta_{\psi}} \over C(\Delta_{\phi}) C(\Delta_{\psi})}S^{J-1}  \int_{- \infty}^{\infty} d \nu \  {1 \over  \nu^2 + (\Delta - {d \over 2})^2 }  \Omega_{i \nu} (L) \ .
}}
Neatly all the double trace operator poles coming from the $\gamma(\nu) \gamma(- \nu)$ are gone. As we explained above without the $\gamma(\nu)\gamma(-\nu)$ pre-factor we cannot expand the integrand for small $L$. Indeed, in this limit the integral diverges at large $\nu$. This signals that the phase shift is singular for small $L$. Similarly, we cannot close the contour at infinity because $\Omega_{i \nu}(L)$ grows exponentially both in the upper and lower half-plane. A better way to do the integral is to use \identity. The advantage of this representation is that $\Pi_{i \nu + {d \over 2}-1}(\rho)$ decays in the lower half-plane. We can use the symmetry of the integrand $\nu \to - \nu$ to see that both terms contribute equally and then we can close the contour in the complex plane by picking the contribution from the pole in the ${1 \over  \nu^2 + (\Delta - {d \over 2})^2 }$.
The result is 
\eqn\resultphaseshift{
\delta(S,L) \propto \lambda_{\phi \phi O_{\Delta,J}} \lambda_{\psi \psi O_{\Delta,J}} S^{J-1}  \Pi_{\Delta-1}(L),
}
which is exactly what one gets in the bulk \CamanhoAPA.

\subsec{Double Trace Operators and The Bulk Phase Shift}

It is natural to pose a question how the asymptotic of the correlator from the previous section is reproduced using the OPE in the $\phi \psi$ $s$-channel. This problem was solved in \CornalbaXM. Let us quickly review their construction. On general grounds we know that the result should be reproduced from an infinite sum over double trace operators that we can schematically denote as $\phi \pa^{2 n} \pa_{\mu_1} ... \pa_{\mu_s} \psi$ which have spin $s$ and dimension $\Delta_{n,s} = \Delta_\phi + \Delta_\psi + 2n + s + \gamma(n,s)$. 

Let us introduce new variables for the quantum numbers of double trace operators
\eqn\variables{
h = {\Delta_{n,s} + s \over 2},~~~ \bar h = {\Delta_{n,s} - s \over 2} \ .
}
The result of \CornalbaXM\ is that the operators that reproduce the result of the previous section have the following property
\eqn\relevantoperators{
h \sim \bar h \sim {1 \over \sqrt \sigma} .
}
In other words the Regge limit is reproduced by the large spin and large twist operators with their ratio related to the impact parameter in the bulk. 
Conformal blocks simplify in this limit and the leading contribution to the correlation function takes the form\foot{The corrections to the three-point functions are subleading in the Regge limit. See for example the discussion in section 6.1 in \MaldacenaIUA.}
\eqn\correlator{
G(z,\bar z) =\int_0^{\infty} d h \int_0^{ h}d \bar h \ {\cal I}_{h,\bar h}(z, \bar z)  \ \left( 1 - i \pi \gamma(h, \bar h)  + ... \right)
}
where $ {\cal I}_{h,\bar h}(z, \bar z) $ is the expression for the conformal block to together with the tree-level three-point couplings in the limit \variables\ see appendix A.
One can show that the correct identification that relates this expression to the previous section is
\eqn\identification{
16 h^2 \bar h^2 = p^2 \bar p^2, ~~~ {h \over \bar h} + {\bar h \over h} =  -{2 p. \bar p \over |p| |\bar p| }  \ .
}
In these variables the $s$-channel conformal block weighted by the three-point function of generalized free field theory takes the following form, see appendix C,
\eqn\blockgeneralized{\eqalign{
{\cal I}_{h, \bar h} (z, \bar z) &= (- x^2)^{\Delta_{\phi \psi}} C(\Delta_\psi) C(\Delta_{\phi}) \int_{M^+} {d p \over (2 \pi)^d} {d \bar p \over (2 \pi)^d} \ (- p^2)^{\Delta_\psi - {d \over 2}} \ (- \bar p^2)^{\Delta_\phi - {d \over 2}} e^{p.x} e^{ \bar p . \bar x} \cr
&4 h \bar h (h^2 - \bar h^2) \delta \left( {p. \bar p \over 2} + h^2 + \bar h^2 \right) \delta \left( {p^2 \bar p^2 \over 16} - h^2 \bar h^2 \right) .
}}

In this form it is trivial to take integrals over $h$ and $\bar h$
\eqn\sumoverblocks{\eqalign{
\int_0^{\infty} d h \int_0^{h} d \bar h ~ 4 h \bar h (h^2 - \bar h^2) \delta \left( {p. \bar p \over 2} + h^2 + \bar h^2 \right) \delta \left( {p^2 \bar p^2 \over 16} - h^2 \bar h^2 \right)  = 1 .
}}

As as result we get the following expression for the correlator
\eqn\expressioncorr{
A(x , \bar x) =C(\Delta_\phi) C(\Delta_{\psi}) \int_{M^+} {d p \over (2 \pi)^d} {d \bar p \over (2 \pi)^d} \ (- p^2)^{\Delta_\phi - {d \over 2}} \ (- \bar p^2)^{\Delta_\psi - {d \over 2}} e^{p.x} e^{ \bar p . \bar x} \left( 1 - i \pi \gamma(S, L) \right)
}
where $S$ and $L$ are related to the original quantum numbers through \identification . To relate to the phase shift we can rotate $p \to - i p$ and $\bar p \to - i \bar p$, so that $S \to e^{- i \pi} S$. In this way we get
\eqn\identification{
\gamma(S, L) = {\delta(e^{i \pi}  S ,L) \over  \pi } .
}
For the simple case of Witten diagram \identification\ simply becomes $\gamma(S,L) =- {\delta(S ,L) \over  \pi}$. A detailed microscopic analysis of AdS exchanges can be found in a recent paper \AldayGDE .

This example gives a nice interpretation for the phase shift in case of a simple exchange in the bulk in terms of anomalous dimensions of double trace operators. Notice that in this particular example the phase shift is purely real which corresponds to the fact that the scattering is elastic. In this case in the $s$-channel it is reproduced via double trace operators only. More generally, the phase shift will be complex and in the $s$-channel new single trace operators will be important as well \PaulosFAP. It corresponds to inelastic effects in the bulk. It would be very interesting to understand the mapping between the phase shift and the $s$-channel data beyond the simple elastic regime considered in this section.

\newsec{Conformal Regge Theory in the Impact Parameter Space}

In this section we start by briefly reviewing conformal Regge theory. 
We take the usual steps reviewed in detail in \refs{\GribovNW,\CaronHuotVEP,\CostaCB}. 
We consider a large $N$ CFT and focus on a planar correlator. We assume that the leading singularity in the $J$ plane is a pole. The outcome of this analysis is an expression for the contribution of the Regge pole to the correlation function.  We then consider spin one operators and generalize the usual conformal Regge analysis to this case. For a previous discussion see \CornalbaAX .

\subsec{Conformal Regge Theory in the $t$-channel: Scalars}

On the principal sheet the $t$-channel blocks have the usual asymptotic, but the continuation $z \to e^{- 2 \pi i} z$ has a dramatic effect on the $t$-channel blocks which develop ${1 \over (1-z)^J}$ singularity in the $z \to 1$ limit. 
The $t$-channel OPE becomes divergent and is not reliable away from the light-cone limit $ z \to 1$ with $\bar z$ held fixed.
The way to deal with this is the subject of Regge theory and we briefly review it here,  using our notations.

Our starting point is the scalar correlator
\eqn\sccorry{  \la  \phi\left( y_4 \right) \psi \left(y_3 \right) \psi \left( y_2 \right) \phi\left(y_1 \right)  \ra  = { \AA(u,v) \over y_{14}^{2 \Delta_\phi} y_{23}^{2 \Delta_\psi}}   \,\,, }
where again
\eqn\defuv{   
u= {y_{12}^2 y_{34}^2\over y_{13}^2 y_{24}^2} =z \bar z, \qquad v= {y_{14}^2 y_{23}^2\over y_{13}^2 y_{24}^2} = (1-z) (1-\bar z)\,\, .
}
As explained above, it will be useful to use different Poincare patches \finiteform\  for each operator insertion.
The conformal mapping  \finiteform\ yields
\eqn\sccorr{  \la \phi\left(-{x \over 2} \right) \psi \left(-{\bar x \over 2} \right) \psi \left({\bar x \over 2} \right) \phi\left({x \over 2} \right)  \ra  = {\AA(u,v) \over (-x^2)^{\Delta_\phi} (-{\bar x}^2)^{\Delta_\OO} }  }
 and the cross ratios take the form
 \eqn\crossratiosB{\eqalign{
 u &= (1 - \sigma e^{- \rho})  (1 - \sigma e^{\rho})  =  \left( {1 + {x^2 \bar x^2 \over 16} - {x.\bar x \over2} \over 1 + {x^2 \bar x^2 \over 16} + {x.\bar x \over2} } \right)^2  , \cr
 v &= \sigma^2 = {x^2 \bar x^2 \over (1 + {x^2 \bar x^2 \over 16} + {x.\bar x \over2})^2} \,.
 }}
 It will be useful to keep in mind that with the specific kinematics, when $x^2$ and $\bar x^2$ are future timelike, $\sigma<0$. The leading term in this equation gives \crossratios, but we will need subleading terms as well which could be easily computed using \crossratiosB.
 
We now quickly review the standard argument of \refs{\GribovNW,\CaronHuotVEP,\CostaCB}. A convenient starting point is the partial wave expansion of the correlation function
\eqn\arelpos{ 
\AA (u, v) {=}  \sum_{J=0}^\infty    \int d\nu b_J(\nu^2) F_{\nu,J}(u,v)    
}
where the partial waves are given by a sum of the corresponding conformal block and its shadow with a particular relative coefficient:
\eqn\fnuj{\eqalign{
F_{\nu,J}(u,\upsilon)  &= \kappa_{\nu,J} g^{(0,0)}_{{d \over 2}+i\nu,J}(1-z, 1-\bar z)  + \kappa_{-\nu,J} g^{(0,0)}_{{d \over 2}-i\nu,J}(1-z, 1-\bar z) , \cr
\kappa_{\nu,J}&={i\nu\over 2\pi K_{{d\over2}+i\nu,J}}\,.
}}
Here $K_{\Delta, J} $ is a known function and its explicit form could be found, for example, in eq (41) of \CostaCB. In the Regge limit the partial waves take the following form
\eqn\reggelimitpartial{\eqalign{
F_{\nu,J}(\sigma,\rho)  &\simeq i 4^{1+J} \pi^{{d \over 2}} \sigma^{1-J} \gamma(\nu,J) \gamma(-\nu,J) \Omega_{i \nu} (\rho) + ...  \ , \cr
\gamma(\nu,J) &=\Gamma \left( 2 \Delta_\psi + J+ i \nu - {d \over 2} \over 2 \right) \Gamma \left( 2 \Delta_\phi + J + i \nu - {d \over 2} \over 2 \right) , 
}}
where we wrote only the leading contribution in the small $\sigma$ limit. Notice that \arelpos\ together with \reggelimitpartial\ are identical to the Regge limit of the Witten diagram \exchangeReggeB\ if we set $b_{J}(\nu^2) \sim {1 \over \nu^2 + (\Delta - {d \over 2})^2}$.  

At the same time the partial wave expansion is closely related to the OPE and integral over $\nu$ is what substitutes the usual sum over dimensions. Indeed, substituting \fnuj\ into \arelpos\ we get
\eqn\intcb{ \AA(u,v)  =  2 \sum_J \int d\nu b_J(\nu^2) \kappa_{\nu,J}g^{(0,0)}_{{d\over2}+i\nu,J}(1-z, 1-\bar z)  ,}
where we used the symmetry of the integrand under $\nu \to - \nu$ which explains the origin of the factor of $2$ in \intcb. Next we notice that $g^{(0,0)}_{{d\over2}+i\nu,J}(1-z, 1-\bar z) $ decays exponentially in the lower $\nu$-plane, which means that we can close there the contour of integration.

\ifig\Contour{Contour Deformation. We switch from the sum over spins $J$ to a contour integral. We then deform the contour and pick the contribution from a singularity with the maximal ${\rm Re}[J]$, which we assume to be a pole.} {\epsfxsize4.5in\epsfbox{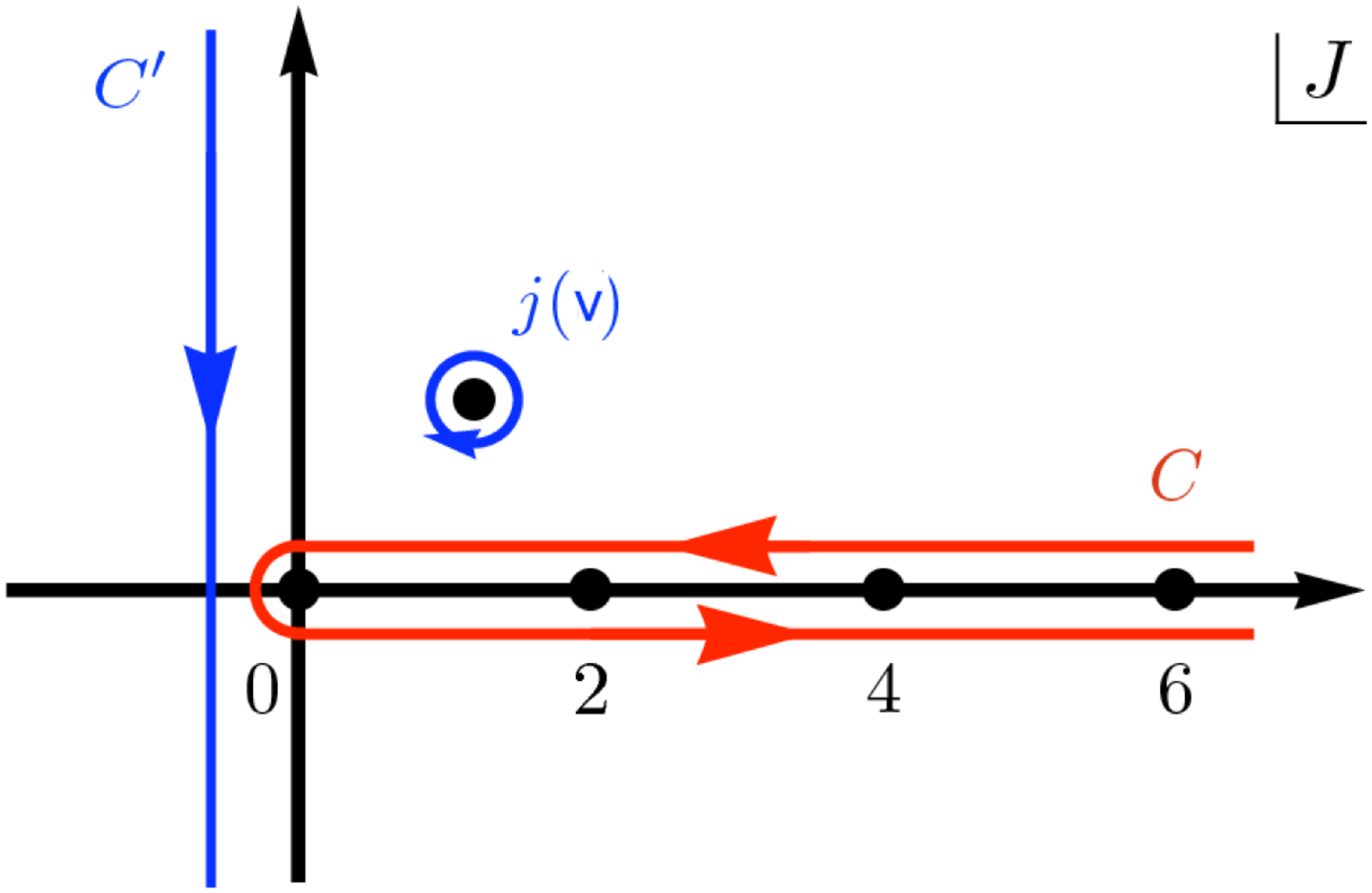}}

We recover the usual OPE, provided that partial amplitudes, $b_J(\nu^2)$, have the following singularities
\eqn\bjapp{  b_J(\nu^2) \approx {r[\Delta,J]\over \nu^2+(\Delta-{d \over 2})^2 }\,, }
with residues given by:
\eqn\resdef{r[\Delta,J]\equiv \lambda_{12\OO_{\Delta, J}} \lambda_{34\OO_{\Delta, J}}  K_{\Delta, J}\,.}
There are also poles in $\kappa_{\nu,J}$ at the following positions\foot{Our discussion of the possible singularities here is far from exhaustive. For more details, see for example \refs{\CostaCB,\CaronHuotVEP}. }
\eqn\followingpositions{
\nu^2 + \left(2 \Delta_\psi + 2 n - {d \over 2}+J\right)^2 = 0,\qquad \nu^2 + \left(2 \Delta_\phi + 2 n - {d \over 2}+J\right)^2=0\,,
}
where $n\geq 0$ is an integer. In finite $N$ CFTs these should be canceled by zeros in $ b_J(\nu^2)$. On the other hand, in large $N$ CFTs poles \followingpositions\ correspond to the contribution of double trace operators and are physical. In this way, for planar correlators in the large $N$ limit, $ b_J(\nu^2)$ is finite for $\nu$ satisfying \followingpositions\ and has poles only at the position of single trace operators. Indeed, this is what we observed in the Witten diagram example.\foot{The partial waves defined here differ from those in \CaronHuotVEP\  by an overall factor of $2 \kappa_{\nu,J}$ and by those defined in \DolanDV\  by a factor equal to $4^{J-1} \pi\gamma(\nu,J)\gamma(-\nu,J)$. In both cases, the overal factor includes the product $\gamma(\nu,J)\gamma(-\nu,J)$ with poles at the positions of double-trace operators. This choice for the partial waves, which is natural from the point of view of the Mellin representation \CostaCB, is more convenient when dealing with large-N theories.}

The starting point of Regge theory is an idea that similarly to the sum over dimensions we can turn the sum over spins into an integral $\sum_{J} \to \int_{ {\cal C}} {d J \over 2 \pi i} {1 \over \sin \pi J}$, where the contour of integration goes around the real axis.\foot{There is a usual subtlety that odd and even spins should be treated separately. Here we have only even spins due to our choice of the correlator.} This requires analytic continuation of the partial amplitudes away from integer spins. In the context of scattering amplitudes this is achieved via the so-called Froissart-Gribov representation \GribovNW. In the context of CFTs, this analytic continuation was recently achieved in \CaronHuotVEP. Here we simply assume that such continuation is possible.\foot{Some subtleties related to this procedure are discussed below in section 5 when we discuss fixed singularities and finite subtractions.}

The crucial step is then to deform the $J$ contour from $\CC$ to $\CC'$ (see \Contour) and notice that in the Regge limit the singularity with the maximal ${\rm Re}[J]$ will dominate.  In this paper we assume that this singularity is a pole. It would be nice to understand how robust is this assumption, at least in the context of large $N$ theories. The common belief states that it is always the case in large $N$ CFTs.  

The leading singularity is related to the so-called leading Regge trajectory, namely the set of operators $\Delta[J]$, with maximal spin for a given dimension. The stress energy tensor operator corresponds to $J=2$. The corresponding singularity in the $J$-plane appears at
\eqn\singularity{
\nu^2 + \left( \Delta(j(\nu)) - {d \over 2} \right)^2 = 0 \ ,
}
where the residue of this pole is again related to the analytically continued three-point coupling due to \bjapp . In the context of large $N$ theories these could be either single trace or double trace operators. As we will see singularities due to double trace operators disappear upon taking the Fourier transform. Henceforth, we focus on the leading contribution coming from single trace operators.

In our case, only even spin operators are being exchanged in the $t$-channel. At the same time, $t$-channel blocks satisfy the identity \DolanDV: 
\eqn\cfid{ g_{\Delta,J} \left({u\over v},{1\over v}\right)=(-1)^J g_{\Delta,J}(u,v) \,.} 
With the help of this identity, we can express the sum over even integer spins as an integral
\eqn\formulaB{
\AA (u, v) {=}   \int_\CC {d J \over 2 \pi i} {\pi \over 2 \sin \pi J}  \int_{-\infty}^\infty d\nu\,  b_J(\nu^2) \left( F_{\nu,J}(u,v)  + F_{\nu,J}\left({u \over v},{1 \over v} \right)\right)\,.
}
A similar expression could be found, e.g., in \CornalbaFS. Close to the pole, the partial wave amplitude becomes
\eqn\closetothepole{
b_{J}(\nu^2) \simeq - {j'(\nu) r[\Delta(j(\nu)),j(\nu)] \over 2 \nu (J - j(\nu))} + ... \,,
}
where $r[\Delta(j(\nu)),j(\nu)]$ represents the analytic continuation of the residue. We deform the contour from $\CC$ to $\CC'$, see \Contour, and pick the contribution of the leading Regge trajectory pole using \closetothepole. The correlation function in the Regge limit takes the following form \CostaCB\
\eqn\exchangeReggeB{\eqalign{
\AA &=1 - 2 \pi i \int_{- \infty}^{\infty} d \nu \, r[\Delta(j(\nu)),j(\nu)]\, \alpha(\nu) \sigma^{1 - j(\nu)}  \Omega_{i \nu} (\rho) + ... \ , \cr
\alpha(\nu) &=- {\pi^{{d \over 2} - 1} 4^{j(\nu)+1} e^{-i \pi j(\nu)/2} \over 2\, \sin \left( {\pi j(\nu) \over 2} \right)} \gamma(\nu,j(\nu)) \gamma(-\nu,j(\nu)) \beta(\nu) \ , \cr
\gamma(\nu,j(\nu)) &=\Gamma \left( 2 \Delta_J + j(\nu) + i \nu - {d \over 2} \over 2 \right) \Gamma \left( 2 \Delta_\psi + j(\nu) + i \nu - {d \over 2} \over 2 \right) , \cr
\beta(\nu) &= {\pi \over 4 \nu} j'(\nu)  \ ,
}}
where by construction $j(\nu) = j(-\nu)$.\foot{Our expression for $\alpha(\nu)$ differs from eq.(57) of \CostaCB\ by an overall factor of $-2^{j(\nu)+2}$.}
This looks very similar to the bulk exchange diagram. The main difference is that instead of a fixed spin $J$ we have $j(\nu)$, namely the spin is a nontrivial function of the scaling dimension which is dictated by the solution of \singularity\ over which we have to integrate.

Using the Fourier transform to the impact parameter space we find that the phase shift is equal to:
\eqn\phaseshift{\eqalign{
\delta(S, L) &=16 \ \tilde{c}_0\,\int_{- \infty}^{\infty} d \nu \, \, {\pi^{{d \over 2} }  e^{-i \pi j(\nu)/2} \over \,\sin \left( {\pi j(\nu) \over 2} \right)} \, r[\Delta(j(\nu)),j(\nu)] \,\beta(\nu)\, S^{j(\nu) - 1} \, \Omega_{i \nu} (L) \cr
\tilde{c}_0&= \Gamma(\Delta_\phi)\Gamma(\Delta_\psi) \Gamma(\Delta_\phi+1-d/2)\Gamma(\Delta_\psi+1-d/2)
}}
Note that the pre-factors $\gamma(\Delta(j(\nu)),j(\nu))$, which contain the contribution from the double trace operators in the Regge limit, do not enter the result for the phase shift.

\subsec{Conformal Regge Theory in the $t$-channel: Vectors}

\noindent We are now ready to compute the Regge limit of the correlator  which involves two vector operators \correlator :
\eqn\corvectory{ 
\la  J^m \left( y_4 \right) \psi \left(y_3 \right) \psi \left( y_2\right) J^n \left(y_1 \right)  \ra  = {\AA^{mn} \over (y_{23}^2)^{\Delta_\psi} (y_{14}^2)^{\Delta_J+1}} \, \,.
 }
As explained in \refs{\CostaMG,\CostaDW} this correlator can be decomposed into conformal blocks which are related to the scalar conformal blocks by the action
of certain differential operators:
\eqn\cbvector{ \AA^{mn}  =\sum_{\Delta,J}        \lambda_{\psi \psi \OO_{\Delta,J}}  \sum_{k=1}^5  \lambda_{JJ\OO_{\Delta,J},k}   
            \,\, \hat D_k^{mn} G_{k,\OO_{\Delta,J}}(1-z,1-z) \,,
           }
where the functions $G_{k,\OO}(u,v)$ are proportional to the scalar conformal blocks (their explicit definitions, together with those of the differential operators $\hat{D}_k^{mn}$, can be found in Appendix D).

The action of the operators $  \hat D_k^{mn}$ can be written in a convenient basis
\eqn\dhatbasis{ \sum_{k=1}^5  \lambda_{JJ\OO_{\Delta,J},k}  \,\,   \hat D_k^{mn} G_{k,\OO_{\Delta, J}}(1-z,1-z)  = \sum_{k=1}^5     Q^{(k)mn}  f_k(u,\upsilon) 
}
so that
\eqn\aareduced{  \AA^{mn}   = \sum_{\Delta,J}     \lambda_{\psi\psi \OO_{\Delta,J}} \sum_{k=1}^5   Q^{(k)mn}     f_k(u,\upsilon)  }
where 
\eqn\defq{   Q^{(k) mn}  =   \{    H_{12}^{mn}, V_1^m V_2^n,  (V_1')^m V_2^n, V_1^m (V_2')^n, (V_1')^m (V_2')^n \}   }
are the conformally covariant structures defined in  \refs{\CostaMG,\CostaDW} and the $f_i$'s are known functions involving derivatives acting on the relevant scalar conformal blocks. Both the $Q^{(k) mn}$ and the $f_i$'s can be found in appendix D. At this point the problem becomes identical to that of the scalar correlator, the only difference being the presence of the differential operators.

Repeating the steps of the previous section, this time for the case of vector operators, we arrive at the following result (for technical details regarding this computation please refer to appendix E)
\eqn\reggecor{\eqalign{
\AA^{mn} (x,\bx) &\simeq  \AA^{mn}_0(x,\bx)+2\pi i\int_{-\infty}^\infty d\nu\, \sum_{k=1}^4 \alpha_k(\nu)\,  \left(x^2\DD_k^{mn}\right) { \Omega_{i\nu}^j(\rho)
                \over  (-x^2)^{\Delta_J+1}  (-\bx^2)^{ \Delta_\psi } (-\sqrt{x^2\bx^2})^{j(\nu)-1}}  \cr
                \alpha_k(\nu)&=r_k[\Delta(j(\nu)),j(\nu)]\,\alpha(\nu)\,.
                }}
Here $\AA^{mn}_0(x,\bx)$ denotes the contribution of the disconnected piece, $\alpha(\nu)$ is defined in \exchangeReggeB\ and the $r_k[\Delta(j(\nu)),j(\nu)]$ represent the analytic continuations of the relevant three-point couplings, $r_k[\Delta,J]=\lambda_{\psi\psi\OO}\lambda_{JJ\OO,k}K_{\Delta,J}$. Explicit formulas for the three-point couplings $r_k[\Delta,J]$ appear in Appendix E. Their detailed form depends on the parametrization used for the three-point functions. Finally, the $ \DD_k^{mn} $ are differential operators defined as follows: 
\eqn\diffdefshort{\eqalign{
x^2\DD_1^{mn}&=x^2 \eta^{mn}-x^m x^n \ , ~~~ x^2\DD_2^{mn}=x^m x^n \ , \cr 
x^2\DD_3^{mn}&=x^2 (x^m \p^n+x^n\p^m) \ , \cr
x^2\DD_4^{mn}&=(x^2)^2 \p^m\p^n +  x^2 (x^m \p^n+x^n\p^m) - {1\over d-1} (x^2\eta^{mn}-x^m x^n) x^2\p^2 \ .
}}

Fourier transforming \reggecor\ to the impact parameter space we obtain an expression for the phase shift
\eqn\phaseshiftvec{
\delta^{mn}(S, L) = 
-4\tilde{c}_0 \,\sum_{k=1}^4 \int_{- \infty}^{\infty} d \nu \,{\pi^{d\over 2}\,e^{-i \pi j(\nu)/2} \over \sin \left( {\pi j(\nu) \over 2} \right)}  \beta_k(\nu) S^{j(\nu) - 1} 
  \hat \DD_k^{m n}  \Omega_{i \nu} (L) \ .
}
where the differential operators $\hat  \DD_k^{mn} $ are 
\eqn\DDdef{\eqalign{\hat{\DD}_1^{mn}&=\eta^{mn}-{p^m p^n\over p^2}, ~~~ \hat{\DD}_2^{mn}={p^m p^n\over p^2} , \cr
\hat{\DD}_3^{mn}&=p^m \p^n+p^n\p^m \ , \cr
\hat{\DD}_4^{mn}&=p^2\p^m\p^n+(p^m \p^n+p^n\p^m)-{1\over d-1}\left(\eta^{mn}-{p^m p^n\over p^2}\right)p^2 \p^2\,,
}}
and the $\beta_k(\nu)$ are linear combinations of the  residues $r_k[\nu, j(\nu)]$ whose explicit form can be found in appendix F.

Formulas \reggecor\ and \phaseshiftvec\ are the main results of this subsection. We will use them to understand what constraints can be obtained in the Regge limit from Rindler positivity and causality.

\subsec{Contribution of the Stress Tensor Pole}

\noindent In this section, we evaluate the contribution to the phase shift $\delta(S,L)$ from the stress-tensor exchange. First, we consider the simple case of the scalar correlator, where the phase shift is given by \phaseshift. 
The starting point is an expansion close to the pole following from
\eqn\minimalansatz{
j(\nu) = 2 + \, c \left(\nu + i {d\over 2} \right) +\cdots
}
consistent with $j( -i {d \over 2}) =2$. In this case, we expect that
\eqn\resanalscalars{
r[\Delta(j(\nu)), j(\nu)]\simeq r[\Delta=d,j(\nu)=2]+\cdots \, \, \simeq \lambda_{\phi\phi T}\lambda_{\psi\psi T} K_{d,2}\,.
} 
Using \minimalansatz\  we find:
\eqn\betaoversin{
{\beta(\nu)\over \sin{\pi \, j(\nu)\over 2}}= -{i \over d} \,{1\over (\nu+i{d\over 2})}+\cdots\,.
}
This leads us to following expression for the phase shift that corresponds to the graviton pole:
 \eqn\phaseshiftsc{\eqalign{
 \delta(S,L)&=- 16 \tilde{c_0} \pi^{{d \over 2} } \,\, \lambda_{\phi\phi T}\lambda_{\psi\psi T} K_{d,2}\,\, S\, \,{- i \over d} {\rm Res}\left[ \ {i \nu\Pi_{i\nu+{d/2}-1}(L)\over \pi } \right]_{\nu = - i {d \over 2}}\, \cr
 &=16 \tilde c_0 \pi^{{d \over 2}}   \lambda_{\phi\phi T}\lambda_{\psi\psi T} K_{d,2} \ S \ \Pi_{d-1}(L) .
 }}
 As expected this precisely coincides with \resultphaseshift\ upon choosing $J=2$ and $\Delta = d$.
To arrive at \phaseshiftsc\ we used the relation between $\Omega_{i\nu}$ and $\Pi_{i\nu+{d\over 2}-1}$ and closed the contour in the lower half-plane.

Let us now consider the correlation function of section 4.2 with vector operators. We can work out this case in precise analogy to the scalar case, the only difference being the presence of four different structures. It is easy to see that instead of \phaseshiftsc\ we now find
 \eqn\phaseshiftvc{\eqalign{
 \delta^{mn}(S,L)=-4 \tilde c_0 \pi^{{d \over 2}} \, S \ \sum_{k=1}^4\, \beta_k(\nu=-i{d\over 2})D_{k}^{mn}\Pi_{d-1}(L) .
 }}
were $\beta_k(\nu)$ are defined in eqs. (F.11-F.15) as linear combinations of the analytically continued residues $r_k[\nu,j(\nu)]$. Here like before, we expect that 
\eqn\residuesvec{
r_k[\Delta(j(\nu)), j(\nu)]\simeq r_k[\Delta=d,j(\nu)=2] \,.
}

We can evaluate the $\beta_k(\nu)$ at the graviton pole, with the help of \minimalansatz\  and \residuesvec\ together with (F.11-F.15) and (E.14). It will be convenient to express the result in terms of the coefficients $(n_s,n_f,n_3)$ which parametrize the three-point function $\vev{JJT}$ in the ``free field'' basis. The first two parameters, correspond to the number of scalars and fermions of a free theory,  whereas $n_3$ is related to the other two via the Ward Identity of the stress-tensor: $n_3={\Delta_J-d+1\over (d-1)(d-2)} (4(d-2)n_f+n_s)$. Clearly, $n_3$ vanishes for conserved currents $J$. This change of parametrization is easily achieved using (D.19). The result is\foot{Recall that $\lambda_{\psi\psi T}<0$.}
\eqn\betaTg{\eqalign{
 \beta_1 &=2 \,{(\Delta_J^2-1)(2(\Delta_J+1)-d)\over (d-1)(d-2)} \, (n_s+4(d-2)n_f) \,\lambda_{\psi\psi T} \,K_{d,2}\cr
 \beta_2&=2 \,{(\Delta_J+1)\,(\Delta_J-d+1)(2(\Delta_J+1)-d)\over (d-1)(d-2)} \, (n_s+4(d-2)n_f)\, \lambda_{\psi\psi T} \,K_{d,2}\cr
 \beta_3&=0\cr
 \beta_4 &=2 \,{1+\Delta_J\over (d-1)(d-2)} \, ( (1-\Delta_J) n_s+4(d-2)n_f)\, \lambda_{\psi\psi T} \,K_{d,2}
 }}
For conserved currents with $\Delta_J=d-1$, eq. \betaTg\ reads:
\eqn\betaTg{\eqalign{
 \beta_1 &={2 d^2\over (d-1)} (n_s+4(d-2)n_f)\, \lambda_{\psi\psi T} \,K_{d,2}\cr
 \beta_4 &=-{2 d\over d-1} (n_s-4 n_f)\, \lambda_{\psi\psi T} \,K_{d,2}\,,
 }}
with $\beta_2=\beta_3=0$ in any $d$. Observe that $\beta_1$ is proportional to the two-point function of the vector currents.
Furthermore, $\beta_1,\beta_2$ are in one-to-one correspondence with the three-point couplings of two gauge fields to the graviton. In particular, the only terms up to field redefinitions which can contibute to the effective graviton-photon-photon couplings are: 
\eqn\efftheory{S_{eff.}\propto\, \beta_1\int \, \sqrt{g_{\mu\nu}}F_{\mu\nu}^2 +\beta_4\int \sqrt{g_{\mu\nu}} R_{\mu\nu\rho\sigma} F^{\mu\nu} F^{\rho\sigma}}
where we ignored constant factors which translate between the terms in the action and the three-point functions.
As shown in \CamanhoAPA\  causality requires $\beta_4=0$ unless there are new states, which is equivalent to the absence of higher derivative terms in the effective action ($a=c$ in the supersymmetric case).

\newsec{Constraints on the Bulk Phase Shift and Leading Regge Trajectory}

In this section we derive bounds on the bulk phase shift and the leading Regge trajectory from causality and unitarity. We use Rindler 
positivity to constrain the correlator in the Regge limit. This includes finite subtractions and fixed singularities. We use causality in the impact parameter space and Wightman functions positivity to place constraints on the phase shift $\delta(S,L)$. As an application we get parametric bounds on the three-point couplings of spinning operators.

\subsec{Finite Subtractions and Particles That Do Not Lie on Regge Trajectories}

In our treatment of the partial wave sum we were not very precise about the analytic continuation away from integer $J$'s. Potentially, this continuation could be subtle. To understand the problem it is instructive to quickly discuss the four-point scattering amplitudes in flat space. 

The simplest example would be scattering in the $\phi^4$ theory. The scattering amplitude, being a constant, has only $J=0$ partial wave. Clearly, this is not part of the Regge trajectory and in the Froissart-Gribov presentation such polynomial terms enter as ``finite subtractions'' terms at infinity.

Similarly, we can consider $\phi^3$ theory. In this case all partial waves are present and the Regge analysis of this scattering amplitude is analogous to section 2 of the present paper. Namely in the $j$ plane it corresponds to the so-called ``fixed singularity'', namely $j(\nu) = {\rm const}$.

In exactly the same way we can have finite subtractions and fixed singularities in the conformal correlator. Polynomial terms correspond to  corrections to the OPE data of purely double trace operators \HeemskerkPN, whereas fixed singularities correspond to single trace operators that do not lie on the Regge trajectories (as in section 2). 

In a recent paper \CaronHuotVEP\ it was shown that in a generic CFT the OPE data that enters the expansion of the scalar correlator that we consider in this paper is analytic in spin of the exchanged operators for ${\rm Re}[J]>1$ which includes all operators with spin greater or equal than two. Generalization of this construction to operators with spin is technically more complicated but conceptually is identical to the one with scalars. In the context of large $N$ CFTs the argument of \CaronHuotVEP\ only assumes UV completion, namely that the large $N$ expansion comes form some finite $N$ well-defined CFT. This result implies that finite subtractions and fixed singularities can only affect $J=0$ and $J=1$ parts of the spectrum. This justifies our use of the Regge theory for the stress tensor which is a part of the leading Regge trajectory.

\subsec{Constraints From Rindler Positivity}

For completeness let us quickly review the form of constraints coming from Rindler positivity (see \refs{\CasiniBF,\HartmanLGU} for more details). We work in the usual Minkowski space coordinates $(y^0,y^1,\vec y)$. The Rindler reflection is defined as follows
\eqn\conjugation{\eqalign{
\overline{(t,z,\vec y)} &= (- t^*, - z^*, \vec y) \ , \cr
\overline{ O_{\mu \nu...}(t,z, \vec y)} &= (-1)^{n_{t} + n_{z}} O^{\dagger}_{\mu \nu ...} (- t^*, - z^*, \vec y).
}}
where $n_t$ and $n_z$ is the number of $t$ or $z$ vector indices correspondingly. Let us define $y^{\pm} = t \pm z$ and complexified Rindler wedges as follows
\eqn\complexifiedRindler{\eqalign{
R_{C} &= \left\{ (y^+,y^-,\vec y): ~~ y^+ y^- < 0 , ~~~ {\rm arg}[y^+] \in (- {\pi \over 2}, {\pi \over 2}) ,~ \vec y \in R^{d-2} \right\} , \cr
L_{C} &= \left\{ (y^+,y^-,\vec y): ~~ y^+ y^- < 0 , ~~~ {\rm arg}[y^+] \in ({\pi \over 2}, {3 \pi \over 2}) ,~ \vec y \in R^{d-2} \right\} .
}}

Positivity takes the form
\eqn\positivity{\eqalign{
\la \overline{O_1 (y_1)} \ \overline{O_2 (y_2)} O_1(y_1) O(y_2) \ra &> 0 , ~~~{\rm Im}[t_1] \leq {\rm Im}[t_2] \ , ~~~ y_{1,2} \in R_{C} \ .
}}
where the reader should take particular attention to the ordering of operators. The indices of spinning operators in the formula above are kept implicit. 

Moreover, in deriving the bounds it is useful to consider correlators of the type $\la \overline{A} B \ra$ to which Cauchy-Schwarz inequality applies
\eqn\cauchyschwarz{
| \la \overline{A} B\ra |^2 \leq \la \overline{A} A \ra \la \overline{B} B \ra .
}

As one can easily check the correlator of the type we considered in the previous sections, namely
\eqn\correlator{
A(x, \bar x) = \eps^*_{m} \eps_{n} \la J^{m}\left({x \over 2} \right) \psi \left(-{\bar x \over 2} \right) \psi \left({\bar x \over 2} \right) J^{n}\left(- {x \over 2} \right)  \ra , 
}
is exactly of the type \cauchyschwarz . For time-like future-oriented $x$ and $\bar x$, we have $y_2$ and $y_4$ lying in the right Rindler wedge. Similarly, $y_1$ and $y_3$ belong to the left Rindler wedge. Moreover, for the configuration of points and polarizations \correlator\ we have $\eps_{\mu}(y_1)   J^{\mu}(y_1) = \overline{\eps_{\nu}(y_4)J^{\nu}(y_4)}$ and $\psi(y_3) = \overline{ \psi(y_2)}$. As a result, applying \cauchyschwarz, we get
\eqn\boundonchaos{
| A(x, \bar x)  | \leq  \eps^*_{m} \eps_{n} \la J^{m}\left({x \over 2} \right) J^{n}\left(- {x \over 2} \right)  \ra \la O \left({\bar x \over 2} \right) O \left(- {\bar x \over 2} \right) \ra \left( 1 + ... \right)
}
where corrections to the disconnected part in the RHS come with positive powers of $x,\bar x$ and are irrelevant for everything we say below. Next we consider \boundonchaos\ in the Regge limit. Let us emphasize again that the argument of this section applies only to $x, \bar x$ in the future Milne wedge.\foot{For example, we cannot use \boundonchaos\ after doing the Fourier transform because it involves integration over spacelike $x$ and $\bar x$.}

If we analytically continue $x \to e^{i \alpha} x$ then $y^+ y^-$ stays intact (and similarly $\bar x$). We would like the points to stay in the complexified Rindler wedge. This constrains possible values of $\alpha$. From  \complexifiedRindler\ together with \substitution\ it follows that $\alpha , \bar \alpha \in  ( - {\pi \over 2} , {\pi \over 2} ) $ . Moreover, for a given ordering as explained in \HartmanLGU\ we would like the complexified point $y_2$ to be in the past light-cone of point $y_3$ (and similarly for points $y_{1,4}$). For small $x$ and $\bar x$ this implies that $\alpha, \bar \alpha < 0$.

First, let us understand the constraints we get from \boundonchaos\ on finite subtractions and fixed singularities in the context of large $N$ CFTs. Without loss of generality we focus on the case of scalar correlators. 
In the Regge limit, we keep $\rho$ fixed and send $\sigma \to 0$. Both fixed $J$ singularity or contact interaction in AdS (finite subtraction in the CFT language) produce\foot{As shown in \MaldacenaWAA\ for ${\rm Arg}[\sigma]= {\pi \over 2}$ the correlator is real.}
\eqn\quoteRegge{
\AA(\sigma, \rho)= 1 - i {f(\rho) \over \sigma^{J_0-1}} ,
}
where an extra minus sign is due to the fact that the correlator \correlator\ is anti time-ordered versus the time-ordered one considered earlier. In terms of $\sigma$ positivity in the complexified Rindler wedges \boundonchaos\ corresponds to
\eqn\positivity{
|\AA(\sigma, \rho) | < 1 , ~~~ {\rm Im}[\sigma] > 0 ,
}
where the argument of $\sigma$ follows from the constraints on $\alpha$ and $\bar \alpha$, together with $\sigma =- \left( x^2 \bar x^2 \right)^{1/2}$.
This implies that 
\eqn\boundonsmth{
J_0 \leq 2 \ \ {\rm and} \ \ f(\rho) \geq 0 \ . 
}
In practice this means that all higher spin particles belong to Regge trajectories and that any finite subtraction terms (e.g. purely double trace solutions to the bootstrap equation in the context of large N theories \HeemskerkPN) must satisfy the same positivity constraint $f(\rho)\geq 0$.

Next let us discuss constraints on the leading Regge trajectory $j(\nu)$. Recall the result for the correlator in the $\sigma \to 0$ limit
\eqn\reggeresult{
\AA(\sigma, \rho) =1 - 2 \pi i \int_{- \infty}^{\infty} d \nu \, r[\nu,j(\nu)]\, \alpha(\nu) \sigma^{1 - j(\nu)}  \Omega_{i \nu} (\rho) .
}

For small $\sigma$ the integral is dominated by the minimal value of $j(\nu)$, which according to \positivity\ satisfies  $j(\nu) \leq 2$. The usual assumption is that the integral is dominated by $\nu = 0$. In this case the asymptotic is controlled by $j(0)$ which is related to the Lyapunov exponent for out-of-time ordered correlators in Rindler space as $\lambda_L = j(0)- 1$ \MaldacenaWAA.

Notice that taking the limit $\rho \to 0$ does not lead to any new constraints because the relevant function $\Omega_{i \nu}(\rho)$ is regular in this limit as explained in section 2. This regularity is the consequence of double trace operators that are not negligible in this limit, see also discussion in appendix D of \AfkhamiJeddiNTF. The same is true for the polynomial corrections which are regular in $\rho$. To make connection with the work of \CamanhoAPA\ we consider causality in the impact parameter space.

For vector operators, the picture is similar. When $\rho\to 0$ there are no new constraints because there is no singular behavior in this limit. At large $\rho$, fixing $\bx=(1,0,0,0)$ Rindler positivity leads to ``half'' of the Hofman-Maldacena constraints, just as in \HartmanDXC.

\subsec{Causality in the Impact Parameter Space}

The Rindler positivity argument is completely rigorous but it does not allow us to probe the scattering at fixed impact parameter. Indeed, to compute the bulk phase shift $\delta(S,L)$ we are making the Fourier transform which involves probing spacelike $x$ and $\bar x$ corresponding to configurations of points that live outside of the original Rindler wedges. This is in accord with the work of \MaldacenaIUA\ where scattering in the bulk was probed by studying the correlator in the regime $x^2, \bar x^2 >0$.

Microscopic CFT causality states that
\eqn\causalityCFT{
[\psi(y_1) , \psi(y_2)] = 0, ~~~ y_{12}^2 >0
}
for spacelike separations $y_{12}^{\mu}$. The condition \causalityCFT\ is an operator equation and, thus, should be true inside any correlation function. In particular, if we consider the correlation function
\eqn\accumulatedshift{
f_{p} (\bar x) =\int d x \ e^{i p x}  \la \phi\left({x \over 2} \right) \psi \left({\bar x \over 2} \right) \psi \left(-{\bar x \over 2} \right) \phi\left(- {x \over 2} \right)  \ra
}
causality \causalityCFT\ implies that $f_p (\bar x)$ should be analytic for time-like $\bar x$ which correspond to spacelike separations of $\psi \psi$. Otherwise, we would get a non-zero commutator. This is not quite the usual $\la \Psi | [\psi, \psi]| | \Psi \ra$ but corresponds to the commutator where we projected on a particular final state.\foot{In \CamanhoAPA\ this projection was achieved by considering coherent states.}

Using the impact parameter representation for the correlator we get
\eqn\fimpactpar{
f_{p} (\bar x) =\int_{M^+} d \bar p \,\, e^{- i \bar p \bar x} B_0 (p, \bar p) (1 + i \delta(S,L)) .
}

The form \fimpactpar\ is not enough to see the problem. To see the problem we assume that there is a regime where the scattering phase exponentiates $\left(1 + i \delta(S,L) \right)^M \to e^{i M \delta(S,L)}$.  The conditions when it is possible were discussed in \CamanhoAPA. In the present context this corresponds to considering a multiple insertions of $\phi$ with properly arranged wave functions
\eqn\exponentiation{
\la \phi_{\bar 1} \phi_{\bar 2} ... \phi_{\bar M} \psi \left({\bar x \over 2} \right) \psi \left(-{\bar x \over 2} \right) \phi_{M} ... \phi_{1} \ra, ~~~ M \gg 1 .
}
Of course, given an abstract CFT a separate argument is needed that the exponentiation could be actually achieved.

The exponentiation in a large N, large gap CFT is natural from the point of view of the $s$-channel due to the relation of the phase shift to the anomalous dimension of the double trace operators $\gamma(n , s)$ which do exponentiate as $e^{- i \pi \gamma(n,s)}$. But in this case to keep $\delta$ small but finite we would have to scale $S$ with $N^2$ since $\delta \sim {S \over N^2}$. This is the so-called eikonal regime and was studied, for example, in \refs{
\CornalbaXK,\CornalbaXM,\CornalbaZB,\CornalbaQF}.

In the regime of our interest, namely when the phase shift is purely real and is given by 
\eqn\phaseshiftCC{
\delta(S,L) = S \hat \delta(L) + ...
}
we get after the exponentiation
\eqn\expphshift{
f_{p} (\bar x) \sim \int_{M^+} d \bar p (-\bar p^2)^{\Delta_O - {d \over 2}}e^{- i \bar p \bar x + i \sqrt{p^2 \bar p^2} \delta(L) }.
}
As usual we write $\bar p = t w$ with $w^2 = -1$ and integrate over $t$. As a result we get
\eqn\modelexpB{
f_{p} (\bar x)  \sim \int {d^{d-1} \vec w \over \sqrt{1 + \vec w^2}} {1 \over \left( \sqrt{1+ \vec w^2} \bar x^0 - \vec w \vec{\bar x} + \sqrt{- p^2} \hat \delta({p.\bar p \over |p| |\bar p|} )\right)^{2 \Delta}}
}

We would like to require that $f_p ( \bar x)$ does not have a discontinuity for time-like $\bar x$ (recall that time-like $\bar x$ correspond to spacelike $y$ in the original coordinates). Without loss of generality we can set $\vec{\bar x} = 0$. Then the discontinuity is given by
\eqn\modelxpC{
\int {d^{d-1} \vec w \over \sqrt{1 + \vec w^2}} {\theta(- \sqrt{- \bar p^2} \hat \delta({p.\bar p \over |p| |\bar p|})-\sqrt{1+ \vec w^2} \bar x^0) \over \left( \sqrt{1+ \vec w^2} \bar x^0 + \sqrt{- p^2} \hat \delta({p.\bar p \over |p| |\bar p|}) \right)^{2 \Delta}} ,
}
namely if for some $\vec w$ we have $\hat \delta(L) < 0$ we get a problem. In this way we get the condition of \CamanhoAPA\
\eqn\conditioncaus{
\hat \delta(L) \geq 0 .
}
In the same manner we can diagonalize the bulk phase shift for vector operators and run the same argument. 

Let us emphasize again that in this section we assumed that the phase shift takes the form \phaseshiftCC , namely that it is dominated by the graviton exchange.

\subsec{Scattering Bound from Unitarity}

It is instructive to see how unitarity constrains the bulk phase shift $\delta(S,L)$. Let us consider a CFT quantized on the Lorentzian cylinder. We imagine $\phi$ and $\psi$ to be real scalar primary operators. 
We would like to consider a state by acting on the vacuum with the pair of operators
\eqn\wavefunction{
| \Psi \ra = \int d x_1 d x_2 f(x_1, x_2) \psi (x_2) \phi (x_1) | 0 \ra .
}

Then positive definiteness condition of Wightman functions state that  \refs{\StreaterVI, \LuscherEZ, \CasiniBF,  \GrinsteinQK}
\eqn\scalarproduct{
\la \Psi | \Psi \ra = \int d y_1 d y_2  \int d x_1 d x_2 f^*(y_1, y_2) f(x_1, x_2) \la \phi(y_1) \psi(y_2)  \psi (x_2) \phi (x_1) \ra \geq 0 ,
}
where operators are ordered as written and the integration is over the Lorentzian cylinder. We would like to insert the operators in the corresponding Poincare patches. The transition between the Poincare patch coordinates and the coordinates on the cylinder involves conformal factors which can be absorbed into $f(x_1, x_2)$. Considering a Poincare patch centered at a point $\tau = 0$ and $\vec n = (1, \vec 0)$ we get
\eqn\mapping{
x^0 = {\sin \tau \over \cos \tau+ n^1}, ~~~ x^i = {n^{i+1} \over \cos \tau + n^1} \ .
}

Let us know consider the transformed state
\eqn\transformed{
| \Psi' \ra = e^{- i \pi D}  U | \Psi \ra ,
}
where $U$ implements inversion on the sphere $\vec n \to - \vec n$ and we evolve the state by global time $\pi$.\foot{For simplicity we can assume that a CFT is parity symmetric.} By considering $\la \Psi' | \Psi \ra$ we get the situation considered in the paper, see \kinematics . 

This correlator can be bounded using the Cauchy-Schwarz inequality
\eqn\cauchyschwarz{
\left| \la \Psi' | \Psi \ra \right| \leq \sqrt{\la \Psi | \Psi \ra \la \Psi' | \Psi' \ra} = \la \Psi | \Psi \ra .
}

We choose the insertion in the corresponding Poincare patches with the wave functions being
\eqn\wavefunction{
f(x, \bar x) = e^{i p x} e^{i \bar p \bar x} e^{- {(x^0)^2 + \sum_i (x^i)^2 \over \sigma^2}} e^{- {(\bar x^0)^2 + \sum_i (\bar x^i)^2 \over \sigma^2}}
}
where we imagine $x$ inserted in Poincare patch $P_1$ and $\bar x$ in Poincare patch $P_2$. The Cauchy-Schwartz inequality \cauchyschwarz\ takes the form

\eqn\inequality{\eqalign{
&\left|  \int d y_1 d y_2 f^*(y_1, y_2) \int d x_1 d x_2 f(x_1, x_2) \la \phi_{P_4}(y_1) \psi_{P_3}(y_2)  \psi_{P_2} (x_2) \phi_{P_1} (x_1) \ra \right| \cr
&\leq  \int d y_1 d y_2 f^*(y_1, y_2) \int d x_1 d x_2 f(x_1, x_2) \la \phi_{P_1}(y_1) \psi_{P_2}(y_2)  \psi_{P_2} (x_2) \phi_{P_1} (x_1) \ra ,
}}
where sub-indices $P_i$ refer to the insertion in the corresponding Poincare patch as in \kinematics.

In the large $p$ and $\bar p$ limit the correlator in the LHS of \inequality\ takes the form\foot{As explained in section 2 of \CornalbaAX\ in the Regge limit this follows from symmetries. Namely the fact that translations in the patches $P_1$ and $P_4$ act as special conformal transformations in the patches $P_2$ and $P_3$ and vice versa.}
\eqn\simplification{\eqalign{
&\la \phi(y_1) \psi(y_2)  \psi(x_2) \phi (x_1) \ra \simeq g(x_1-y_1, x_2- y_2) \cr
&=\int_{M^+} d q d \bar q e^{- i q (x_1 - y_1)} e^{- i \bar q (x_2 - y_2)} B_0(q, \bar q) (1 + i \delta(S,L)) ,
}}
whereas in the RHS we have the correlator which is dominated by the disconnected piece plus corrections suppressed by $p^{-1}, \bar p^{-1}$.

The bound then takes the form
\eqn\inequalitysG{\eqalign{
&\left| \int_{M^+} d q d \bar q \ |f(q, \bar q)|^2 B_0(q,\bar q) \left( 1 + i \delta (S,L) \right) \right| \leq   \int_{M^+} d q d \bar q \ |f (q, \bar q)|^2 B_0(q,\bar q) \left( 1 + ... \right) \ .
}}
where by $...$ we denotes terms that are suppressed in the large momentum limit. With the wave functions being wave packets centered around $p, \bar p$ the bound becomes
\eqn\boundonscattering{
\left| 1 + i \delta(S,L) \right| \leq 1 + ... \ ,
}
where $...$ denotes the terms suppressed in the large $S \gg 1$ limit. We got the bound for real $S$ and it is only useful for terms in $\delta(S,L)$ which are growing with $S$ because this terms are absent in the RHS of \boundonscattering.\foot{The bound \boundonscattering\ agrees with the discussion of \CamanhoAPA\ where the terms that do not grow with $S$ failed to exponentiate. Here terms that do not grow with $S$ are not constrained due to the analogous terms in the RHS.}

In contrast to \conditioncaus\ the bound \boundonscattering\ should be true for any $L$ and does not require $\delta(S,L)$ to be real or linear in $S$. 

In \CamanhoAPA\ using a semi-classical model of signal propagation it was observed that \boundonscattering\ should hold in the upper $S$ plane. We could not derive this from the Cauchy-Schwarz inequalities above but it looks plausible that  \boundonscattering\ is true in any CFT in the upper $S$ plane. Before doing the Fourier transform this was achieved by means of Rindler positivity in the complexified Rindler wedge or using OPE, see the recent discussion in \HartmanLGU. It is not immediately clear to us how to run these arguments after doing the Fourier transform. We leave this problem for the future.

\subsec{Bound on the Three-point Coupling}

Let us now discuss the corrections to \phaseshiftCC . First, we notice that when going to the impact parameter space the finite subtraction polynomial terms are localized at $L=0$, see appendix G. Second, when we deform the $\nu$ integral to compute $\delta(S,L)$ notice that double trace operator poles do not contribute. These poles are eliminated after doing the Fourier transform. In the bulk this corresponds to the usual intuition that exchanging particles correspond to single trace operators. Third, to evaluate the contribution of heavy single trace operators we can use the Regge theory.

As in previous works we are not able to put a precise bound on the three-point coupling in terms of the gap of the theory. The reason is that in deriving the bound some non-universal parameters appear that do not seem to be possible to fix on general grounds.
To make it precise we start with the expression for the bulk phase shift
\eqn\phaseshiftB{
\delta(S, L) =16 \ \tilde{c}_0\,\int_{- \infty}^{\infty} d \nu \, \, {\pi^{{d \over 2} }  e^{-i \pi j(\nu)/2} \over \,\sin \left( {\pi j(\nu) \over 2} \right)} \, r[\Delta(j(\nu)),j(\nu)] \,\beta(\nu)\, S^{j(\nu) - 1} \, \Omega_{i \nu} (L)  \ .
}
Notice that in the limit $S \to \infty$ the integral is expected to be dominated by the $\nu = 0$ region. This happens in the known examples both at weak and strong coupling \CostaCB.

Instead we are considering a different limit. We take $S = S_* \gg 1$ to be large and fixed and study the integral as a function of $L$. Linear dependence on $S$ is a transient phenomenon which does not happen when $S$ is either too large or too small.\foot{This is analogous to the discussion of Lyapunov exponents in the context of chaos. Similarly, in a finite $N$ theory Regge poles are expected to turn into ``Regge bumps.''}  To do the integral we use the symmetry under $\nu \to - \nu$ to rewrite it as follows
\eqn\phaseshiftB{
\delta(S, L) =16 \ \tilde{c}_0\,\int_{- \infty}^{\infty} d \nu \, \, {\pi^{{d \over 2} }  e^{-i \pi j(\nu)/2} \over \,\sin \left( {\pi j(\nu) \over 2} \right)} \, r[\Delta(j(\nu)),j(\nu)] \,\beta(\nu)\, S^{j(\nu) - 1} \  {i \nu \over \pi}\Pi_{i \nu + {d \over 2}-1}(L)\ .
}
We can deform the contour in the lower half-plane using the fact that $\Pi_{i \nu + {d \over 2}-1}(L) \sim e^{- i \nu L}$. To understand which $\nu$'s dominate we
can use the saddle point equation
\eqn\saddlepoint{\eqalign{
&j'(\nu_*) \log S_* - i L  = 0 \ . \cr
}}
When we deform the contour so that it passes through the saddle we may encounter some poles on the way when $j(\nu)$ crosses integer. These are particle exchanges in the bulk (see similar discussion in appendix E of  \CamanhoAPA). Taking the limit $L \gg 1$ pushes the contour further and as expected the Regge limit is simply dominated by the graviton exchange at large impact parameters. 

We are interested in impact parameters $L > L_*$ for which the contribution from the saddle is sub-leading compared to the contribution from the graviton, namely
\eqn\gravitonpolesaddle{
S_*^{j(\nu_*) - 2} e^{- i \nu_* L}= \eps \ll 1 \ ,
} 
where $\eps$ denotes the error to \phaseshiftCC . Given $S_*$ we can consider \gravitonpolesaddle\ as an equation for $\nu_*$. Then plugged in \saddlepoint\ it gives $L_*$ as a function of $S_*$. By construction $\nu_* < - i {d \over 2}$ so that the saddle lies above the graviton pole.

Next we evaluate the phase shift at the graviton pole, find its eigenvalues and impose that they are positive for $L > L_*$ which corresponds to the causality condition \conditioncaus.
For external vector operators the phase shift is a matrix $\delta^{mn}(p,\bp)$ hence causality constrains its eigenvalues to be positive definite. 
To diagonalize the matrix, it is convenient to set $p^0=1,\,\,\vec{p}=0$ and then use the definition of the impact parameter $\cosh{L}=-{p\cdot\bp\over\sqrt{p^2\bp^2}}$ and rotational invariance to express $(\bp^0,\vec{\bp})$ as:
\eqn\pnotpvec{\bp^0=\sqrt{-\bp^2}\,\cosh{L},\qquad \bp^1=\sqrt{-\bp^2}\, \sinh{L},\qquad \bp^i=0,\quad i=2,3,\cdots,d-1\,\,.}
With these choices, the phase shift matrix is completely determined in terms of the hyperbolic space coordinate $L$ and the OPE coefficients of $JJ\sim T$ (see appendix H for further details).

For large impact parameters, the positivity of the phase shift is equivalent to the Hofman-Maldacena constraints \HofmanAR.
On the other hand, when the impact parameter is small $L \ll 1$, the positivity condition requires $\beta_4$ to vanish both for conserved and non-conserved vector operators $J$. Hence, in the infinite gap limit the only consistent coupling to the graviton is the one derived from Yang-Mills theory.

More precisely, we get an equation of the type
\eqn\constraint{
 |\beta_4|  \leq a_0 L_*^{2}
}
where $a_0$ is a $d$-dependent constant. As before $L_*$ is the minimal impact parameter for which we can trust the computation based on the graviton pole. Expression \constraint\ assumes that $L_* \ll 1$. For very large $L_*$ the expression is different, see appendix H. The minimal length $L_*$ depends on the details of the theory. In a theory with a small gap we expect $L_* \gg 1$ and the constraint being close to the Hofman-Maldacena bound.  

Next we study theories with the large gap in the spectrum of higher spin operators. As we see below this leads to stronger bounds. For theories with a large gap $\Delta_{gap} \gg 1$ we expect the simple model 
\eqn\trajectory{
j(\nu) = 2 -  2 {\nu^2 + {d^2 \over 4} \over \Delta_{gap}^2} + ... 
} 
to be a good approximation. This is a minimal ansatz that follows from $j(\nu)=j(- \nu)$, $j(\pm i{d \over 2}) = 2$ and $j(\pm i \Delta_{gap}) \simeq 4$. The general discussion regarding this form of the Regge trajectory could be found section 3.4 in \BrowerEA . Notice that this ansatz connects two different notions of bulk locality, namely maximality of chaos $j(0) \simeq 2$ and large gap $\Delta_{gap} \gg 1$. Indeed, for the form \trajectory\ one follows from the other.  For this ansatz we get 
\eqn\saddle{
\nu_*= -{i L \Delta_{gap}^2 \over 4 \log S_*} .
}

Given $L$ and $\nu_*$, we want \gravitonpolesaddle\ to hold. We get
\eqn\saddle{
\eps = e^{- {L^2 \Delta_{gap}^2 \over 8  \log S_*}} \ll 1 .
}

Thus, we can only trust the graviton pole for $L_*^2 = 8 { 1 \over \Delta_{gap}^2 } \log S_* \log{1 \over \eps}$. In this way the bound becomes
\eqn\finalbound{
 |\beta_4|  \lesssim  { 1 \over \Delta_{gap}^2 } \log S_*
}
where $\lesssim$ stands for parametric dependance on the gap $\Delta_{gap} \gg 1$.

To get a precise bound on the three-point coupling we, thus, need to understand better the energies $S>S_*$ for which we can run the present argument. Compared to the leading Regge trajectory the further corrections are suppressed by powers of $S_*$. Moreover, to discuss the scattering at impact parameters $L$ we should localize the scatterers at distances smaller than $L$. This requires $S_* L^2 \gg 1$.\foot{See also the discussion in \CamanhoAPA\ where the same conclusion was reached.} This translates into the polynomial dependence of $S_*$ on $\Delta_{gap}$. In this way we get\foot{Recall that in  the discussion of the eikonal approximation $\log S_* \sim \log c_T$. So if we are to use eikonal reasoning the bound would be slightly weaker.}
\eqn\boundB{
 |\beta_4| \lesssim {\log \Delta_{gap} \over \Delta_{gap}^2} .
}

\subsec{Bound on the Three-point Coupling from Inelastic Effects}

Another way is to derive the same bound is to consider \boundonscattering\ for real $S$. In the context of large $N$ theories when the phase shift is suppressed by ${1 \over N}$ this bound becomes
\eqn\boundreal{
{\rm Im}[\delta(S,L)] > 0 .
}

We consider the case when the gap is large. A minimal ansatz is again
\eqn\ansatz{\eqalign{
j(\nu) &= 2 - 2{\nu^2 + {d^2 \over 4} \over \Delta_{gap}^2} + ... \ , \cr
r(\nu) &= r_{T_{\mu \nu}} - (r_{j_4} -  r_{T_{\mu \nu}}){\nu^2 + {d^2 \over 4} \over \Delta_{gap}^2} + ... \ ,
}}
which correctly captures the spin two and four microscopic data. The ansatz \ansatz\ is enough to compute the imaginary part of the phase shift for $L \Delta_{gap} \sim 1$. It is easy to see that the result is 
\eqn\resultimaginary{
{\rm Im}[\delta(S,L)] \sim r_{T_{\mu \nu}} e^{- {L^2 \Delta_{gap}^2 \over 8 \log S}} .
}
where we omitted dependence on $L$ that does not involve $\Delta_{gap}$ and will be subleading for the purpose of imposing \boundreal .
This result looks identical to the usual transverse spreading of the string \refs{\AmatiUF\AmatiTN\GiddingsBW-\BrowerEA}. In our case it came out form the minimal ansatz \ansatz. 

We can now repeat the same exercise for external vector operators. Notice that taking the derivative with respect to the impact parameter $\pa_{L}$ brings down a large factor ${\Delta^2_{gap} \over \log S}$. Thus, again by choosing the proper polarizations we get from \boundreal
\eqn\boundP{
| \beta_4 | \lesssim {\log S_* \over \Delta_{gap}^2 } 
}
as in the previous section. 

\subsec{Other Regge Trajectories}

In the section above we neglected the contribution from the saddle as compared to the contribution from the graviton pole or, equivalently, the stress tensor. We should, however, discuss the contribution from poles other than the leading Regge trajectory. These other Regge trajectories correspond to non-conserved spin two single trace operators, $\tilde{T}_{mn}$ and its higher spin cousins. We can imagine that these new massive spin two particles will fix the problem. 

As in \CamanhoAPA\ to address this possibility, we consider the four-point function of vector operators $\la J_{m} J_{a} J_{b} J_{n} \ra$. The phase shift in this case takes the form
\eqn\phaseshiftvec{\eqalign{
\delta^{mn,ab}(S, L) = 
16\tilde{c}_0 \,\sum_{k,\ell}^4 \int_{- \infty}^{\infty} d \nu \,{\pi^{d\over 2}\,e^{-i \pi j(\nu)/2} \over \sin \left( {\pi j(\nu) \over 2} \right)} S^{j(\nu) - 1} \,\, \beta_k(\nu)\beta_\ell(\nu)     \hat \DD_k^{m n}  \hat \DD_\ell^{a b}\Omega_{i \nu} (L)  \ .
}}
To distinguish the contributions of two different, single trace, spin-2 operators -- the stress-tensor and a non-conserved one -- to the phase shift in the infinite gap limit, we denote:
\eqn\betatwotwomas{\eqalign{
\hat{\beta}_4&=\beta_4\left(\nu=-i{d/2}+\cdots,j(\nu)=2+\cdots\right),\quad\cr
\widetilde{\beta}_4&=\beta_4\left(\nu=-i(\Delta_{\tilde{T}}-{d\over 2})+\cdots, j(\nu)=2+\cdots\right)
}}
where the dots represent terms suppressed by the $\Delta_{gap}^2$. The first line corresponds to the stress-tensor operator and the second to that of the non-conserved one.
Focussing on the small impact parameter limit and considering various polarizations while imposing positivity of the phase shift leads to:
\eqn\betamassiveargument{
\hat{\beta}_4^2+\widetilde{\beta}_4^2\geq 0 , \ \  \ -\hat{\beta}_4^2-\widetilde{\beta}_4^2\geq 0\,,
}
which implies that $\beta_4=\tilde{\beta}_4=0$. In other words, the contribution of non-conserved, spin-two, single trace operators can only make the problem worse and, thus, all the non-minimal couplings to massive spin two particles are bounded by the gap as well. Obviously, the same is true for non-minimal couplings to massive higher spin particles as well.
 
\newsec{Conclusions}

In this paper we studied a four-point correlation function of local operators in a CFT. The external wave functions for operators were chosen 
to describe  fixed impact parameter scattering in AdS. This corresponds to a particular Fourier transform of the correlator \transformimpact , which defines the phase shift $\delta(S,L)$ through \definitionphase . 
Here $S$ is the energy of the scattering and $L$ is the impact parameter. In principle, by changing $S$ and $L$ we can explore different physical regimes where stringy or gravitational effects become strong. We focused on the simplest regime of large but fixed energies $S \gg 1$ and large impact parameters $L > L_*$ for which the phase shift is dominated by the graviton exchange. 
In the bulk it is a familiar statement that at large distances massless particles with maximal spin dominate the interactions. From the CFT point of view it is due to the fact that stress tensor is generically a minimal twist operator of maximal spin that appear in the OPE.

By evaluating the contribution of the stress tensor to the phase shift we recover the bulk expressions of  \CamanhoAPA . The fact that these expressions followed from considering the stress tensor exchange in a generic CFT is in accord with the bulk picture, where the shock wave computation is only sensitive to the three-point couplings of the probe to gravity. Our computation in this paper is the CFT analog of the scattering amplitude computation in \CamanhoAPA .

Equipped with a CFT way to compute the phase shift we analyzed  corrections to the gravity computation using Regge theory \CostaCB. Assuming that the leading singularity in the $J$ plane is a pole, the phase shift is given by \phaseshiftvec, where $\beta_k(\nu)$ are related in a known way to the three-point couplings of the leading Regge trajectory operators to the external operators. Importantly, the phase shift  $\delta(S,L)$ does not acquire contributions from the double trace operators which are important in the  Regge limit. The leading Regge trajectory re-sums the higher spin states that were discussed in \CamanhoAPA. In particular, the phase shift $\delta(S,L)$ develops an imaginary part which corresponds to the production of stringy states which are dual to heavy single trace operators in a CFT. 

We then analyzed constraints on the phase shift due to causality and unitarity. 
First, we focussed  on the regime where the phase shift takes the simple form \phaseshiftCC.
Eq. \phaseshiftCC\ is expected to be true in a generic CFT for large enough impact parameters $L$ and is dual to a single graviton exchange in the bulk.
Assuming  that we can exponentiate the phase shift, we derived the bound \conditioncaus .
 This bound is identical to the one discussed in \CamanhoAPA. 
 Next, we considered the bound on the phase shift due to unitarity. The result is \boundonscattering\ obtained for real $S$. 
 This bound should be valid in any CFT$_{d>2}$ and for arbitrary impact parameters.  This agrees with the semi-classical unitarity analysis of \CamanhoAPA . In \CamanhoAPA\ it was, moreover, concluded that \boundonscattering\ should hold in the upper $S$ half-plane. It would be very interesting to see if the argument that led us to \boundonscattering\ could be extended in this way. 

We used these bounds on the phase shift to constrain three-point functions of spinning operators  which led to the parametric bound \boundB . The conclusion is that any non-minimal coupling to the graviton, as well as any other particle of spin two or higher, should be suppressed by the gap in the spectrum of higher spin particles.\foot{These could be either classical stable resonances as in  tree-level string theory, or two-particle states that appear in the loops as in QED. }  

Having a non-perturbative, purely CFT  definition of the bulk phase shift may lead to new insights into the gravitational physics at energies $S$ and impact parameters $L$ that are not easily accessible using bulk effective theory. Conversely, having a simple bulk intuition for $\delta(S,L)$ may shed new light into microscopic properties of CFTs. For example, one would like to study the bulk phase shift at smaller impact parameters (see e.g. the discussion in \GiddingsBW, \CornalbaAX). At smaller impact parameters  inelastic effects become important. The imaginary part of the phase shift that we found above corresponds to  tidal excitations of the string. At even smaller impact parameters, gravitational or ${1 \over N}$ corrections become important and a black hole is potentially produced in the bulk \GiddingsXY.\foot{In theories with $j(0)<{3 \over 2}$ this regime is believed to be absent \refs{\BrowerEA ,\AharonyBM }.}  It would be interesting to see if  bootstrap methods, see e.g. \refs{\RychkovIQZ,\SimmonsDuffinGJK} and references therein, could be used to chart various physical regimes of the bulk phase shift $\delta(S,L)$ in an abstract CFT.

\newsec{Acknowledgments}

We are grateful to S. Caron-Huot, M. Costa, T. Dumitrescu, T. Hartman, Z. Komargodski, J. Penedones, M. Strassler,  and Xi Yin for useful discussions. M.K. is supported in part by a Marie-Curie fellowship with project no 203972 within the European Research and Innovation Programme EU H2020/2014-2020. The work of AZ is supported by a Simons Investigator Award from the Simons Foundation. We would like to thank GGI for hospitality during the workshop ``Conformal Field Theories and Renormalization Group Flows in Dimensions $d>2$." MK and AP are also grateful to the Simons Center for Geometry and Physics and NORDITA for hospitality.

\appendix{A}{Harmonic Functions}

Here we collect some useful formulas for harmonic function on $H^{d-1}$ \PenedonesNS. The harmonic function $\Omega_{i \nu}(\rho)$ in the hyperbolic space $H^{d-1}$ can be written as follows
\eqn\propagator{\eqalign{
\Omega_{i \nu}(\rho) &= {\nu  \ \sinh( \nu ) \over 2^{d-1} \pi^{{d +1\over 2}} \Gamma({d-1 \over 2})}\ \Gamma \left({d \over 2} - 1 + i \nu \right) \Gamma \left({d \over 2} - 1 - i \nu \right) \cr
& \ _2 F_1 \left({d \over 2} -1 + i \nu, {d \over 2} - 1 - i \nu , {d \over 2} - {1 \over 2} , - \sinh^2 {\rho \over 2} \right) \ .
}}

$\Omega_{i \nu}(\rho)$ satisfies the equation
\eqn\correctequ{
\ddot \Omega_{i \nu} (\rho) + (d-2) \coth \rho \ \dot \Omega_{i \nu} (\rho) + \left(\nu^2 + \left({d -2\over 2} \right)^2 \right) \Omega_{i \nu} (\rho) = 0 .
}
and is regular for $\rho =0$.

Consider the product of two time-like unit vectors which define the scalar product in $H^{d-1}$ $\cosh r =- e. \bar e$. It is useful to consider the following decomposition
\eqn\expansion{
{1 \over (-e.w)^a} = \int_{- \infty}^{\infty} d \nu \ {\pi^{d-2 \over 2} \over 2^{1-a}} {\Gamma({a-{d-2 \over 2}+ i \nu \over 2}) \Gamma({a-{d-2 \over 2} -  i \nu \over 2}) \over \Gamma(a)} \ \Omega_{i \nu} (e.w) .
}

Using this formula and some formulas from harmonic analysis \PenedonesNS\ one can check the following identity
\eqn\identity{\eqalign{
\gamma(\nu) \gamma(- \nu)  \Omega_{i \nu} (e. \bar e)  &={1 \over \pi^{d-2}} 2^{2-a-b} \int_{H^{d-1}} d w d \bar w {\Gamma(a) \Gamma(b) \over (-w. e)^{a} (-\bar w . \bar e)^{b}} \Omega_{i \nu} (w . \bar w) , \cr
a &= 2 \Delta_O + J - 1, ~~~ b = 2 \Delta_{\psi} + J - 1 .
}}

We can introduce an additional transformation 
\eqn\additional{
\int_{M^+} d p d \bar p {e^{p.x } e^{\bar p . \bar x} \over (-p^2)^{{d -a \over 2} } (-\bar p^2)^{{d-b \over 2} }} = \int_{H^{d-1}} d w d \bar w  {\Gamma(a) \Gamma(b) \over (-w. x )^{a} (-\bar w . \bar x)^{b} },
}
where we  substituted $p = t  w$ used that $w^0 > 0$ (and similarly for $\bar x$) and integrated $\int_0^\infty d  t \ t ^{d-1}$. 

In this way we get 
\eqn\relation{
{\gamma(\nu) \gamma(- \nu)  \Omega_{i \nu} (e. \bar e) \over (-x^2 )^{\Delta_O + {J-1 \over 2}} (-\bar x^2  )^{\Delta_\phi + {J-1 \over 2}}} ={2^{2-a-b} \over \pi^{d-2}} \int_{M^+} d p d \bar p {e^{ x.p } e^{ \bar x. \bar p } \over (-p^2)^{{d -a \over 2} } (-\bar p^2)^{{d-b \over 2} }} \Omega_{i \nu} (\hat p . \hat{\bar{p}}) \
}
which we derived for $x^2$ future-directed and timelike. We can rescale and rotate momenta in the RHS $p \to - i p $, $\bar p \to - i \bar p $ to get 
\eqn\identityJA{
{\gamma(\nu) \gamma(- \nu)  \Omega_{i \nu} (e. \bar e) \over (-x^2 )^{\Delta_O + {J-1 \over 2}} (-\bar x^2  )^{\Delta_\phi + {J-1 \over 2}}} ={2^{2-a-b} e^{- i {\pi \over 2}(a + b)} \over \pi^{d-2}} \int_{M^+} d^d p d^d \bar p {e^{- i x.p } e^{- i \bar x. \bar p } \over (-p^2)^{{d -a \over 2} } (-\bar p^2)^{{d-b \over 2} }} \Omega_{i \nu} (L) \ .
}

Let us briefly comment in the $i \eps$ prescription in the formula above. We derived the formula above first for timelike $x^2, \bar x^2$ when analytically continuing to spacelike distances we have to specify the prescription how to go around the branch points. As reviewed in the bulk of the paper we shift $x^0 \to x^0 + i \eps$, $\bar x^{0} \to \bar x^0 + i \eps$. In the RHS the effect of this shift is $e^{- \eps p^0}$ (and similarly for $\bar p^0$) which makes the integral convergent when we integrate over $p$'s in the future Milne wedge $M^+$ for any $x$ and $\bar x$.

\appendix{B}{Choice of Coordinates}

It is convenient to think of a CFT as defined on a universal cover of the projective space in $R^{2,d}$
\eqn\universalcover{
- (Z^{-1})^2 - (Z^0)^2 + \sum_{i=1}^{d} (Z^i)^2 = 0 , ~~~ Z \sim \lambda Z .
}

Let us describe different coordinate systems. The global coordinate system corresponds to 
\eqn\globalcoord{
Z^{-1} + i Z^0 = e^{i \tau}, ~ Z^i = n^i, ~~~ (Z^0)^2 + (Z^1)^2 = 1 ,
}
where $\vec n$ is the unit vector $\vec n^2 = 1$ on the $S^d$.

Let us review different coordinate systems that we used in the bulk of the paper \CornalbaAX. The original $y$ coordinates can be described as follows
\eqn\ycoord{
y^{\mu} = {Z^{\mu} \over Z^{0} + Z^d} , ~~~ \mu = 0,1,..,d-1.
} 
In other words, we can think of this patch as gauge fixing $Z^{0} + Z^{d} = 1$. Different ``gauges'' describe CFTs on spaces that differ by Weyl rescaling. We also introduce $y^{\pm} = y^0 \pm y^1 = {Z^{-1} \pm Z^1 \over Z^{0} + Z^d}$. This choice corresponds to 
\eqn\choice{\eqalign{
\bar K_y^A &=  \delta_{0}^{A} + \delta_{d}^A  , \cr
K_y^A &= {1 \over 2} \left( \delta_{0}^{A} - \delta_{d}^A \right) .
}}

The four Poincare patches discussed in the main text correspond to the following coordinates. For the transverse coordinates we get
\eqn\fourpatches{\eqalign{
x_{1,4}^{i} &= - {Z^{i}\over Z^{-1} + Z^1} , ~~~ i = 2, ..,d-1 , \cr
x_{2,3}^{i} &= - {Z^{i}\over Z^{-1} - Z^1} , ~~~ i = 2, ..,d-1 ,
}}

Whereas the light-cone coordinates take the following form
\eqn\lightconecord{\eqalign{
x_{1,4}^+ &= - {Z^{0} + Z^d \over Z^{-1} + Z^1} , ~~~ x_{1,4}^- =-  {Z^{0} - Z^d \over Z^{-1} + Z^1} , \cr
x_{2,3}^+ &= - {Z^{0} + Z^d \over Z^{-1} - Z^1} , ~~~ x_{2,3}^- =-  {Z^{0} - Z^d \over Z^{-1} - Z^1} .
}}

Let us understand how symmetries act. Clearly Lorentz transformations $x \to \Lambda x$ are the same in all four Poincare patches. They are $M_{\hat \mu \hat \nu}$ where $\hat \mu , \hat \nu = 0 , d, 2, .. , d-1$. In the original patch, these are different generators. For example boost $M_{\hat + \hat -}$ is a dilatation operator in $y$-coordinates.
Similarly, boost $M_{+ -}$ in the $y$-coordinates acts as a dilatation operator in each patch. Translations in patches $1,3$ become special conformal transformations in patches $2,4$ and conversely. 

Let us also recall how the AdS coordinate is introduced in each patch. In the embedding coordinates the AdS is given by
\eqn\embeddingAdS{
- (Z^{-1})^2 - (Z^0)^2 + \sum_{i=1}^{d} (Z^i)^2 = -1 .
} 

In the original Poincare patch the radial coordinate is given by
\eqn\originalz{
z = {1 \over Z^{0} + Z^{d}}
}
and in each Poincare patch we have
\eqn\poincarez{\eqalign{
z_{1} &=- {1 \over Z^{-1} + Z^{1} } = - {z \over y^+}, ~~~ z_3 = {1 \over Z^{-1} - Z^1} = {z \over y^+} \ , \cr
z_2 &=- {1 \over Z^{-1} - Z^1} = - {z \over y^-}, ~~~ z_4= {1 \over Z^{-1} + Z^1} = {z \over y^-} \ .
}}

\appendix{C}{Appendix Generalized Free Fields in $d=2$ in the Regge Limit}

Here we derive necessary formulas to understand the Regge limit for a small correction to generalized free fields. We have for double trace operators of spin $s$
\eqn\definitions{\eqalign{
\Delta_{n,s} &= \Delta_\phi + \Delta_{\psi} + 2 n + s  , \cr
h &={\Delta_{n,s} + s \over 2}, ~~~ \bar h = {\Delta_{n,s} - s \over 2}.
}}

We would like to understand the limit of the $s$-channel blocks when $h, \bar h \to \infty$ and $z,\bar z \to 1$ with $(1-z) h^2$ and $(1-\bar z) \bar h^2$ held fixed. A useful identity in this context is the following
\eqn\limit{
\lim_{z \to 1, (1-z) h^2 - {\rm fixed} } \ _2  F_1 \left( h + x, h + y, 2 h , z \right) ={1 \over \sqrt{\pi}}2^{2 h}\sqrt{h}(1-z)^{-{x+y \over 2}} K_{x+y}\left(2 h \sqrt{1-z}\right) + O({1 \over h}).
}

Let us now consider conformal blocks in two dimensions. These are explicitly known and could be found for example in \HijanoZSA . We take $\Delta$ and $J$ of the exchanged operator to be \definitions .  In the limit $z e^{- 2 \pi i},\bar z \to 1$ such that $(1-z) h^2$, ${h \over \bar h}$ and ${1-z \over 1 - \bar z}$ are fixed conformal blocks take the form
\eqn\twodblock{\eqalign{
&\lim_{z e^{- 2 \pi i},\bar z \to 1, h, \bar h \to \infty} g_{\Delta = \Delta_{n,s}, J = s}^{\Delta_{\phi \psi}, - \Delta_{\phi \psi}}(z, \bar z) = e^{- 2 \pi i \bar h} { 2^{2h+ 2 \bar h} \over \pi } \sqrt{h} \sqrt{\bar h} (1-z)^{{\Delta_{\phi \psi} \over 2} }  (1- \bar z)^{{ \Delta_{\phi \psi} \over 2} } \cr
 &\left( K_{\Delta_{\phi \psi} }\left(2 h \sqrt{1-z}\right)  K_{\Delta_{\phi \psi} }\left(2 \bar h \sqrt{1-\bar z}\right)  + K_{\Delta_{\phi \psi} }\left(2 h \sqrt{1-\bar z}\right)  K_{\Delta_{\phi \psi} }\left(2 \bar h \sqrt{1- z}\right)  \right) + ... \  .
}}

We next apply it to the limit of the generalized free fields. In this case $e^{- 2 \pi i \bar h} = e^{- i \pi (\Delta_O + \Delta_{\phi})} (-1)^s$ and the three-point coupling coefficients take the form, see formula (43) in  \FitzpatrickDM,
\eqn\threepoint{
c_{h, \bar h} =(-1)^s {16 \pi \ 2^{-2 (h + \bar h)} \over \Gamma(\Delta_{\phi})^2 \Gamma(\Delta_{\psi})^2} h^{\Delta_\phi + \Delta_\psi - {3 \over 2}} \bar h^{\Delta_\phi + \Delta_\psi - {3 \over 2}}  (1- {1 \over 2} \delta(h - \bar h)) + ... \ .
}
The $s$-channel OPE takes the form
\eqn\schannelOPE{\eqalign{
G(z,\bar z) &= \sum_{\bar h<h, h} {\cal I}_{h, \bar h} + ... \ , \cr
{\cal I}_{h, \bar h} &=\left( {4 \over  \Gamma(\Delta_{\phi}) \Gamma(\Delta_{\psi}) } \right)^2 h^{\Delta_\phi + \Delta_\psi - 1} \bar h^{\Delta_\phi + \Delta_\psi - 1} (1-z)^{{\Delta_{\phi \psi} \over 2} }  (1- \bar z)^{{ \Delta_{\phi \psi} \over 2} } \cr
 &\left( K_{\Delta_{\phi \psi} }\left(2 h \sqrt{1-z}\right)  K_{\Delta_{\phi \psi} }\left(2 \bar h \sqrt{1-\bar z}\right)  + K_{\Delta_{\phi \psi} }\left(2 h \sqrt{1-\bar z}\right)  K_{\Delta_{\phi \psi} }\left(2 \bar h \sqrt{1- z}\right)  \right) .
}}
We can now go from the sum to the integral to get
\eqn\sumoverblocksApp{\eqalign{
\int_0^{\infty} d h \int_0^{h} d \bar h ~ {\cal I}_{h, \bar h} (z, \bar z)={1 \over 2} \int_0^{\infty} d h \int_0^{\infty} d \bar h ~ {\cal I}_{h, \bar h} (z, \bar z) =  {1 \over \left[ (1-z)(1-\bar z) \right]^{\Delta_\psi} } \ .
}}
We also got that the contribution from the $\delta(h - \bar h)$ which is the spin zero operators is subleading in the $z,\bar z \to 1$ limit as expected.

It is instructive to rewrite the two-dimensional result \schannelOPE\ as an integral over the future Milne wedge $M^+$.  It is easy to check that
\eqn\blockgeneralizedB{\eqalign{
{\cal I}_{h, \bar h} (z, \bar z) &= (- x^2)^{\Delta_{\phi \psi}} C(\Delta_\phi) C(\Delta_{\psi}) \int_{M^+} {d p \over (2 \pi)^2} {d \bar p \over (2 \pi)^2} \ (- p^2)^{\Delta_\phi -1} \ (- \bar p^2)^{\Delta_\psi - 1} e^{p.x} e^{ \bar p . \bar x} \cr
&h \bar h (h^2 - \bar h^2) \delta \left( {p. \bar p \over 2} + h^2 + \bar h^2 \right) \delta \left( {p^2 \bar p^2 \over 16} - h^2 \bar h^2 \right) \ .
}}
Indeed, let us write $\int_{M^+}d p =\int_0^{\infty} d p^+ d p^- $ and $- p^2 = p^+ p^-$, we have then
\eqn\integraldelta{\eqalign{
&\delta \left( {p. \bar p \over 2} + h^2 + \bar h^2 \right) \delta \left( {p^2 \bar p^2 \over 16} - h^2 \bar h^2 \right) \cr
&={16\over -\bar p^2} {1 \over h^2 - \bar h^2} \left( \delta(p^+ - {4 h^2 \over \bar p^-}) \delta(p^- - {4 \bar h^2 \over \bar p^+}) + \delta(p^+ - {4 \bar h^2 \over \bar p^-}) \delta(p^- - {4 h^2 \over \bar p^+})  \right)
}}

A beautiful result of \CornalbaXM\ is that the formula \blockgeneralizedB\ could be trivially generalized to $d$ dimensions
\eqn\blockgeneralizedB{\eqalign{
{\cal I}_{h, \bar h} (z, \bar z) &= (- x^2)^{\Delta_{\phi \psi}} C(\Delta_\psi) C(\Delta_{\phi}) \int_{M^+} {d p \over (2 \pi)^d} {d \bar p \over (2 \pi)^d} \ (- p^2)^{\Delta_\phi - {d \over 2}} \ (- \bar p^2)^{\Delta_\psi - {d \over 2}} e^{p.x} e^{ \bar p . \bar x} \cr
&4 h \bar h (h^2 - \bar h^2) \delta \left( {p. \bar p \over 2} + h^2 + \bar h^2 \right) \delta \left( {p^2 \bar p^2 \over 16} - h^2 \bar h^2 \right) \ ,
}}
where the integral is now over the $d$-dimensional future Milne wedge.

Indeed, noting that
\eqn\anotheridentity{
\int_0^{\infty} d h \int_0^{h} d \bar h \ h \bar h (h^2 - \bar h^2) \ \delta \left( {p. \bar p \over 2} + h^2 + \bar h^2 \right) \delta \left( {p^2 \bar p^2 \over 16} - h^2 \bar h^2 \right) = 1
}
we arrived at
\eqn\laststep{\eqalign{
&\int_0^{\infty} d h \int_0^{h} d \bar h ~ {\cal I}_{h, \bar h} (z, \bar z) =(- x^2)^{\Delta_{\phi \psi}} C(\Delta_\phi) C(\Delta_{\psi}) \int_{M^+} {d p \over (2 \pi)^d} {d \bar p \over (2 \pi)^d} \ (- p^2)^{\Delta_\phi - {d \over 2}} \ (- \bar p^2)^{\Delta_\psi - {d \over 2}} e^{p.x} e^{ \bar p . \bar x} \cr
&= {(- x^2)^{\Delta_{\phi O}}  \over (-x^2)^{\Delta_{\phi}} (-\bar x^2)^{\Delta_{\psi}}} = {1 \over ( x^2 \bar x^2)^{\Delta_{\psi}}} =  {1 \over \left[ (1-z)(1-\bar z) \right]^{\Delta_\psi} } \ .
}}

\appendix{D}{Vectors From Scalars}

\noindent We start by considering the following  correlator of vector operator $J^\mu$ of dimension $\Delta_J$ and a scalar operator of dimension $\Delta_\psi$.
\eqn\cor{ {\vev{  J(\eps_1, y_1)J(\epsilon_2, y_2)\psi(y_3)\psi(y_4)  }}    = A^{\mu\nu}(y_1,y_2,y_3,y_4) \epsilon_{1\mu} \epsilon_{2\nu}}
This correllator can be expressed as a sum over conformal blocks in the s-channel as follows:
\eqn\schanneldec{ {\vev{  J(\epsilon_1, y_1)J(\eps_2, y_2)\psi(y_3)\psi(y_4)  }}    =\sum_\OO  \WW_\OO(y_1,y_2,y_3,y_4,\epsilon_1,\eps_2) }
where the sum runs over operators $\OO$ of specific conformal dimension and spin, and $\WW_\OO$ denotes the partial wave of the exchanged operator, including all the prefactors and the conformal blocks, normalized according to \cor. The partial wave\foot{Here we only consider the contribution of symmetric and traceless operators in the OPE.} is generally given by a sum of five terms\foot{Recall that $D_{ij} {1\over (y_{12}^2)^{\Delta_J+1}(y_{34}^2)^{\Delta_\psi}}=0$ for any $i,j=1,2$.}
\eqn\cbdec{   \WW_\OO(y_1,y_2,y_3,y_4,\epsilon_1,\eps_2)  =   {1\over (y_{12}^2)^{\Delta_J+1} (y_{34}^2)^{\Delta_\psi}}\sum_{k=1}^5  \lambda_{JJ\OO,k}\lambda_{\psi\psi\OO}  {\hat D_k}  G_{k,\OO } \,, }
where
\eqn\Dkdef{\hat{D_1}\equiv D_{11} D_{22},\quad \hat{D_2}=D_{12}D_{21},\quad \hat{D_3}=-H_{12},\quad \hat{D_4}=D_{21}D_{11},\quad \hat{D_5}=D_{12}D_{22}  }
with the $D_{ij}$ as defined in \refs{\CostaMG-\CostaDW}, and  
\eqn\WWdef{G_{1,\OO}=G_{2,\OO}=G_{3,\OO}\equiv g_\OO(u,v), \quad G_{4,\OO}={y_{24}^2\over y_{14}^2} g_\OO^{+2,0} (u,v),\quad G_{5,\OO}={y_{14}^2\over y_{24}^2} g_\OO^{-2,0}}
with $g_\OO^{\Delta_{12},\Delta_{34}}$ the corresponding conformal block for scalar operators and $\Delta_{ij}\equiv \Delta_i-\Delta_j$. Here $(u, v)$ denote the conformal cross-ratios defined by:
\eqn\defuv{u = {y_{12}^2 y_{34}^2 \over y_{13}^2 y_{24^2}      },    \qquad     u = {y_{14}^2 y_{23}^2 \over y_{13}^2 y_{24}^2      } , \qquad y_{ij}=y_i-y_j}

Every term in the sum of \cbdec\ can be expanded as follows
\eqn\expDG{\hat D_k G_{k,\OO}=\sum_{i=1}^5 Q^i \hat{D}_k^i g_\OO^{a,b}\, ,} 
where the $Q^i$ denote the conformally covariant tensor structures
\eqn\basisfive{  Q^{i} =   \{ H_{12},  V_1 V_2,  V_1'V_2, V_1 V_2', V_1' V_2'   \,\} \,,}
defined as follows:
\eqn\Qidef{\eqalign{H_{12}&=-2 \eps_j\cdot y_{12}\, \eps_i\cdot y_{ij}+y_{ij}^2 \eps_i\cdot\eps_j\cr
V_{i,jk}&={y_{ij}^2 \eps_i\cdot y_{ik}-y_{ik}^2 \eps_i\cdot y_{ij}\over y_{jk}^2} 
}}
with 
\eqn\Vprimesdef{V_1=V_{1,23},\quad V_1'=V_{1,24},\quad V_2=V_{2,31},\quad V_2'=V_{2,14} }

Generally, only four out of five coefficients $\lambda_{JJ\OO,k}$ 
are independent. In particular, $\lambda_{JJ\OO,4}=\lambda_{JJ\OO,5}$ as a result of considering identical vector operators. To simplify the notation in what follows we will denote by $\lambda_k$ the OPE coefficients $\lambda_{JJ\OO,k}$. We hope that this will not create any confusion.
When the operator $\OO$ in the OPE is conserved, where by unitarity its conformal dimension and spin are related: $\Delta-J=d-2$, then $\lambda_4=\lambda_5=0$ and we are left with only three independent parameters.
On the other hand when the external vector operators are conserved, $\Delta_J=d-1$ and only two independent coefficients remain \CostaDW :
\eqn\Nkconserved{\eqalign{\lambda_1&=-{ (\Delta-j-d)(\Delta+j-2) \lambda_4 -2 (d-2)\lambda_2-4 \lambda_3\over \Delta(\Delta-d)+j(j+d-2) }\cr 
\lambda_4&=\lambda_5=-{(\Delta+j)(\Delta-j-d+2)\over \Delta(\Delta-d)+j(j+d-2)  } \lambda_2\,.
}}
The parameters $(\lambda_1,\lambda_2,\lambda_3,\lambda_4)$ introduced here are linearly related to the $(\alpha,\beta,\gamma,\eta)$ which define the three-point function $\vev{JJ\OO}$ in the basis of the conformally covariant tensor structures defined in \Qidef :
\eqn\JJO{\vev{JJ\OO_{\Delta,j}}=V_3^{j-2} {\alpha V_1 V_2 V_3^2+\beta V_3\left(V_1 H_{23}+V_2 H_{13}\right)+\gamma H_{12} V_3^2+\eta H_{13}H_{23}\over (y_{12}^2)^{\xi+1-{\Delta+j\over 2}} (y_{23}^2)^{\Delta+j\over 2} (y_{13}^2)^{\Delta+j\over 2} } \,.}
The precise mapping between the parameters is given below:
\eqn\NtoJJO{\eqalign{\alpha&=(j-{1\over 4}(j+\Delta)^2) \lambda_1-{1\over 4} (j-\Delta)^2 \lambda_2+{1\over 2} (j-\Delta) (j+\Delta-2)\lambda_4\cr
\beta&={1\over 4}j (\Delta+j-2) \lambda_1+{1\over 4}j \, (j-\Delta)\lambda_2-{1\over 2} j(j-1)\lambda_4\cr
\gamma&={1\over 4} (j-\Delta)\lambda_1+{1\over 4} (j-\Delta)\lambda_2-\lambda_3-{1\over 2} (j+\Delta-2)\lambda_4\cr 
\eta&=-{1\over 4}j(j-1)(\lambda_1+\lambda_2-2 \lambda_4)
}}

In particular, for the exchange of an operator $\OO$ of conformal dimension $\Delta$ and spin $j\geq 2$, the conformal partial wave takes the following form 
 \eqn\confblocka{\WW_\OO=   {1\over (y_{12}^2)^{\Delta_J+1} (y_{34}^2)^{\Delta_\psi} } \left( f_1 H_{12} +f_2 V_1 V_2+f_3 V_1'V_2+f_4 V_1 V_2'+f_5 V_1' V_2'\right)\,.}
with coefficient functions $f_i$ 
\eqn\fdefa{\eqalign{{f_1\over \lambda_{\psi\psi\OO}}&={1\over 2}\left(-2 \lambda_3 +(\lambda_1+\lambda_2) u\p_u\right) g_\OO +\cr
&+{1\over 2} \, \lambda_5 \left(1-v(v-1) \p_v-u v \p_u\right) g_{\OO,(-2)}+{1\over 2}\, \lambda_4 \left(-1+(1-v)\p_v -u \p_u\right)g_{\OO,(+2)} \cr
{f_2\over \lambda_{\psi\psi\OO}}&= \left( (\lambda_1+\lambda_2) v\p_v +(\lambda_1+\lambda_2) v^2 \p_v^2 + (\lambda_1+\lambda_2) v u \p_u \p_v \right) g_\OO+ \cr
&+\lambda_5 v\left(-2 v\p_v-v^2\p_v^2 -2 u \p_u-2 u v \p_v\p_u-u^2\p_u^2 \right) g_{\OO,(-2)} -\lambda_4\, v \p_v^2  g_{\OO,(+2)} \cr
{f_3 \over \lambda_{\psi\psi\OO}}&=\left((v \lambda_1+\lambda_2) (-\p_v-v\p_v^2)-\lambda_1 (u\p_u+2 u v \p_u\p_v+u^2\p_u^2)   \right )g_\OO+\cr
&+ \lambda_5 \left(2 v \p_v+v^2 \p_v^2 +u\p_u+u v\p_u \p_v\right) g_{\OO,(-2)}+\lambda_4 \left( v \p_v^2+u \p_u\p_v \right)g_{\OO,(+2)}\cr
{f_4  \over \lambda_{\psi\psi\OO}}&=\left(v (\lambda_1+\lambda_2 v)\p_v+ v^2 (\lambda_1+v \lambda_2) \p_v^2+ \lambda_2 (u v\p_u+2 u v^2\p_u \p_v+u^2 v \p_u^2 \right)g_\OO+\cr
&+ \lambda_5 v\left( -2 v \p_v-v^2 \p_v^2-u \p_u-u v \p_u\p_v \right)  g_{\OO,(-2)}+\lambda_4  v  \left( -v \p_v^2-u \p_u\p_v     \right )g_{\OO,(+2)}\cr
{f_5  \over \lambda_{\psi\psi\OO}}&=-\left( (\lambda_1+\lambda_2) v\p_v +(\lambda_1+\lambda_2) v^2 \p_v^2 + (\lambda_1+\lambda_2) v u \p_u \p_v \right) g_\OO+\cr
&+ \lambda_5 v\left(2 \p_v+v\p_v^2 \right) g_{\OO,(-2)} +\lambda_4  \left(v^2 \p_v^2 +2 u v \p_u \p_v+u^2\p_u^2  \right)g_{\OO,(+2)} \,.
}}
and $\lambda_4=\lambda_5$. 
For the specific case of interest, where the exchanged operator is the stress-tensor operator, \confblocka\ can be also written as:
\eqn\confblockb{\WW_\OO=   {1\over (y_{12}^2)^{\Delta_J+1} (y_{34}^2)^{\Delta_\psi} } \left( f_1 H_{12} +f_2 V_1 V_2+f_3 V_1'V_2+f_4 V_1 V_2'+f_5 V_1' V_2'\right)\,.}
with coefficient functions $f_i$ 
\eqn\fdefb{\eqalign{f_1&=-{1\over ( d-1)} \left[(d-1)(\lambda_3+4 n_f)+n_s - {n_s\over d-2} u\p_u \right] g_\OO\cr
f_2&=-{1\over (d-1)(d-2)} 2 n_s \left(v\p_v+v^2\p_v^2+u v\p_u\p_v\right) g_\OO \cr
f_3&= {2\,n_s\over (d-1)(d-2)} \left(\p_v+v\p^2_v\right)g_\OO +{4\, n_f\over d-1} \left[ (1-v) (\p_v+v \p_v^2) -u\p_u-2 u v\p_u \p_v-u^2\p_u^2\right]g_\OO\cr
f_4&=-{2\, n_s\over (d-1)(d-2)}\, v \left(v\p_v +v^2 \p_v^2+u\p_u+2 u v\p_u\p_v+u^2\p_u^2\right)g_\OO+\cr
&\,+{4\over d-1}n_f v\left[(1-v) (\p_v+v \p_v^2)-u\p_u-2 u v\p_u\p_v-u^2\p_u^2 \right]g_\OO \cr
f_5&=-f_2,\,.
}}
In \confblockb\ we found it convenient to use as independent parameters, the constants $n_s$ and $n_f$ which determine the three point function $\vev{JJ\OO}$ in the ``free field'' basis \ZhiboedovBM. $\lambda_3$ is an additional parameter, related via the Ward Identity of the stress-tensor in $\vev{JJT}$ to the other two parameters $n_s,n_f$ as follows \KomargodskiGCI :
\eqn\nthree{n_3={\Delta_J-d+1\over (d-1)(d-2)} \left(4(d-2) n_f+n_s\right)}
which clearly vanishes when the vector operators are conserved.  To be specific, $(n_s, \, n_f)$ here are defined as in Appendix B.1 of \HartmanDXC, where the partial wave for the stress-tensor operator in $d=4$ was also quoted. The OPE coefficients $\lambda_i$ are linearly related to the $(n_s,n_f,\lambda_3)$:
\eqn\Ntons{\eqalign{\lambda_1={4 n_f\over (d-1)},\quad &\lambda_2=-{ 2(n_s+(2 d-4)n_f)\over (d-1)(d-2)},\quad \lambda_3={(d-1) n_3+4 (d-1) n_f+ n_s\over (d-1)}\cr
 &\lambda_4=\lambda_5=0 \,.}}
Note that $f_3$ and $f_4$ in \fdefb\ differ from those quoted in (C.3) of \HartmanDXC, which contain a couple of small typos.

\appendix{E}{The Regge Limit of the Correlation Function $\vev{J_{\mu}\psi\psi J_{\nu}}$}

\noindent In this appendix we consider the correlation function $\vev{J_{\mu}\psi\psi J_{\nu}}$ expressed as a sum over conformal partial waves and follow the steps outlined in section 4.1 to arrive at eq. \reggecor\ of section 4.2. Our starting point will be the correlation function \cor. To avoid confursion due to the different ordering, we will restate here all the relevant transformations and notations. The final result will of course be the same.

Starting from \cor\ we introduce the Poincare patches 
\eqn\finiteformb{\eqalign{
x_i &= (x_i^+ , x_i^-, \vec x_i) = - {1 \over y_i^+} (1 , y_i^2, \vec y_i ), ~~~ i = 1,2 \  , \cr
x_j &= (x_j^+ , x_j^-, \vec x_j) = - {1 \over y_j^-} (1 , y_j^2, \vec y_j ), ~~~ j = 3,4 ~ ,
}}
where the subscript stands for the operator insertion $y_i$ with $i= 1, .., 4$.  After doing the coordinate transform we set $x_{1,2} =\mp {x \over 2}$, $x_{3,4} = \mp {\bar x \over 2}$. Note that this is a time-ordered correlator. The explicit transformation formulas are 
\eqn\ytox{\eqalign{
y_1^+ &= - y_2^+ ={2 \over x^+}  , ~~~y_3^- = - y_4^- ={2 \over \bar x^+} \ ,\cr
y_1^- &= - y_2^- = {x^2 \over 2 x^+} ,~~~y_3^+ = - y_4^+ = {\bar x^2 \over 2 \bar x^+}  \ , \cr
\vec y_1 &= \vec y_2 = {\vec x \over x^+} ,~~~ \vec y_3 = \vec y_4 = {\vec{\bar{x}} \over \bar x^+} .
}}
We take $x^{\mu}$ and $\bar x^{\mu}$ to be future-directed time-like vectors, {\it i.e.}, $x^+, \bar x^+ > 0$ and $x^2 < 0$ and $\bar x^2 < 0$. 
The conformal mapping leads to 
\eqn\sccorrAb{\eqalign{
\eps_{1m} \eps_{2n} A^{mn}(x, \bar x)&= \la J^m\left(-{x \over 2} \right)\psi \left(- {\bar x \over 2} \right) \psi \left({\bar x \over 2}\right) J^n\left({x \over 2} \right)     \ra= \cr
&= \left( {x^+ \over 2} \right)^{-2 (\Delta_{J}+1) } \left( {\bar x^+ \over 2} \right)^{-2 \Delta_\psi}  {\p x^m\over \p y_1^\mu} {\p x^n \over \p y_2^\nu} \,\, \la J^\mu (y_1)  \psi(y_3) \psi(y_4) J^\nu (y_2)\ra =\cr
&={\eps_m^\ast \eps_n  \AA^{mn}(u,v) \over (-x^2)^{ \Delta_J+1} (-{\bar x}^2)^{\Delta_\psi} }  \,.
}}
The cross ratios in terms of new variables take the form
 \eqn\crossratiosC{\eqalign{
 u &= {x^2 \bar x^2 \over \left(1 + {x^2 \bar x^2 \over 16} - {x.\bar x \over2} \right)^2} =\sigma^2 , \cr
 v &= \left( {1 + {x^2 \bar x^2 \over 16} + {x.\bar x \over2} \over 1 + {x^2 \bar x^2 \over 16} - {x.\bar x \over2} } \right)^2 =1-2\sigma\cosh{\rho}+\sigma^2 \,.
 }}
Notice that here $\sigma>0$ as opposed to the main text.

Using \schanneldec\ we can express \sccorrAb\ as:
\eqn\amnpre{
\eps_{1m}\eps_{2n} A^{mn}(x,\bx)= \eps_{1m}\eps_{2n} \sum_{\Delta, J} {\sum_{k=1}^5 {\tilde{Q}}^{(k)mn}\,f_\OO^i (u,v)\over (-x^2)^{\Delta_J+1} (-\bx^2)^{\Delta_\psi}}\,,
} 
where the $f^i_\OO$ are defined in \fdefa\ and $(u,v)$ are the conformal cross-ratios expressed in terms of $(x,\bx)$ in eq \crossratiosC. 
Here 
\eqn\defq{   {\tilde Q}^{(k)mn}  =   Q^{(k)\mu\nu}  {\p x^m\over \p y_1^\mu} {\p x^n \over \p y_2^\nu}  = \{    \tilde H_{12}^{\mu\nu}, \tilde V_1^\mu \tilde V_2^\nu,  (\tilde V_1')^\mu \tilde V_2^\nu, 
                    \tilde V_1^\mu (\tilde V_2')^\nu, (\tilde V_1')^\mu (\tilde V_2')^\nu \}   \,.}

To proceed we should rewrite the sum over the conformal dimensions of the exchanged operators as an integral over $\nu$. To this end, one needs to define the corresponding partial waves $F_{\nu,J}$ for vector operators. This is a straightfoward exercise; one simply replaces the scalar conformal blocks  in \fdefa\ with the respective partial waves $F_{\nu,J}(u,v)$. The corresponding partial amplitudes $b_j(\nu^2)$ have the same exact form as in \bjapp. For notational simplicity, we will henceforth denote by $\tilde{f}^i_\OO$ the $f^i_\OO$ resulting from replacing the conformal blocks by $r_k[\Delta, J]\sigma^{1-J}\Omega_{i\nu}(\rho)$. Repeating the procedure outlined in section 4.1 we arrive at:
\eqn\amna{ 
A^{mn}(x,\bx) \simeq A_0^{mn}+2\pi i 
   \sum_{i}   \int d\nu {  \, {\alpha}(\nu)\, e^{i\pi j(\nu)}  \left( \tilde{Q}^{(k)mn} \tilde{f}_\OO^i (\sigma,\rho)\right)_{Regge} \over (-x^2)^{\Delta_J+1} (-\bx^2)^{\Delta_\psi}}\,,  
} 
where $\alpha(\nu)$ is defined in \exchangeReggeB\ and the index $Regge$ denotes the Regge limit of the corresponding expression. We also seperated the contribution of the disconnected part which cannot be studied using Regge theory. Note the overall factor of $e^{-i\pi(1-j(\nu))}$ compared to the main text, which can be traced to the opposite sign of $\sigma$. The factor can be completely fixed from the $i\eps$-prescription in $\sigma$ and the identity \cfid\ evaluated for the analytically continued blocks.
   
To obtain the explicit form of $\tilde{Q}^{(k)mn} \tilde{f}_\OO^i (\sigma,\rho)$ in the Regge limit, we first replace the scalar partial waves with their leading Regge behavior, eq. \fnuj, obtained after the appropriate analytic continuation\foot{Note that the leading behavior of the scalar conformal blocks and their shadows is independent of the external operators' dimensions. In other words, the Regge behavior of $g_\OO^{a,b}$ is the same for any $a,b$.}. The correct analytic continuation here involves transporting $z$ around 1 counter-clockwise.
The behavior of the scalar blocks and the partial waves will therefore be opposite to eq. \analyticcont. This is another way to understand the plus sign in front of the second term of \amna. 
Next we evaluate $\tilde{Q}^{(k)mn}$ in the Regge limit, or equivalently
\eqn\defh{ \eqalign{ 
\tilde H_{12} &=-4\left( x^2 \eta^{mn} - 2 x^m x^n\right)  \cr
 \tilde V_1^m &= \tilde {V'_2}^m =-2 x^m  \left(1-\sigma \cosh\rho +\sigma^2 \left({3\over 8}-{1\over 2}\cosh^2\rho \right) \right) + \cr
 &+\bar x^m x^2 \left(1-{1\over 2}\sigma \cosh\rho+\sigma^2({3\over 16}-{1\over 4}\cosh^2\rho) \right)  +\ldots \cr
 \tilde {V'_1}^m &=- \tilde {V_2}^m =-2\, x^m  \left( 1+\sigma\cosh\rho-\sigma^2({5\over 8}-{3\over 2}\cosh^2\rho)   \right) + \cr
 &+\bar x^m x^2 \left(-1-{1\over 2} \sigma\cosh\rho +\sigma^2 ({5\over 16}-{3\over 4}\cosh^2\rho)\right)  +\ldots 
 }}
We must keep subleading terms in ${\tilde Q}^{(k)mn}$ up to $\OO(\sigma^2)$ because generally, $\tilde{f}^i_\OO\sim \sigma^{-1-j}$ in the Regge limit.

A straightforward computation then yields
\eqn\reggecorb{\eqalign{   
A^{mn} (x,\bx) &\simeq A_0^{mn}+\,2\pi i \int d\nu\, \sum_{k=1}^4  \,{\hat\alpha}_k(\nu) \, e^{i\pi j(\nu)} \,  \sigma^{1-j(\nu)}\,{  {\LL}_k^{mn} \Omega_{i\nu}(\rho)
                \over  (-x^2)^{\Delta_J}  (-\bx^2)^{ \Delta_\psi } }\cr  
{\hat\alpha}_k(\nu)&=\hat{r}_k[\nu,j(\nu)] \alpha(\nu)\,     \,.           
                }}              
The differential operators $\LL^{mn}_k$ are defined as:
\eqn\tildediffdef{\eqalign{x^2 \LL_1^{mn}=x^2\eta^{mn}\cr
x^2 \LL_2^{mn}=x^m x^n\cr
x^2 \LL_3^{mn}=x^2 (x^m\p^n+x^n\p^m)\cr
x^2 \LL_4^{mn}=(x^2)^2\p^m\p^n\,\,,
} }     
and the $\hat{r}_k[\nu,j(\nu)]$ denote the analytic continuation of the residues $\hat{r}_k[\Delta,J]$ which are explicitly given by:
\eqn\alhpatkdef{\eqalign{\hat{r}_1[\Delta,J] &=\left( (J-1)(\lambda_1+\lambda_2) -4 \lambda_3+2(J-1)\lambda_4\right)\lambda_{\psi\psi\OO}\,\,K_{\Delta, J}\cr  
\hat{r}_2[\Delta,J]&= -\left((J^2-1)(\lambda_1+\lambda_2) -8 \lambda_3+2(J-1)(J+3)\lambda_4  \right) \lambda_{\psi\psi\OO}\,\, K_{\Delta,J}\cr
\hat{r}_3[\Delta,J]&= \left(-(J+1)  \lambda_1+(J-1) \lambda_2+2 \lambda_4\right)\lambda_{\psi\psi\OO}\,K_{\Delta,J} \cr
\hat{r}_4[\Delta,J]&= (-\lambda_1-\lambda_2+2 \lambda_4)\lambda_{\psi\psi\OO}\, K_{\Delta,J}\,.
} }          
Observe that the differential operators only depend on $x$ since $\left({x\over 2},-{x\over 2}\right)$ denote the positions of the vector operators.

Moving to the impact parameter representation, requires taking the Fourier transform of \reggecorb. This is most conveniently done after a change of basis such that:
\eqn\reggecorc{\eqalign{
A^{mn} (x,\bx) &\simeq  A_0^{mn}(x,\bx)-\,2\pi i \int d\nu\, \sum_{k=1}^4  \alpha_k(\nu)\,e^{i\pi \,j(\nu)}\, \left(x^2 \DD_k^{mn}\right) { \Omega_{i\nu}(\rho)
                \over  (-x^2)^{\Delta_J+(j(\nu)+1)/2}  (-\bx^2)^{ \Delta_\psi+(j(\nu)-1)/2 }}  \cr
 &\alpha_k(\nu)=r_k[\nu,j(\nu)]\,\alpha(\nu) \,              
                }  }           
where the new differential operators $x^2\DD_k^{mn}$ commute with $x^2$ and are equal to:
\eqn\diffdef{\eqalign{x^2\DD_1^{mn}&=x^2 \eta^{mn}-x^m x^n\cr
x^2\DD_2^{mn}&=x^m x^n\cr
x^2\DD_3^{mn}&=x^2 (x^m \p^n+x^n\p^m) -2 x^m x^n x^s\p_s \cr
x^2\DD_4^{mn}&=(x^2)^2 \p^m\p^n- \,\, x^2\, x^q (x^m\p^n+x^n\p^m) \p_q-\cr
&- {1\over d-1} (x^2\eta^{mn}-x^m x^n) x^2\p^2+{1\over d-1}(x^2\eta^{mn} +(d-2) x^m x^n) x^q x^s \p_q\p_s\,,
}}       
and the $r_k[\nu,j(\nu)]$ are the analytic continuations of the $r_k[\Delta,J]$ given by:
\eqn\alhpakdef{\eqalign{ r_1[\Delta,J] &=\left({ (d-2)^2+4 \nu^2 \over 4(d-1)}-4 \lambda_3-4 \lambda_4+J(\lambda_1+\lambda_2+2 \lambda_4)\right) \,\lambda_{\psi\psi\OO}\,K_{\Delta,J}\cr  
r_2[\Delta,J] &=\left[ J(1-J) (\lambda_1+\lambda_2)-2 J(1+J ) \lambda_4+4(\lambda_3+\lambda_4)) \right]  \lambda_{\psi\psi\OO}\,K_{\Delta,J}\,\cr
r_3[\Delta,J]&= J(\lambda_2-\lambda_1) \lambda_{\psi\psi\OO}\,\,K_{\Delta,J} \cr
r_4 [\Delta,J]&= (-\lambda_1-\lambda_2+2 \lambda_4) \lambda_{\psi\psi\OO}\,K_{\Delta,J}\,.
}}

Technically, the above change of basis is possible due to the following identities:
\eqn\iddiff{\eqalign{x^m\p_m \Omega_{i\nu}(\rho)&=0 \cr
x^2\p^2\Omega_{i\nu}(\rho)&=-\ddot\Omega_{i\nu}-(d-2)  \coth{\rho} \,\, \dot\Omega_{i\nu}(\rho)\cr
x^2\, x^q (x^m\p^n+x^n\p^m) \p_q\Omega_{i\nu}&=-x^2 (x^m \p^n+x^n\p^m) \Omega_{i\nu}\cr
x^q x^s \p_q\p_s\Omega_{i\nu}&=0
}}
and the differential equation $\Omega_{i\nu}$ satisfies (see also appendix A):
\eqn\diffeqnomega{
\ddot\Omega_{i\nu}(\rho)+(d-2) \coth{\rho} \,\, \dot\Omega_{i\nu}(\rho)+\left(\nu^2+\left({d-2\over 2}\right)^2\right)\Omega_{i\nu}=0
}                
where the dots denote differentiation with respect to $\rho$. The equations above allow us to find a simple relation relation between the differential operators $\LL^{mn}$ and $\DD^{mn}$:
\eqn\DDLLrel{\eqalign{\DD^{mn}_1=\LL_1^{mn}+\LL_2^{mn}\cr
\DD_2^{mn}=\LL_2^{mn}\cr
\DD_3^{mn}\Omega_{i\nu}(\rho)=\LL_3^{mn}\Omega_{i\nu}(\rho)\cr
\DD_4^{mn}\Omega_{i\nu}(\rho)=\left(\LL_4^{mn}+\DD_3^{mn}-{1\over d-1}\left[\nu^2+\left({d-2\over 2}\right)^2\right] \DD_1^{mn}\right)\Omega_{i\nu}(\rho)
}}         
We are now ready to discuss the Fourier transform to the impact parameter space. Before doing so, it would be useful to revert to notations of the main text. The correlator in the main text is time-ordered as here, the only difference being the definition of $\sigma$. To obtain the correct expression for the position space correlator in the Regge limit, we thus need to replace $\sigma_{here}$ by $ e^{i\pi(1-j(\nu))}\sigma_{there}$. For the correlator in the main text we thus have:
\eqn\reggecord{\eqalign{
A^{mn} (x,\bx) &\simeq  A_0^{mn}(x,\bx)+\,2\pi i \int d\nu\, \sum_{k=1}^4 \, \,\alpha_k(\nu)\, \left(x^2 \DD_k^{mn}\right)  {(-\sqrt{x^2\bx^2})^{1-j(\nu)}\,\Omega_{i\nu}(\rho)
                \over  (-x^2)^{\Delta_J+1}  (-\bx^2)^{ \Delta_\psi} }  \cr
 &\alpha_k(\nu)=r_k[\nu,j(\nu)]\,\alpha(\nu)               
                }  }

\appendix{F}{The Regge Limit of $\vev{J_{\mu} \psi\psi J_{\nu}}$ in the Impact Parameter Space}

\noindent Here we discuss how to obtain the Fourier transform of the correlator $A^{mn}(x,\bar{x})$ in the impact parameter space:
\eqn\trimpact{B^{mn}(p,\bp)=\int d^dx\,d^d\bx e^{ip\cdot x}\,e^{i\bp\cdot\bx} \, A^{mn}(x,\bx)\,\,.}
Our starting point is eq.\reggecor\ from section 4.
Substituting \reggecor\ into \trimpact, one first replaces all the powers of $x$ with derivatives in $p$
\eqn\replacea{x_\mu\rightarrow -i{\p\over\p p^\mu}\,\,,}
and then integrates by parts to effectively replace all the derivatives in $x$ with 
\eqn\replaceb{{\p\over\p x^m}\rightarrow -ip_m\,\,.} 
Afterwards, one can use the impact parameter representation of $\Omega_{i\nu}(\rho)$ eq. (A.7) with $a\rightarrow a+2$ to write:
\eqn\reggeB{\eqalign{
\delta^{mn}&(p,\bar{p})\simeq 
(-p^2)^{d/2-\Delta_J}(-\bp^2)^{d/2-\Delta_\psi}\,\times\cr
&\times 4\,\tilde{c}_0 \, \sum_{k=1}^4  \,\hat{\LL}_k^{mn}\int d\nu \,\, {\pi^{d\over 2} \,e^{-{i\pi j(\nu)\over 2}}\over \sin{\pi j(\nu)\over 2}}r_k[\nu,j(\nu)]\beta(\nu) {\Omega_{i\nu}(L)\over (-p^2)^{d/2-\Delta_J-{j(\nu)+1\over 2}}(-\bar{p}^2)^{d/2-\Delta_\psi-{j(\nu)-1\over 2}}}  \,   ,
}}
where the differential operators $\hat{\LL}_k$ are given by:
\eqn\hatLLdef{\eqalign{-\hat{\LL}_1^{mn}&=\eta^{mn}\p^2-\p^m\p^n\cr
-\hat{\LL}_2^{mn}&=\p^m \p^n  \cr
-\hat{\LL}_3^{mn}&=-\p^2 (\p^m p^n+\p^n p^m)+2 \p^m \p^n (\p\cdot p)\cr
-\hat{\LL}_4^{mn}&=\p^4 p^m p^n-  \p^2 \p_s (\p^m p^n+\p^n p^m) p^s-\cr
&-{1\over d-1} \p^2 \left(\eta^{mn} \p^2-\p^m \p^n \right) p^2 +{1\over d-1} \p_s\p_t \left(\eta^{mn} \p^2+ (d-2) \p^m \p^n\right) p^s p^t
}}
Note that here the derivatives $\p_m$ denote differentiation with respect to $p^m$.
To proceed, it is convenient to rewrite the differential operators above in the following form:
\eqn\hatLLdef{\eqalign{-p^2 \hat{\LL}_1^{mn}&=\eta^{mn} (p^2\p^2)-p^2 \p^m\p^n\cr
- p^2 \hat{\LL}_2^{mn}&=p^2\p^m \p^n  \cr
- p^2\hat{\LL}_3^{mn}&=- (p^m \p^n+p^n \p^m) (p^2 \p^2)- 2 \left(\eta^{mn}-2 {p^m p^n\over p^2} \right) p^2 \p^2+2 p^2 \p^m \p^n \left(p\cdot \p +d-2\right)\cr
- p^2 \hat{\LL}_4^{mn}&={p^m p^n\over p^2} \left( {12\over d-1} -{2(d-4)\over d-1} p\cdot\p + p^2 \p^2 \right) (p^2 \p^2)-\cr
 &-(p^m \p^n+p^n \p^m) p^2 \p^2 \left({d^2-5 d+10\over 2(d-1)} +{d+2\over 2(d-1)} p\cdot \p \right)+\cr
&+{1\over d-1} p^2 \p^m \p^n \left[6 (d-2)+p^2\p^2+2(d+1) p\cdot \p+2 (p\cdot\p)^2\right]-\cr
&-{1\over d-1} \eta^{mn}p^2\p^2\left[6+ (5-d ) p\cdot \p+p^2\p^2-(p\cdot \p)^2  \right]\,.
}}
so that the commutting operators $p\cdot \p$ and $p^2\p^2$ can be replaced by:
\eqn\pdpnumbers{\eqalign{p\cdot \p&\rightarrow 2 \Delta_J+j(\nu)-(d-1)\cr
p^2\p^2&\rightarrow \left(2\Delta_J+j(\nu)-(d-1)\right)\,\,\left(2\Delta_J+j(\nu)-1\right)+\nu^2+\left({d-2\over 2}\right)^2
}}
The impact parameter representation then takes the form
\eqn\reggeB{
\delta^{mn}(p,\bar{p})\simeq \,-4\tilde{c}_0 \int d\nu \,  {\pi^{d\over 2} \,e^{-{i\pi j(\nu)\over 2}}\over \sin{\pi j(\nu)\over 2}}  \,s^{j(\nu)-1}\,\, \sum_{k=1}^4 \beta_k(\nu) \, {\hat{\DD}}_k^{mn} \Omega_{i\nu}(L)\, ,
} 
with 
\eqn\sLdef{s=\sqrt{p^2\bar{p}^2} ,\qquad \cosh{L}=-{p.\bar{p} \over \sqrt{p^2\bar{p}^2 } } \,.}
The differential operators $\hat{\DD}_k^{mn}$ are equal to 
\eqn\DDdef{\eqalign{\hat{\DD}_1^{mn}&=\eta^{mn}-{p^m p^n\over p^2}\cr
\hat{\DD}_2^{mn}&={p^m p^n\over p^2}\cr
\hat{\DD}_3^{mn}&=p^m \p^n+p^n\p^m\cr
\hat{\DD}_4^{mn}&=p^2\p^m\p^n+(p^m \p^n+p^n\p^m)-{1\over d-1}\left(\eta^{mn}-{p^m p^n\over p^2}\right)p^2 \p^2\,,
}}
and the $\beta_k(\nu)$ denote the linear combinations
\eqn\betakdef{\beta_k(\nu)=-\beta(\nu)\sum b_k^i \,r_i[\nu,j(\nu)]\,}
with:

\eqn\betakdefone{\eqalign{b_1^1&= \,{d^3+d^2 (-4 j-8 \Delta_J +2)+4 d \left((j+2 \Delta_J )^2+\nu ^2-1\right)+4 \left(j-2 \nu ^2+2 \Delta_J \right)-4 (j+2 \Delta_J )^2\over 4 (d-1)}\cr
b_1^2&={d (-3 d+4 j+8 \Delta_J +4)-4 \left(j-\nu ^2+2 \Delta_J \right)\over 4 (d-1)},\qquad b_1^3=-\,{\left((d-2)^2+4 \nu ^2\right) (d-j-2 \Delta_J )\over 2 (d-1)}\cr
b_1^4&=- {\left((d-2)^2+4 \nu ^2\right)\over 16(d-1)^2}\times\cr
&\times \left\{ d^3-2 d^2 (2 j+4 \Delta_J +5)+4 d \left(j^2+j (4 \Delta_J +6)+\nu ^2+4 (\Delta_J +1) (\Delta_J +2)\right)-\right.\cr
&\left. -8 \left(4 j+\nu ^2+8 \Delta_J \right)-16 (j+2 \Delta_J )^2\right\}
}}

\eqn\betakdeftwo{\eqalign{b_2^1&= \left(-{3 d^2\over 4}+d (j+2 \Delta_J+1)-j+\nu ^2-2 \Delta_J\right),\qquad b_2^2=\left((d-j-2 \Delta_J -1) (d-j-2 \Delta_J )\right)\cr
b_2^3&= {1\over 2} \left((d-2)^2+4 \nu ^2\right) (d-j-2 \Delta_J ),\qquad b_2^4=-(d-1)\, b_1^4
}}

\eqn\betakdefthree{\eqalign{ b_3^1& = \left( d-j-2 \Delta_J\right),\quad b_3^2=-b_3^1,\quad b_3^3={7 d^2\over 4}-3 d (j+2 \Delta_J +2)+(j+2 \Delta_J )^2+6 j-\nu ^2+12 \Delta_J  \cr
b_3^4&=-\left[ {d^3 (j+2 \Delta_J -4)-4 d^2 \left((j+2 \Delta_J )^2-7\right)+4 d \left(j^3+j^2 (6 \Delta_J +4)+j \left(\nu ^2+4 (\Delta_J  (3 \Delta_J+4)-1)\right)  \right)\over 8(d-1) }+\right. \cr
&\quad \left. + {4 d\left(2 \nu ^2 (\Delta_J -2)+8 (\Delta_J -1) (\Delta_J+1) (\Delta_J +2)\right)+16 \left(4 j+3 \nu ^2+8 \Delta_J \right)-16 (j+2 \Delta_J )^3 \over 8(d-1)} \right]  }}

\eqn\betakdeffour{\eqalign{b_4^1&= -1,\qquad b_4^2=1,\qquad b_4^3=-2(d-3-j-2\Delta_J)\cr
b_4^4&={d^2-4 d (3 j+6 \Delta_J -2)+4 \left(3 j^2+6 j (2 \Delta_J +1)+\nu ^2+12 \Delta_J^2+12 \Delta_J -8\right)\over 4 (d-1)}\,.
}}

\appendix{G}{Polynomial Terms in the Mellin Amplitude}

\noindent 

Here we consider how do polynomial corrections to the Mellin amplitude affect the bulk phase shift. Due to the chaos bound (or Rindler positivity), we will restrict ourselves to considering terms of the form 
\eqn\polymel{
M_{pol}(s,t)=M_{m n} s^m t^n \  {\rm with} \ m,n\leq 2 \ ,
}
where $M_0$ is some arbitrary coefficient. We will see that polynomial terms in the Mellin amplitude yield $\delta$-function-terms in the phase shift as expected.
For simplicity, we will focus on the scalar correlator $\vev{\phi\phi\psi\psi}$ and closely follow the Appendix C of \CostaCB. The case of spinning operators is a straightforward generalization which will not alter the main result. 

Starting from the expression for the reduced amplitude $\AA(u,v)$ in the Mellin representation
\eqn\AMdef{\AA(u,v)=\int {dt \,ds\over (4\pi i)^2}\, M(s, t) u^{t\over 2} v^{-{(s+t)\over 2}} \Gamma\left({2\Delta_\phi-t\over 2}\right)  \Gamma\left({2\Delta_\psi-t\over 2}\right) \Gamma^2\left(-{s\over 2}\right) \Gamma^2\left({t+s\over 2}\right) }
we rotate
\eqn\vrot{v^{-{(s+t)\over 2}} \rightarrow v^{-{(s+t)\over 2}}  e^{-i\pi(s+t)}\,,}
and use the following approximation for $s=ix$, with $x\rightarrow +\infty$,
\eqn\gammaaprox{\Gamma(a+i{x\over 2})   \Gamma(b-i{x\over 2})\simeq 2\pi e^{i{\pi\over 2}(a-b)} \left({x\over 2}\right)^{a+b-1} e^{-{\pi\over 2}x}\,, }
to express \AMdef\ as:
\eqn\AAone{\AA(u,v)\simeq \int_{-i\infty}^{i\infty}  {dt\over 4 i} u^{t\over 2}v^{-t\over 2} \Gamma\left({2\Delta_\phi-t\over 2}\right)\Gamma\left({2\Delta_\psi-t\over 2}\right) e^{-i\pi{t\over 2}} M_{m,n} \int_0^\infty dx\, (ix)^m\, t^n\, \left({x\over 2}\right)^{t-2} v^{-{i\over 2} x}\,. }
Further approximating $v\simeq 1-2\sigma\cosh{\rho}$ and recalling that $u=\sigma^2$, we proceed to evaluate the integral 
\eqn\AAtwo{\AA\simeq \int_{-i\infty}^{i\infty}  {dt\over 4 i} \sigma^t  \Gamma\left({2\Delta_\phi-t\over 2}\right)\Gamma\left({2\Delta_\psi-t\over 2}\right) e^{-i\pi{t\over 2}+im{\pi\over 2}} \,2^m\, t^n M_{m,n} \int_0^\infty dx  \left({x\over 2}\right)^{m+t-2} \, e^{ix\sigma\cosh{\rho}} \,,}
which leads to
\eqn\AAthree{\AA\simeq \int_{-i\infty}^{i\infty}  {dt\over 4 i} \sigma^{1-m}  M_{m,n} \Gamma\left({2\Delta_\phi-t\over 2}\right)\Gamma\left({2\Delta_\psi-t\over 2}\right) e^{-i {\pi\over 2}} (-1)^m \,2^{m+1}\, t^n  {\Gamma(m+t-1)\over (2\cosh{\rho})^{t+m-1}} \,. }
To simplify the presentation, we henceforth set $n=0$. We will discuss the $n\neq 0$ case separately at the end.

We are primarily interested in the impact parameter representation of these terms, we will thus procceed by using the identity:
\eqn\identitycosh{\eqalign{{1\over (2 \cosh{\rho})^z} = {\pi^{d-2\over 2}\over 2}\int_{-\infty}^\infty d\nu\, g(\nu,z) \Omega_{i\nu}(\rho) \cr
g(\nu,z)= {\Gamma\left({z-{d-2\over 2}+i\nu\over 2}\right)\Gamma\left({z-{d-2\over 2}-i\nu\over 2}\right) \over\Gamma(z)}\,, 
}}
to rewrite \AAthree\ as:
\eqn\AAfour{\eqalign{&\AA_{n=0}\simeq (-1)^{m+1} 2^{m-2} \pi^{d-2\over 2}\, (2\pi i ) M_{m,0}\,\sigma^{1-m} \,\,\times\,\cr
&\int d\nu \int_{-i\infty}^{i\infty} {dt\over 2\pi i}  \Gamma\left({t+m-1-{d-2\over 2}-i\nu\over 2}\right) \Gamma\left({t+m-1-{d-2\over 2}+i\nu\over 2}\right) \Gamma\left({2\Delta_\phi-t\over 2}\right)\Gamma\left({2\Delta_\psi-t\over 2}\right) \,.}}
Let us recall an integral representation of the hypergeometric function, i.e.,
\eqn\inthyper{\,  {\Gamma(a)\Gamma(b)\Gamma(c-a)\Gamma(c-b)\over\Gamma(c)} \, _2F_1[a,b,c,z]=\int_{-i\infty}^{i\infty} {dt\over 2\pi i} \Gamma(t)\Gamma(c-a-b+t)\Gamma(a-t)\Gamma(b-t) \,(1-z)^{-t}\,.}
Changing variables in \AAfour\ according to $t\rightarrow {1 \over 2} (2\Delta_\phi-t)$ allows us to make use of the identity above by setting:
\eqn\abcdef{a={-{d\over 2}+2\Delta_\phi+m- i\nu\over 2},\quad b={-{d\over 2}+2\Delta_\phi+m-i\nu\over 2},\quad c=\Delta_\phi+\Delta_\psi+m-{d\over 2}\,\,.}
The result is 
\eqn\AAfive{\eqalign{\AA_{n=0}&= (-1)^{m+1} 2^{m-2} \pi^{d-2\over 2}\,(2\pi i ) M_{m,0}\,{\Gamma(2\Delta_\phi+m-1)\Gamma(2\Delta_\psi+m-1)\over \Gamma(\Delta_\phi+\Delta_\psi+m-2)}\times\cr
&\times\sigma^{1-m}  \int d\nu\, g(2\Delta_\phi+m-1,\nu)g(2\Delta_\psi+m-1,\nu)\,\Omega_{i\nu}(\rho)\,}}
where we used the fact that $\,_2F_1[a,b,c,0]=1$.

To obtain the impact parameter representation of \AAfive\ we use the following identity
\eqn\ideimpact{ {g(2\Delta_\phi+m-1,\nu)g(2\Delta_\psi+m-1,\nu)\Omega_{i\nu}(e\cdot \be)\over (-x^2)^{\Delta_\phi+{m-1\over 2}} (-x^2)^{\Delta_\psi+{m-1\over 2}} }={4\over \pi^{d-2} } \int_M dp d\bp {e^{2x\cdot p} e^{2\bx\cdot\bp} \Omega_{i\nu}({p\over\sqrt{-p^2}}, {\bp\over \sqrt{-\bp^2}}) \over (-p^2)^{d-2\Delta_1-m+1\over 2} (-\bp^2)^{d-2\Delta_3-m+1\over 2}} \,,  }
which leads to:
\eqn\AAsix{\eqalign{ &{\AA_{n=0}(x,\bx)\over  (-x^2)^{\Delta_\phi+{m-1\over 2}} (-x^2)^{\Delta_\psi+{m-1\over 2}}} =  (-1)^{m+1} 2^{m} \pi^{-d+2\over 2}\,(2\pi i ) M_{m,0}\, {\Gamma(2\Delta_\phi+m-1)\Gamma(2\Delta_\psi+m-1)\over \Gamma(\Delta_\phi+\Delta_\psi+m-2)} \times \cr
&\times \int_M dp d\bp  { e^{2x\cdot p} e^{2\bx\cdot\bp} \over  (-p^2)^{d-2\Delta_\psi\over 2} (-\bp^2)^{d-2\Delta_\psi\over 2} } \,\, s^{m-1} \,\, \int_{-\infty}^\infty d\nu \, \Omega_{i\nu}\left({p\over\sqrt{-p^2}}, {\bp\over \sqrt{-\bp^2}}\right) \, . }}
From eq. \AAfive\ we can read off the impact parameter representation of the reduced correlation function:
\eqn\Bone{
\delta_{n=0}(S,L) \sim M_{m,0} S^{m-1} \delta_{H^{d-1}} \left( L  \right) \, \ .
}

We see that the addition of an $s^m$ polynomial term in the Mellin amplitude, produces a phase shift which is proportional to a $\delta$-function as expected on general grounds \HeemskerkPN. Using the identification between anomalous dimensions of double trace operators in the s-channel and the phase shift we see that the $\delta$-function in the impact parameter space corresponds to $h=\bar{h}$, or equivalently to the s-channel exchange of spin zero double trace operators. 

Let us now move onto the case $n\neq 0$. Starting from \AAthree\ for $n=0$ and using the identity 
\eqn\Gammaid{\Gamma\left({2\Delta_\phi-t\over 2}\right)=\left({\Delta_\phi-1-{t\over 2}}\right) \,\Gamma\left({2\Delta_\phi-2-t\over 2}\right)} 
allows us to simply express the result for $n=1$ as:
\eqn\Atonea{\AA_{n=1}^{\Delta_\phi ,\Delta_\psi}=2 \Delta_\phi \,\AA_{n=0}^{\Delta_\phi,\Delta_\psi}- 2 \AA_{n=0}^{\Delta_\phi+1,\Delta_\psi}\,.}
To obtain the impact parameter representation of $\AA_{n=1}$, we use the $\Gamma$ function identities to relate $g(2\Delta_1+m-1,\nu)$ to $g(2\Delta_1 +m,\nu)$. This yields:
\eqn\Atoneb{\eqalign{ &{\AA_{n=1}(x,\bx)\over  (-x^2)^{\Delta_\phi+{m-1\over 2}} (-x^2)^{\Delta_\psi+{m-1\over 2}}} =  (-1)^{m+1} 2^{m} \pi^{-d+2\over 2}\,(2\pi i ) M_{m,1}\, {\Gamma(2\Delta_\phi+m-1)\Gamma(2\Delta_\psi+m-1)\over 2\Gamma(\Delta_\phi+\Delta_\psi+m-1)} \times \cr
&\qquad\qquad \times \int_M dp d\bp  { e^{2x\cdot p} e^{2\bx\cdot\bp} \over  (-p^2)^{d-2\Delta_\phi\over 2} (-\bp^2)^{d-2\Delta_\psi\over 2} } \,\, s^{m-1} \,\, \int_{-\infty}^\infty d\nu \, (\hat{a}-\nu^2)\Omega_{i\nu}\left(w,\bw\right) \, , }}
with
\eqn\hatadef{\hat{a}= 4\Delta_\phi(\Delta_\phi+\Delta_\psi+m-2)-\left(2\Delta_\phi+m-1-{d-2\over 2}\right)^2\,.}

Eq. \Atoneb\ leads to the following impact parameter representation
\eqn\Btone{
\delta_{n=1}(S,L) \propto M_{m,1} \, S^{m-1} \, \left[\hat{a}+\left({d-2\over 2}\right)^2+\nabla^2_H\right] \,\delta_{H}(L)\,
}
where to arrive at \Btone\ we used the differential equation $\Omega_{i\nu}$ satisfies. Here $\nabla^2_H$ represent the Laplacian on the Hyperbolic space $H_{d-1}$. One can easily work out higher powers of $n$ the same way. They all lead to $\delta$-function-like terms, as expected.

\appendix{H}{Diagonalization of the Bulk Phase Shift.}

\noindent In this appendix we will address the implications of imposing positivity on the phase shift $\delta^{mn}(S, L)$. Our focus will be on a large-N theory with an infinite gap in the spectrum. As discussed in section 3.4, to perform the impact parameter integral \phaseshiftvec\ one must express it in terms of the propagator in hyperbolic space, $\Pi_{i\nu+d/2-1}$, defined in \propagator, and close the contour below the real axis. In the infinite gap limit, the integral picks up a pole from the contribution of the stress-tensor operator for $\nu=-i{d\over 2}$. The result for the phase shift is then proportional to the matrix 
\eqn\Cmatrixdef{\sum_{k=1}^4 \,\beta_k(\nu) \hat{\DD}_k^{mn}\Pi_{d-1}(L)\,,}
and positivity requires that its eigenvalues are positive definite. 

To perform the diagonalization, one needs explicit expressions for \Cmatrixdef, which can be obtained by differentiation. 
In particular, we have that:
\eqn\evaldcd{\eqalign{ \hat\DD_3^{mn}\Pi_{d-1} &=-\dot{\Pi}_{d-1} \csch{L} \left({p^m\bp^n+\bp^m p^n\over\sqrt{p^2\bp^2} }\right)-\coth{L}\,\dot{\Pi}_{d-1}  {2 p^m p^n\over p^2}\cr
p^2\p^m\p^n \Pi_{d-1}&=-\csch^2{L} \left({\dot\Pi}_{d-1} \,\coth{L}-\ddot{\Pi}_{d-1} \right)\, {\bp^m\bp^n\over \bp^2} -\eta^{mn}\,\dot{\Pi}_{i\nu}\coth{L} \cr
&+\csch{L}\, \left(\coth{L}\ddot{\Pi}_{d-1}-\csch^2{L} \dot{\Pi}_{i\nu}\right) \left({p^m\bp^n+\bp^m p^n\over\sqrt{p^2\bp^2} }\right) \cr
&+ \coth{L} \left( \ddot{\Pi}_{i\nu} \coth{L}+\dot{\Pi}_{i\nu} (2-\csch^2{L})  \right) {p^m p^n\over p^2}
}}
With the help of \evaldcd\ the phase shift can be expressed as follows:
\eqn\phsha{\eqalign{&\sum_{k=1}^4 \,\beta_k(\nu) \hat{\DD}_k^{mn}\Pi_{d-1}(L)=\eta^{mn} A_1+{p^m p^n\over p^2} A_2+A_3\left({p^m\bp^n+\bp^m p^n\over\sqrt{p^2\bp^2} }\right)+A_4 {\bp^m\bp^n\over \bp^2} 
}}
with:
\eqn\Asdef{\eqalign{A_1&=\beta_1\Pi_{d-1}-\beta_4 \left(\coth{L}\, \dot{\Pi}_{d-1}-\Pi_{d-1}\right)\cr
A_2&=-\beta_1\Pi_{d-1}+\beta_2\Pi_{d-1}-2\beta_3\, \coth{L}\dot{\Pi}_{d-1}+\beta_4\left(\coth^2{L}\, \ddot{\Pi}_{d-1}-\coth{L}\,\csch^2{L}\dot{\Pi}_{d-1}-\Pi_{d-1}\right) \cr
A_3&=-\beta_3\, \csch{L}\, \dot{\Pi}_{d-1}+\beta_4\, \csch{L}\,\left(\coth{L} \ddot{\Pi}_{d-1} -\coth^2{L}\, \dot{\Pi}_{d-1}\right)\cr
A_4&=\beta_4 \, \csch^2{L} \,\left(\ddot{\Pi}_{d-1} -\coth{L}\,\dot{\Pi}_{d-1} \right)
 }}
Without loss of generality, we set $\bp^0=1,\,\, \vec{\bp}=0$ and use $\cosh{L}=-{p\cdot\bp\over\sqrt{p^2\bp^2}}$ to express $(p^0,\vec{p})$ as:
\eqn\pnotpvec{p^0=\sqrt{-p^2}\,\cosh{L},\qquad p^1=\sqrt{-p^2}\, \sinh{L},\qquad p^2=p^3=\cdots=p^{d-1}=0\,,}
The phase shift is completely determined in terms of the hyperbolic space coordinate $L$,
\eqn\phshb{\eqalign{&\sum_{k=1}^4 \,\beta_k(\nu) \hat{\DD}_k^{mn}\Pi_{d-1}(L)\cr
&=\pmatrix{-A_1-A_2\, \cosh^2{L}+2\,A_3\, \cosh{L}-A_4   & -A_2 \sinh{L}\,\cosh{L} + A_3 \,\sinh{L}& 0 &\cdots &0\cr
-A_2 \sinh{ L}\,\cosh{L}+ A_3 \,\sinh{L}  &  A_1-A_2 \,\sinh^2{L} &0 & \cdots & 0 \cr
0& \cdots& A_1&0 &0 \cr
\cdots\cr
0 & 0 &\cdots & 0&A_1}
}}
and its eigenvalues are:
\eqn\eigenvalues{\eqalign{\lambda_{0,1}&={1\over 2} \left(a_1\pm \sqrt{a_1^2-4 (a_2-a_3^2)}\right),\quad \lambda_i=A_1(L),\quad i=2,\cdots, d\cr
&a_1=-A_4-A_2 \cosh{2 L} +2 A_3 \cosh{L}, \cr
&a_2=(A_1-A_2 \sinh^2{L})(-A_1-A_2\cosh^2{L}+2A_3\cosh{L}-A_4),\cr
& a_3=(A_3-{A_2\over 2} \cosh{L})\, \sinh{L}
}}

We proceed by analyzing the bahaviour of the eigenvalues for small and large impact parameters. To do that, it is useful to recall that $\Pi_{d-1}$ behaves as:
\eqn\Pileading{\Pi_{d-1}(L)=\left\{ \eqalign{2^{-d}\pi^{1-d/2}{ d(d-2)\Gamma(d-3)\over \Gamma(d/2+1)} L^{-d+3}+\cdots,&\quad L<<1\cr
u^d\left(\pi^{1-d/2}{ \Gamma(d-1)\over d\Gamma({d\over 2}) }  {1\over u}+ \pi^{1-d/2} {(d-1)\Gamma(d)\over 4\Gamma(d/2+2)}u+\cdots\right) ,\quad u=e^{-L}&\quad L>>1   }\right.
} 

When the impact parameter is small $L<<1$, then the leading behaviour of the eigenvalues turns out to be:
\eqn\eigenLsmall{\lambda_{1,2}\propto -c_0{\beta_4 (d-2)\over L^{d-1}},\qquad \lambda_i\propto c_0{\beta_4 \over L^{d-1}}}
where $c_0=2^{-d}\pi^{1-d/2}{ d \Gamma(d-1)\over \Gamma(d/2+1)}$ is equal to the proportionality coefficient of the leading term in the small $L$ expansion and is positive definite.
Causality then requires that $\beta_4=0$ for both conserved and non-conserved vector operators $J$. 
Setting $\beta_4=0$ is enough to guarantee positivity of the phase shift at small impact parameters for unitary theories satisfying $\Delta_J\geq d-1$, when $n_s\geq 0$.

It is interesting to check whether the positivity of the phase shift reduces to the Hofman-Maldacena bounds for large values of the impact parameter. The leading behaviour of the eigenvalues  \eigenvalues\ for large impact parameters $L$, or equivalently, $0<u <<1$ is:
\eqn\eigenLlarge{\lambda_{1,2} =c_1 u^{d-3} \left[(\beta_1-\beta_2-d(d-2)\beta_4)+\OO(u^2)\right], \quad\lambda_i=c_2 u^{d-5} (\beta_1+d\beta_4)+\cdots \,,}   
where $(c_1,c_2)$ are two positive constants and we have used the fact that $\beta_3=0$ when evaluated at the stress-tensor Regge pole $\nu=-i{d\over 2}$. Positivity of the phase shift leads to a single condition, namely that,
\eqn\largeLcona{\beta_1-\beta_2-d(d-2)\beta_4\geq 0\,.}
For conserved currents, \largeLcona\ simply reduces to $n_f\geq 0$ for $n_s\in R$. 
For non-conserved vector operators, the result is the inequality obtained in deep inelastic scattering (eq. (4.29) of \KomargodskiGCI ) but for any real $n_s$.\foot{$(n_s,n_f)$ are related to the $(a_2,\,a_3)$ of \KomargodskiGCI\ as follows:
\eqn\aitonsf{a_2= {n_s\over (d-1)(d-2)},\qquad a_3={4 n_d\over d-1}\,. }}

In search of the energy-flux constraints we repeat the analysis above, this time fixing $p^m=(1,\vec{0})$ and solving for $\bp^m$ through 
$\cosh{L}=-{p\cdot\bp\over\sqrt{p^2 \bp^2}}$. The diagonalization problem is exactly the same as before, with the replacement: $A_2 \rightarrow A_4$ and $A_4\rightarrow A_2$. For large impact parameters the eigenvalues read:
\eqn\eigenLlargeb{\lambda_1\propto -d\,\beta_2\, u^{d-1},\quad \lambda_2\propto d(\beta_1-d(d-2)\beta)\, u^{d-1},\quad \lambda_i\propto (\beta_1+d\beta_4)\, u^{d-1}\,,}
with positive proportionality constants.
Substituting (expressions for betas) into \eigenLlargeb\ we obtain the energy flux constraints for both conserved and non-conserved vector operators. In this way the usual Hofman-Maldacena constraints are reproduced.

\listrefs

\bye